\documentclass[11pt,3p,authoryear]{elsarticle}

\usepackage{amsmath}
\usepackage{amssymb}
\usepackage{dsfont}
\usepackage{xcolor}
\usepackage{hyperref}

\journal{arXiv}
\DeclareMathOperator*{\argmin}{arg\,min}
\def\ens{\mathrm{ens}}
\def\WIS{\mathrm{WIS}}
\def\rWIS{\mathrm{rWIS}}

\begin{document}

\begin{frontmatter}

\title{Comparing trained and untrained probabilistic ensemble forecasts of COVID-19 cases and deaths in the United States}

\author[umass]{Evan L. Ray\corref{cor}}
\cortext[cor]{Corresponding author}
\ead{elray@umass.edu}

\author[cmu]{Logan C. Brooks}

\author[usc]{Jacob Bien}

\author[cdc]{Matthew Biggerstaff}

\author[lshtm]{Nikos I. Bosse}

\author[kit,hits]{Johannes Bracher}

\author[umass]{Estee Y. Cramer}

\author[lshtm]{Sebastian Funk}

\author[umass]{Aaron Gerding}

\author[cdc]{Michael A. Johansson}

\author[cmu]{Aaron Rumack}

\author[umass]{Yijin Wang}

\author[umass]{Martha Zorn}

\author[cmu]{Ryan J. Tibshirani}

\author[umass]{Nicholas G. Reich}

\address[umass]{School of Public Health and Health Sciences, University of Massachusetts Amherst}
\address[cmu]{Machine Learning Department, Carnegie Mellon University}
\address[usc]{Department of Data Sciences and Operations, University of Southern California}
\address[cdc]{COVID-19 Response, U.S. Centers for Disease Control and Prevention}
\address[lshtm]{London School of Hygiene \& Tropical Medicine}
\address[kit]{Chair of Statistical Methods and Econometrics, Karlsruhe Institute of Technology}
\address[hits]{Computational Statistics Group, Heidelberg Institute for Theoretical Studies}


\begin{abstract}
The U.S. COVID-19 Forecast Hub aggregates forecasts of the short-term burden of
COVID-19 in the United States from many contributing teams. We study methods for
building an ensemble that combines forecasts from these teams. These experiments
have informed the ensemble methods used by the Hub. To be most useful to policy
makers, ensemble forecasts must have stable performance in the presence of two
key characteristics of the component forecasts: (1) occasional misalignment with
the reported data, and (2) instability in the relative performance of component
forecasters over time. Our results indicate that in the presence of these
challenges, an untrained and robust approach to ensembling using an equally
weighted median of all component forecasts is a good choice to support public
health decision makers. In settings where some contributing forecasters have a
stable record of good performance, trained ensembles that give those forecasters
higher weight can also be helpful.
\end{abstract}


\end{frontmatter}

\section{Introduction}

Accurate short-term forecasts of infectious disease indicators (i.e., disease surveillance signals) can inform public health decision-making and outbreak response activities such as non-pharmaceutical interventions, site selection for clinical trials of pharmaceutical treatments, or the distribution of limited health care resources \citep{wallinga_optimizing_2010, lipsitch_improving_2011, dean_ensemble_2020}.
Epidemic forecasts have been incorporated into public health decision-making in a wide variety of situations, including outbreaks of dengue fever in Brazil, Vietnam, and Thailand \citep{Rlowe2016,colon-gonzalez_probabilistic_2021,reich_challenges_2016} and
influenza in the U.S. \citep{mcgowan_collaborative_2019}.

These efforts frequently use ensemble forecasts that combine predictions from many models.
In a wide array of fields, ensemble approaches have provided consistent improvements in accuracy and robustness relative to stand-alone forecasts \citep{gneiting_weather_2005, polikar_ensemble_2006}.
The usefulness of ensemble forecasts has also been demonstrated repeatedly in multiple infectious disease settings, including influenza, Ebola, dengue, RSV, and others \citep{yamana_superensemble_2016, viboud_rapidd_2018, mcgowan_collaborative_2019, johansson_open_2019, reis_superensemble_2019, reich_accuracy_2019}.
In light of this record of strong performance, ensembles are natural candidates for forecasts used as an input to high-stakes public health decision-making processes.

This paper describes ensemble modeling efforts at the U.S.\ COVID-19 Forecast Hub (\url{https://covid19forecasthub.org/}, hereafter the ``U.S.\ Hub''), from spring 2020 through spring 2022.
Starting in April 2020, the U.S.\ Hub created ensemble forecasts of reported incident deaths one through four weeks ahead in the 50 states, Washington, D.C., and 6 territories as well as at the national level by combining forecasts submitted by a large and variable number of contributing teams using different modeling techniques and data sources.
In July 2020, forecasts of incident reported COVID-19 cases were added.
Of note, the U.S.\ Hub produces \textit{probabilistic} forecasts in which uncertainty about future disease incidence is quantified through the specification of a predictive distribution that is represented by a collection of predictive quantiles.
Since the inception of the U.S.\ Hub, these ensemble forecasts have been provided to the U.S.\ Centers for Disease Control and Prevention (CDC) and have been the basis of official CDC forecasting communications \citep{cdc_covid_modeling}.

\subsection{Related literature}

A wide variety of standalone methodological approaches have been shown to be able to make forecasts of short-term outbreak activity that are more accurate than naive baseline forecasts in various epidemiological settings.
Some approaches have used existing statistical frameworks to model associations between outcomes of interest and known or hypothesized drivers of outbreaks, such as recent trends in transmission or environmental factors.
To cite just a few examples, methods used include multiscale probabilistic Bayesian random walk models \citep{osthus_multiscale_2021}, Gaussian processes \citep{johnson_phenomenological_2018}, kernel conditional density estimation \citep{ray_infectious_2017,brooks_nonmechanistic_2018}, and generalized additive models \citep{lauer_prospective_2018}.
Other models have an implicit or explicit representation of a disease transmission process, such as variations on the susceptible-infectious-recovered (SIR) compartmental model \citep{shaman_forecasting_2012,lega_data-driven_2016,osthus_forecasting_2017,pei_forecasting_2018,turtle_accurate_2021}. Aspects of these modeling frameworks can also be combined, for instance using time series methods to build models that have a compartmental structure or incorporate key epidemiological parameters such as the effective reproduction number $R_t$, or models that use a time series process to capture systematic deviations from a compartmental core \citep{bartolucci_multivariate_2021, agosto_monitoring_2021, osthus_dynamic_2019}.

There is a large literature on ensemble forecasting, but of particular relevance to the present work is the research on combining, calibrating and evaluating distributional forecasts \citep{gneiting_strictly_2007,gneiting_probabilistic_2007,ranjan_combining_2010,claeskens_forecast_2016}.
We note that prior work on forecast combination has mostly focused on combining forecasts represented as probability densities or probability mass functions rather than forecasts parameterized by a set of discrete quantile levels, which is the format of the forecasts in the present study.
However, in psychological studies there is a long history of combining quantiles from multiple distributions as a mechanism for summarizing distributions of response times, error rates, and similar quantities across many subjects \citep{vincent1912functionsOfVibrissae, ratcliff1979vincentization}.
More recently, this approach has also been used to combine probabilistic assessments from multiple subject matter experts or statistical models in fields such as security threat detection and economic forecasting \citep{hora2013medianAggregationDistribution, lichtendahl2013betterAveProbQuant, gaba2017combiningIntervalForecasts,
busetti2017quantileAggregationDensityForecasts}.
In the context of infectious disease forecasting, \cite{bracher_preregistered_covid_de_pl_2021} conducted a similar but less extensive analysis to the one presented here using data from a related forecast hub focusing on Germany and Poland. \cite{taylor2021combiningForecastsCOVID} recently explored several approaches to constructing quantile-based ensemble forecasts of cumulative deaths due to COVID-19 using the data from the U.S. Hub, although they did not generate ensemble forecasts in real time or appear to have used the specific versions of ground truth data that were available for construncting ensembles in real time.

As was mentioned earlier, ensemble forecasts have also been used in a variety of other applications in real-time forecasting of infectious diseases, often with seasonal transmission dynamics where many years of training data are available \citep{yamana_superensemble_2016,reich_accuracy_2019,reis_superensemble_2019,colon-gonzalez_probabilistic_2021}.
In such applications, simple combination approaches have generally been favored over complex ones, with equal-weighted approaches often performing similarly to trained approaches that assign weights to different models based on past performance \citep{ray_feature_weighted_ensembles, bracher_preregistered_covid_de_pl_2021}.
These results align with theory suggesting that the uncertainty in weight estimation can pose a challenge in applications with a low signal-to-noise ratio \citep{claeskens_forecast_2016}.

\subsection{Contributions of this article}


This paper is focused on explaining the careful considerations that have gone into building a relatively simple ``production'' ensemble model for a difficult, high-stakes, real-time prediction problem: forecasting COVID-19 cases and deaths in the United States, to support public health decision-making.
We do not empirically investigate the performance of complex forecast combination strategies from the online prediction literature, which generally require richer and larger training data sets.

\begin{figure}
\centering
\includegraphics[width=\textwidth]{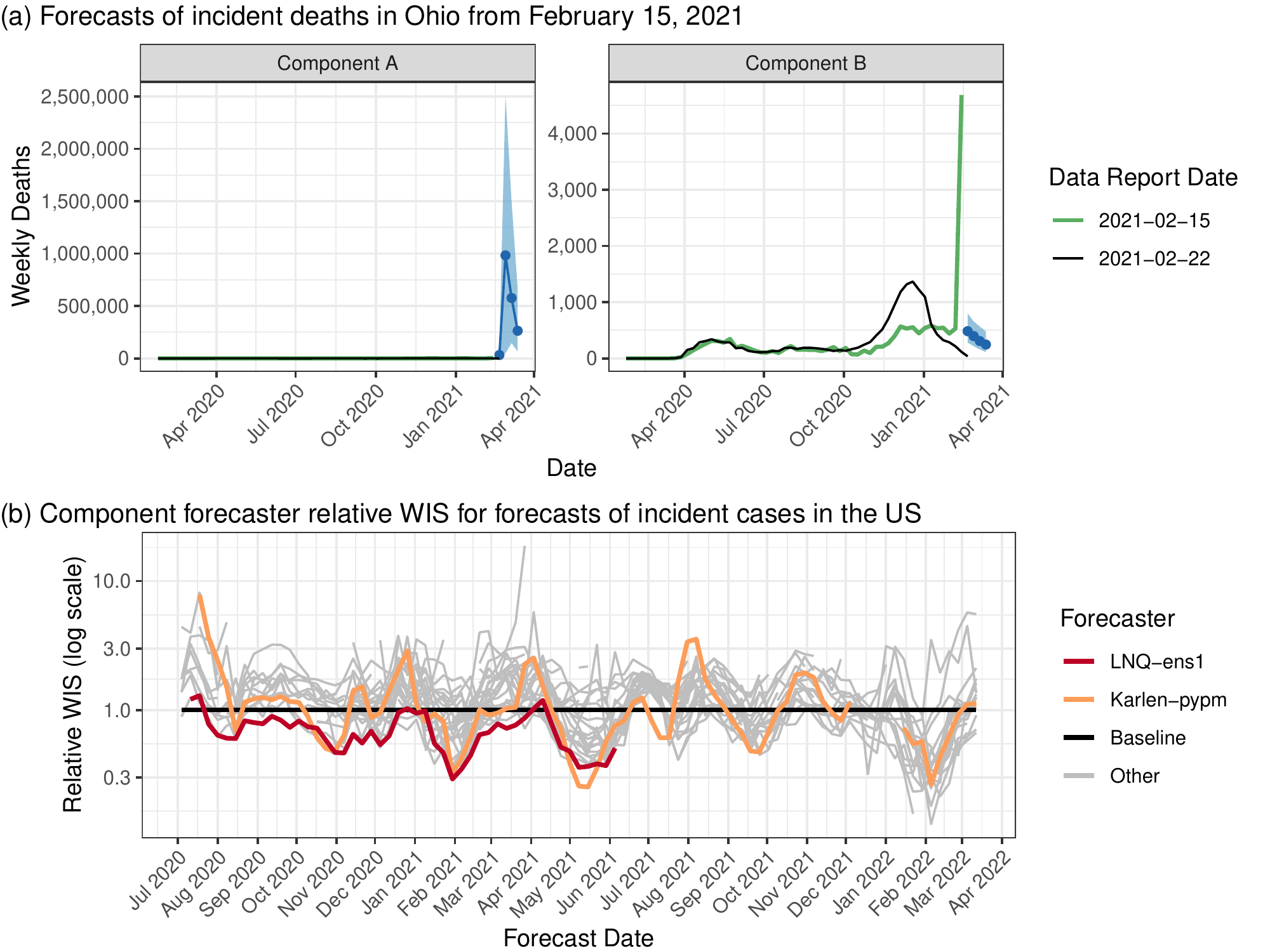}
\caption{(a) Predictive medians and 95\% prediction intervals for incident deaths in Ohio generated on February 15, 2021 by two example component forecasters. The vertical axis scale is different in each facet, reflecting differences across several orders of magnitude in forecasts from different forecasters; the reference data are the same in each plot. The data that were available as of Monday, February 15, 2021 included a large spike in reported deaths that had been redistributed into the history of the time series in the version of the data available as of Monday, February 22, 2021. In this panel, forecaster names are anonymized to avoid calling undue attention to individual teams; similar behavior has been exhibited by many forecasters. (b) Illustration of the relative weighted interval score (WIS, defined in Section \ref{subsec:methods_evaluation}) of component forecasters over time; lower scores indicate better performance. Each point summarizes the skill of forecasts made at a given date for the one through four week ahead forecasts of incident cases across all state-level locations. }
\label{fig:ensemble_motivations}
\end{figure}

The goal of the U.S.\ Hub in developing an operational ensemble was to produce forecasts of the short-term trajectory of COVID-19 that had good performance on average and stable performance across time and different locations.
Real-time forecasting for an emerging pathogen in an open, collaborative setting introduces important challenges that an ensemble combination method must be able to handle.
First, teams occasionally submitted outlying component forecasts due to software errors, incorrect model assumptions, or a lack of robustness to input data anomalies (Figure~\ref{fig:ensemble_motivations} (a), Supplemental Figures 1 and 2).
Second, some component models were generally better than others, but the relative performance of different models was somewhat unstable across time (Figure~\ref{fig:ensemble_motivations} (b), Supplemental Figures 3 and 4).
In particular, some forecasters alternated between being among the best-performing models and among the worst-performing models within a span of a few weeks, which introduced a challenge for ensemble methods that attempted to weight component forecasters based on their past performance.
In this manuscript, we explore and compare variations on ensemble methods designed to address these challenges and produce real-time forecasts that are as accurate as possible to support public health decision-makers.

We give detailed results from experiments that were run concurrently with the weekly releases of ensemble forecasts from the start of the U.S. Hub in 2020 through the spring of 2022, as documented in preliminary reports \citep{2020BrooksRay_covid_ensemble_blog, 2021RayBrooks_covid_ensemble_blog}.
These experiments provided the evidence for decisions (a) to move to a median-based ensemble from one based on means in July 2020; (b) to switch to a trained ensemble for forecasts of deaths in November 2021; and (c) to implement a weight regularization strategy for that trained ensemble starting in January 2022.
In a secondary analysis, we also consider the prospective performance of these methods in the closely related setting of forecasting cases and deaths in Europe, to examine the generalizability of the results from our experiments using data from the U.S.

The following sections document the format and general characteristics of COVID-19 forecasts under consideration, the ensemble approaches studied, and the results of comparing different approaches both during model development and during a prospective evaluation of selected methods.

\section{Methods}

We give an overview of the U.S.\ and European Forecast Hubs and the high-level structure of our experiments in Sections~\ref{subsec:methods_context} through \ref{subsec:methods_evaluation}, and then describe the ensemble methods that we consider in Section~\ref{subsec:methods_ensembles}.

\subsection{Problem context: forecasting short-term COVID-19 burden}
\label{subsec:methods_context}

Starting in April 2020, the U.S.\ Hub collected probabilistic forecasts of the short-term burden of COVID-19 in the U.S.\ at the national, state/territory, and county levels \citep{cramer_united_2021}; a similar effort began
in February 2021 for forecasts of disease burden in 32 European countries \citep{EuropeanCovid19Forecast}.
In this manuscript, we focus on constructing probabilistic ensemble forecasts of weekly counts of reported cases and deaths due to COVID-19 at forecast horizons of one to four weeks for states and territories in the U.S.\ and for countries in Europe.
A maximum horizon of four weeks was set by collaborators at CDC as a horizon at which forecasts would be useful to public health practitioners while maintaining reasonable expectations of a minimum standard of forecast accuracy and reliability.
Probabilistic forecasts were contributed to the Hubs in a quantile-based format by teams in academia, government, and industry.
The Hubs produced ensemble forecasts each week on Monday using forecasts from teams contributing that week.
In the U.S.\ Hub, seven quantile levels were used for forecasts of cases and 23 quantile levels were used for forecasts of deaths; in the European Hub, 23 quantile levels were used for both target variables. 

\begin{figure}
\includegraphics[width=\textwidth]{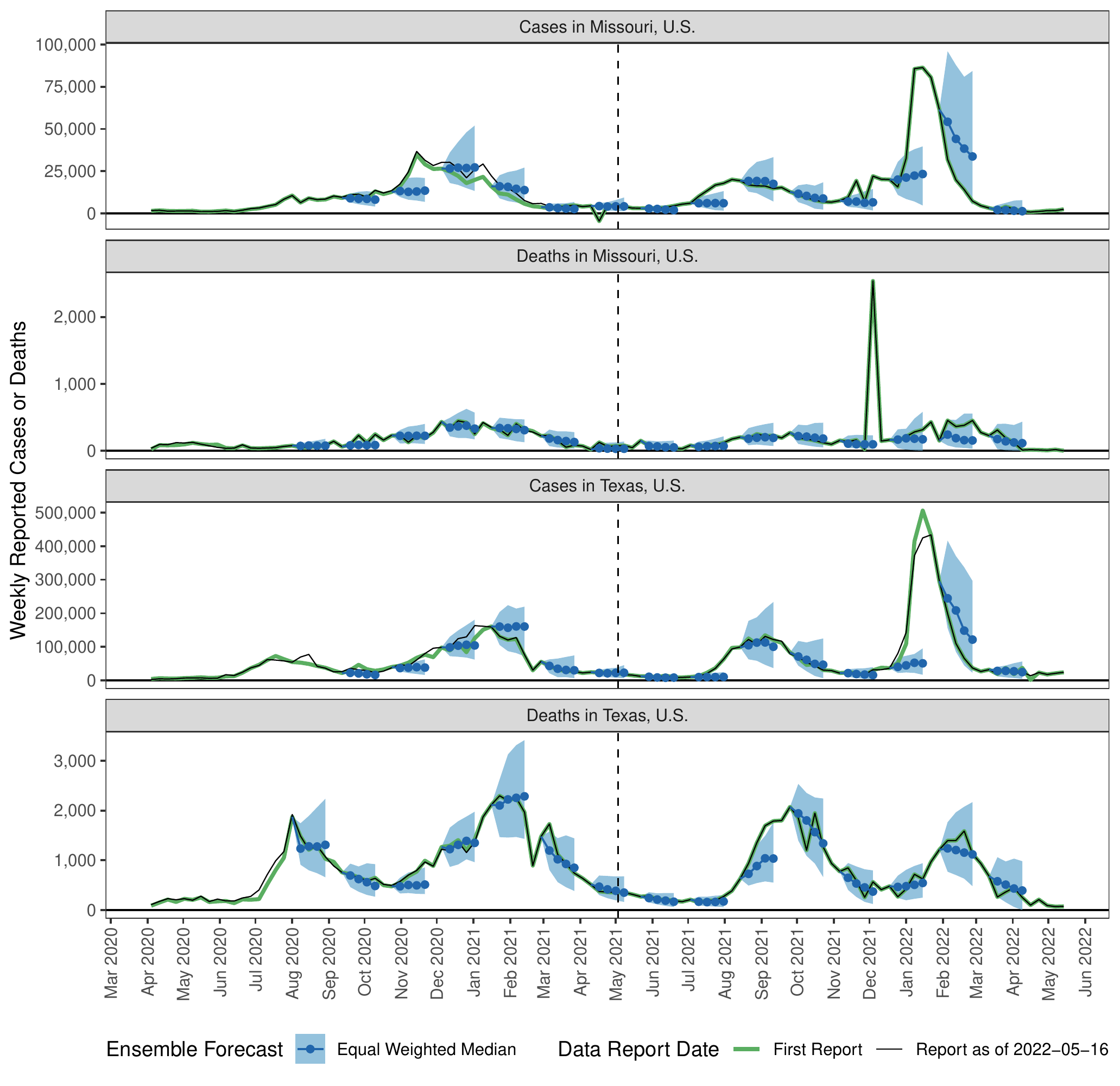}
\caption{Weekly reported cases and deaths and example equally weighted median ensemble forecasts (predictive median and 95\% interval) for selected U.S.\ states. Forecasts were produced each week, but for legibility, only forecasts originating from every sixth week are displayed. Data providers occasionally change initial reports (green lines) leading to revised values (black lines). Vertical dashed lines indicate the start of the prospective ensemble evaluation phase.}
\label{fig:eval_phase_setup}
\end{figure}

Weekly reported cases and deaths were calculated as the difference in cumulative counts on consecutive Saturdays, using data assembled by the Johns Hopkins University Center for Systems Science and Engineering as the ground truth \citep{dong_interactive_2020}.
Due to changes in the definitions of reportable cases and deaths, as well as errors in reporting and backlogs in data collection, there were some instances in which the ground truth data included outlying values, or were revised. Most outliers and revisions were inconsequential, but some were quite substantial in the U.S.\ as well as in Europe (Figure~\ref{fig:eval_phase_setup}).
When fitting retrospective ensembles, we fit to the data that would have been available in real time.
This is critical because the relative performance of different component forecasters may shift dramatically depending on whether originally-reported or subsequently-revised data were used to measure forecast skill.
An ensemble trained using revised data can therefore have a substantial advantage over one trained using only data that were available in real time, and its performance is not a reliable gauge of how that ensemble method might have done in real time.

The U.S.\ Hub conducted extensive ensemble model development in real time from late July 2020 through the end of April 2021, with smaller focused experiments ongoing thereafter. We present results for the model development phase as well as a prospective evaluation of a subset of ensemble methods in the U.S.\ starting with forecasts created on May 3, 2021 and continuing through March 14, 2022. We note that we continued examining a wider range of methods to inform weekly operational forecasting tasks, but the methods that we chose to evaluate prospectively were selected by May 3, 2021, the beginning of the prospective evaluation period, with no alterations thereafter. Real-time submissions of the relative WIS weighted median ensemble described below are on record in the U.S. Hub for the duration of the prospective evaluation period. In one section of the results below, we present a small post hoc exploration of the effects of regularizing the component forecaster weights; these results should be interpreted with caution as they do not constitute a prospective evaluation. To examine how well our findings generalize, we also evaluated the performance of a subset of ensemble methods for prospective forecasts of cases and deaths at the national level for countries in Europe from May 3, 2021 to March 14, 2022.

\subsection{Eligibility criteria}
\label{subsec:methods_eligibility}

In the Forecast Hubs, not all forecasts from contributing models are available for all weeks. For example, forecasters may have started submitting forecasts in different weeks, and some forecasters submitted forecasts for only a subset of locations in one or more weeks.

The ensemble forecast for a particular location and forecast date included all component forecasts with a complete set of predictive quantiles (i.e., 7 predictive quantiles for incident cases, 23 for deaths) for all 4 forecast horizons. Teams were not required to submit forecasts for all locations to be included in the ensemble. Some ensemble methods that we considered require historical forecasts to inform component model selection or weighting; for these methods, at least one prior submission was required. The Forecast Hubs enforced other validation criteria, including that predictions of incident cases and deaths were non-negative and predictive quantiles were properly ordered across quantile levels.

\subsection{Notation}
\label{subsec:methods_notation}

We denote the reported number of cases or deaths for location $l$ and week $t$ by $y_{l,t}$. A single predictive quantile from component forecaster $m$ is denoted by $q^m_{l,s,t,k}$, where $s$ indexes the week the forecast was created, $t$ indexes the target week of the forecast, and $k$ indexes the quantile level. The forecast horizon is the difference between the target date $t$ and the forecast date $s$. There are a total of $K = 7$ quantile levels for forecasts of cases in the U.S., and $K = 23$ quantile levels otherwise. The quantile levels are denoted by $\tau_k$ (e.g., if $\tau_k = 0.5$ then $q^m_{l,s,t,k}$ is a predictive median). We collect the full set of predictive quantiles for a single model, location, forecast date, and target date in the vector $q^m_{l,s,t,1:K}$. We denote the total number of available forecasters by $M$; this changes for different locations and weeks, but we suppress that in the notation.

\subsection{Baseline forecaster}
\label{subsec:methods_baseline}

In the results below, many comparisons are made with reference to an epidemiologically naive baseline forecaster that projects forward the most recent observed value with growing uncertainty at larger horizons. This baseline forecaster was a random walk model on weekly counts of cases or deaths, with $Y_{l,t} \mid Y_{l,t-1} = Y_{l,t-1} + \varepsilon_{l,t}$. The model used a non-parametric estimate of the distribution of the innovations $\varepsilon_{l,t}$ based on the observed differences in weekly counts $d_{l,s} = y_{l,s} - y_{l,s-1}$ over all past weeks $s$ for the specified location $l$.
Predictive quantiles were based on the quantiles of the collection of these differences and their negations, using the default method for calculating quantiles in R.
The inclusion of negative differences ensured that the predictive distributions were symmetric and the predictive median was equal to the most recent observed value.
Forecasts at horizons greater than one were obtained by iterating one-step-ahead forecasts. Any resulting predictive quantiles that were less than zero were truncated to equal zero.

\subsection{Evaluation metrics}
\label{subsec:methods_evaluation}

To evaluate forecasts, we adopted the \emph{weighted interval score} (WIS) \citep{bracherEvaluatingEpidemicForecasts2021}.
Let $q_{1:K}$ be predictive quantiles for the observed quantity $y$.
The WIS is calculated as
$$\WIS(q_{1:K},y) = \frac{1}{K} \sum_{k=1}^K 2 \left\{ \mathds{1}_{(-\infty, q_k]}(y) - \tau_k \right\} (q_k - y),$$
where $\mathds{1}_{(-\infty, q_k]}(y)$ is the indicator function that takes the value $1$ when $y \in (-\infty, q_k]$ and $0$ otherwise.
This is a negatively-oriented proper score, meaning that negative scores are better and its expected value according to a given data generating process is minimized by reporting the predictive quantiles from that process. WIS was designed as a discrete approximation to the continuous ranked probability score, and is equivalent to pinball loss, which is commonly used in quantile regression \citep{bracherEvaluatingEpidemicForecasts2021}. We note that some other commonly used scores such as the logarithmic score and the continuous ranked probability score are not suitable for use with predictive distributions that are specified in terms of a set of predictive quantiles, since a full predictive density or distribution function is not directly available (see Supplemental Section 3 for further discussion).

To compare the skill of forecasters that submitted different subsets of forecasts, we used \emph{relative WIS}, as done in \cite{cramerEvaluationIndividualEnsemble2022}.
The ensemble forecasters developed and evaluated in this manuscript provided all relevant forecasts; missingness pertains only to the component forecasters, and in the present work the relative WIS is primarily used to summarize component forecaster skill as an input to some of the trained ensemble methods described below.
Let $\mathcal{I}$ denote a set of combinations of location $l$ and forecast creation date $s$ over which we desire to summarize model performance, and $\mathcal{I}_{m, m'} \subseteq \mathcal{I}$ be the subset of those locations and dates for which both models $m$ and $m'$ provided forecasts through a forecast horizon of at least four weeks. The relative WIS of model $m$ over the set $\mathcal{I}$ is calculated as
\begin{align*}
\rWIS^{m}_{\mathcal{I}} &= \frac{\theta^m}{\theta^{\text{baseline}}} \text{, where } \\
\theta^m &= \left(\prod_{m'=1}^M \frac{(4 \cdot \vert \mathcal{I}_{m, m'} \vert)^{-1} \sum_{(l, s) \in \mathcal{I}_{m, m'}} \sum_{t=s+1}^{s+4} \WIS(q^m_{l,s,t,1:K},y_{l,t}) }{ (4 \cdot \vert \mathcal{I}_{m, m'} \vert)^{-1} \sum_{(l, s) \in \mathcal{I}_{m, m'}} \sum_{t=s+1}^{s+4} \WIS(q^{m'}_{l,s,t,1:K},y_{l,t})} \right)^{\frac{1}{M}}.
\end{align*}
In words, we computed the ratio of the mean WIS scores for model $m$ and each other model $m'$, averaging across the subset of forecasts shared by both models. $\theta^m$ was calculated as the geometric mean of these pairwise ratios of matched mean scores, and summarized how model $m$ did relative to all other models on the forecasts they had in common. These geometric means were then scaled such that the baseline forecaster had a relative WIS of 1; a relative WIS less than 1 indicated forecast skill that was better than the baseline model.
We note that if no forecasts were missing, $\mathcal{I}_{m, m'}$ would be the same for all model pairs, so that the denominators of each $\theta^m$ and of $\theta^{\text{baseline}}$ would cancel when normalizing relative to the baseline and the relative WIS for model $m$ would reduce to the mean WIS for model $m$ divided by the mean WIS for the baseline model.
We used the geometric mean to aggregate across model pairs to match the convention set in \cite{cramerEvaluationIndividualEnsemble2022}, but this detail is not critical: Supplemental Figure 5 illustrates that the relative WIS changes very little if an arithmetic mean is used instead.

We also assessed probabilistic calibration of the models with the one-sided coverage rates of predictive quantiles, calculated as the proportion of observed values that were less than or equal to the predicted quantile value.
For a well-calibrated model, the empirical one-sided coverage rate is equal to the nominal quantile level.
A method that generates conservative two-sided intervals would have an empirical coverage rate that is less than the nominal rate for quantile levels less than 0.5 and empirical coverage greater than the nominal rate for quantile levels greater than 0.5.

\subsection{Ensemble model formulations}
\label{subsec:methods_ensembles}

All of the ensemble formulations that we considered obtain a predictive quantile at level $k$ by combining the component forecaster predictions at that quantile level:
$$
q^\ens_{l,s,t,k} = f(q^{1}_{l,s,t,k}, \ldots, q^{M}_{l,s,t,k}).
$$
We conceptually organize the ensemble methods considered according to two factors.
First, \textit{trained} ensemble methods use the past performance of the component forecasters to select a subset of components for inclusion in the ensemble and/or assign the components different weights, whereas \textit{untrained} methods assign all component forecasters equal weight.
Second, we varied the robustness of the combination function $f$ to outlying component forecasts.
Specifically, we considered methods based on either a (weighted) mean, which can be sensitive to outlying forecasts, or a (weighted) median, which may be more robust to these outliers.
The weighted mean calculates the ensemble quantiles as 
$$
q^\ens_{l,s,t,k} = \sum_{m = 1}^M w^m_{s} q^{m}_{l,s,t,k}.
$$
The weighted median is defined to be the smallest value $q$ for which the combined weight of all component forecasters with predictions less than or equal to $q$ is at least 0.5; the ensemble forecast quantiles are calculated as:
$$
q^\ens_{l,s,t,k} = \inf\left\{q \in \mathbb{R}: \sum_{m = 1}^M w^m_s \mathds{1}_{(-\infty, q]}(q^{m}_{l,s,t,k}) \geq 0.5\right\}.
$$
In practice, we used the implementation of the weighted median in the \verb`matrixStats` package for R, which linearly interpolates between the central weighted sample quantiles \citep{matrixStats}.
Graphically, these ensembles can be interpreted as computing a horizontal mean or median of the CDFs of component forecasters (Supplemental Figure 7).


In trained ensemble methods that weight the component forecasters, the weights were calculated as a sigmoidal transformation of the forecasters' relative WIS (see Section \ref{subsec:methods_evaluation}) over a rolling window of weeks leading up to the ensemble forecast date $s$, denoted by $\rWIS^{m}_{s}$:
$$
w^m_{s} = \frac{\exp(-\theta_s \cdot \rWIS^{m}_{s} )}{\sum_{m' = 1}^M \exp(-\theta_s \cdot \rWIS^{m'}_{s} )}.
$$
This formulation requires estimating the nonnegative parameter $\theta_s$, which was updated each week.
If $\theta_s = 0$, the procedure reduces to an equal weighting scheme.
However, if $\theta_s$ is large, better-performing component forecasters (with low relative WIS scores) are assigned higher weight.
We selected $\theta_s$ by using a grid search to optimize the weighted interval score of the ensemble forecast over the training window, summing across all locations and relevant target weeks on or before time $s$:
$$
\theta_s = \argmin_\theta \sum_{l} \sum_{r = s - 1}^{s-a} \sum_{t = r + 1}^{\min(r + 4, s)} \WIS(q^{\ens, \theta}_{l,r,t,1:K}, y_{l,t}).
$$
The size of the training window, $a$, is a tuning parameter that must be selected; we considered several possible values during model development, discussed further below.
In a post hoc analysis, we considered regularizing the weights by setting a limit on the weight that could be assigned to any one model.
We implemented this regularization strategy by restricting the grid of values for $\theta_s$ to those values for which the largest component forecaster weight was less than the maximum weight limit.

In this parameterization, the component forecaster weights are by construction nonnegative and sum to 1. When forecasts were missing for one or more component forecasters in a particular location and forecast date, we set the weights for those forecasters to 0 and renormalized the weights for the remaining forecasters so that they summed to 1.

Some trained ensembles that we considered used a preliminary component selection step, where the top few individual forecasters were selected for inclusion in the ensemble based on their relative WIS during the training window.
The number of component forecasters selected is a tuning parameter that we explored during model development.
This component selection step may be used either in combination with the continuous weighting scheme described above, or with an equally-weighted combination of selected forecasters.
Throughout the text below, we use the term ``trained" ensemble to refer generically to a method that uses component selection and/or weighting based on historical component forecaster performance.

There are many other weighted ensembling schemes that could be formulated. For example, separate weights could be estimated for different forecast horizons, for different quantile levels, or for subsets of locations. 
As another example, the weights could be estimated by directly minimizing the WIS associated with look-ahead ensemble forecasts \citep{taylor2021combiningForecastsCOVID}.
We explored these and other ideas during model development, but our analyses did not show them to lead to substantial gains, and thus we settled on the simpler weighting schemes presented above. Further discussion of alternative schemes is deferred to the supplement.

\subsection{Data and code accessibility}

All component model forecasts and code used for fitting ensemble models and conducting the analyses presented in this manuscript are available in public GitHub repositories \citep{cramer_reichlabcovid19-forecast-hub_2021, ray_reichlabcovidensembles_2020, ray_covid-19_ensemble_manuscript_2021}.

\section{Results}

We discuss the decisions that we made during model development in Section \ref{subsec:results_development} before turning to a more focused discussion of the impact on ensemble forecast skill of using robust or non-robust combination mechanisms in Section \ref{subsec:results_robust}, and trained or untrained methods in Section \ref{subsec:results_trained}.
Section \ref{subsec:results_regularization} presents a post hoc evaluation of a variation on ensemble methods that regularizes the component forecaster weights.
Results for the evaluation using forecasts in Europe are presented in Section \ref{subsec:results_eu}.

Throughout this section, scores were calculated using the ground truth data that were available as of May 16, 2022 unless otherwise noted. This allowed five weeks of revisions to accrue between the last target end date that was evaluated and the date of the data used for evaluation. When reporting measures of forecast skill, we dropped forecasts for which the corresponding reported value of weekly cases or deaths was negative. This could occur when an error in data collection was identified and corrected, or when the definition of a reportable case or death was changed. We included scores for all other outlying and revised data in the primary analysis because it was difficult to define objective standards for what should be omitted. However, a supplemental analysis indicated that the results about the relative performance of different ensemble methods were not sensitive to these reporting anomalies (Supplemental Section 5.4, Supplemental Figures 14 through 16).

\subsection{Model development}
\label{subsec:results_development}

\begin{figure}
\includegraphics[width=\textwidth]{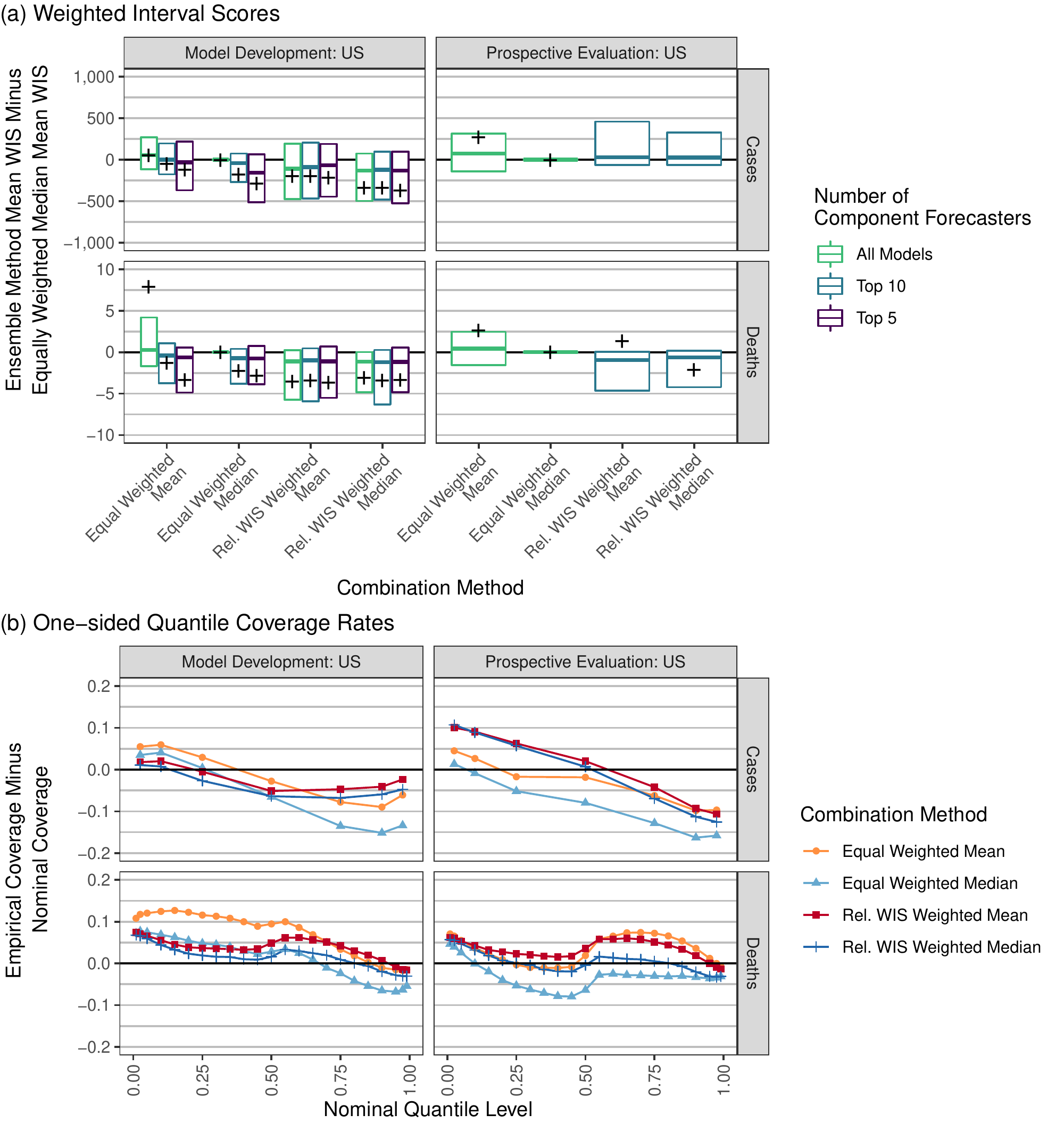}
\caption{Performance measures for ensemble forecasts of weekly cases and deaths at the state level in the U.S. In panel (a) the vertical axis is the difference in mean WIS for the given ensemble method and the equally weighted median ensemble.
Boxes show the 25th percentile, 50th percentile, and 75th percentile of these differences, averaging across all locations for each combination of forecast date and horizon.
For legibility, outliers are suppressed here; Supplemental Figure 8 shows the full distribution.
A cross is displayed at the difference in overall mean scores for the specified combination method and the equally weighted median averaging across all locations, forecast dates, and horizons.
Large mean score differences of approximately 2,005 and 2,387 are suppressed for the Rel. WIS Weighted Mean and Rel. WIS Weighted Median ensembles respectively in the prospective phase forecasts of cases.
A negative value indicates that the given method outperformed the equally weighted median.
The vertical axis of panel (b) shows the probabilistic calibration of the ensemble forecasts through the one-sided empirical coverage rates of the predictive quantiles.
A well-calibrated forecaster has a difference of 0 between the empirical and nominal coverage rates, while a forecaster with conservative (wide) two-sided intervals has negative differences for nominal quantile levels less than 0.5 and positive differences for quantile levels greater than 0.5.
}
\label{fig:wis_boxplots_and_calibration_by_phase_US_central_only}
\end{figure}

During model development, we evaluated many variations on trained ensemble methods. In these comparisons we take the equally weighted median ensemble as a reference approach because this is the method used for the production ensemble produced by the U.S.\ Hub during most of the time that we were running these experiments. As measured by mean WIS over the model development phase, the equally weighted median ensemble was better than the equally weighted mean ensemble, but both were outperformed by the trained ensemble variations using component forecaster selection and/or weighting (Figure~\ref{fig:wis_boxplots_and_calibration_by_phase_US_central_only}). The weighted approaches had stable performance no matter how many component forecasters were included. Approaches using an equally weighted combination of selected component forecasters were generally better only when top-performing component forecasters were included.

We also considered varying other tuning parameters such as the length of the training window and whether component forecaster weights were shared across different quantile levels or across forecast horizons. However, we did not find strong and consistent gains in performance when varying these other factors (Supplemental Figures 17 through 22). Finally, we evaluated other possible formulations of weighted ensembles, with weights that were not directly dependent on the relative WIS of the component forecasters but were instead estimated by optimizing the look-ahead ensemble WIS over the training set. As measured by mean WIS, the best versions of these other variations on weighted ensembles had similar performance to the best versions of the relative WIS weighted median considered in the primary analysis. However, they were more sensitive to settings like the number of component forecasters included and the training set window size (Supplemental Figures 17 and 18).

Based on these results, on May 3, 2021 we selected the relative WIS weighted ensemble variations for use in the prospective evaluation, as these methods had similar mean WIS as the best of the other variations considered, but were more consistent across different training set window sizes and numbers of component forecasters included. We used intermediate values for these tuning parameter settings, including 10 component forecasters with a training set window size of 12 weeks. We also included the equally weighted mean and median of all models in the prospective evaluation as reference methods. The following sections give a more detailed evaluation of these selected methods, describing how they performed during both the model development phase and the prospective evaluation phase.

\subsection{Comparing robust and non-robust ensemble methods}
\label{subsec:results_robust}

We found that for equally weighted ensemble approaches, robust combination methods were helpful for limiting the effects of outlying component forecasts. For most combinations of evaluation phase (model development or prospective evaluation) and target variable (cases or deaths), the equally weighted median had better mean and worst-case WIS than the equally weighted mean, often by a large margin (Figure \ref{fig:wis_boxplots_and_calibration_by_phase_US_central_only}, Supplemental Figure 8). Results broken down by forecast date show that the methods achieved similar scores most of the time, but the equally weighted mean ensemble occasionally had serious failures (Supplemental Figure 10). These failures were generally associated with instances where a component forecaster issued extreme, outlying forecasts, e.g., forecasts of deaths issued the week of February 15th in Ohio (Figure \ref{fig:ensemble_motivations}).

There were fewer consistent differences between the trained mean and trained median ensemble approaches. This suggests that both trained approaches that we considered had similar robustness to outlying forecasts (if the outliers were produced by component forecasters that were down weighted or not selected for inclusion due to poor historical performance) or sensitivity to outlying forecasts (if they were produced by component forecasters that were selected and given high weight).

Panel (b) of Figure \ref{fig:wis_boxplots_and_calibration_by_phase_US_central_only} summarizes probabilistic calibration of the ensemble forecasts with one-sided quantile coverage rates. The median-based ensemble approaches generally had lower one-sided quantile coverage rates than the mean-based approaches, indicating a downward shift of the forecast distributions. This was associated with poorer probabilistic calibration for forecasts of cases, where the ensemble forecast distributions tended to be too low. For forecasts of deaths, which were better centered but tended to be too narrow, the calibration of the median-based methods was not consistently better or worse than the calibration of the corresponding mean-based methods.

\subsection{Comparing trained and untrained ensemble methods}
\label{subsec:results_trained}

\begin{figure}
\includegraphics[width=\textwidth]{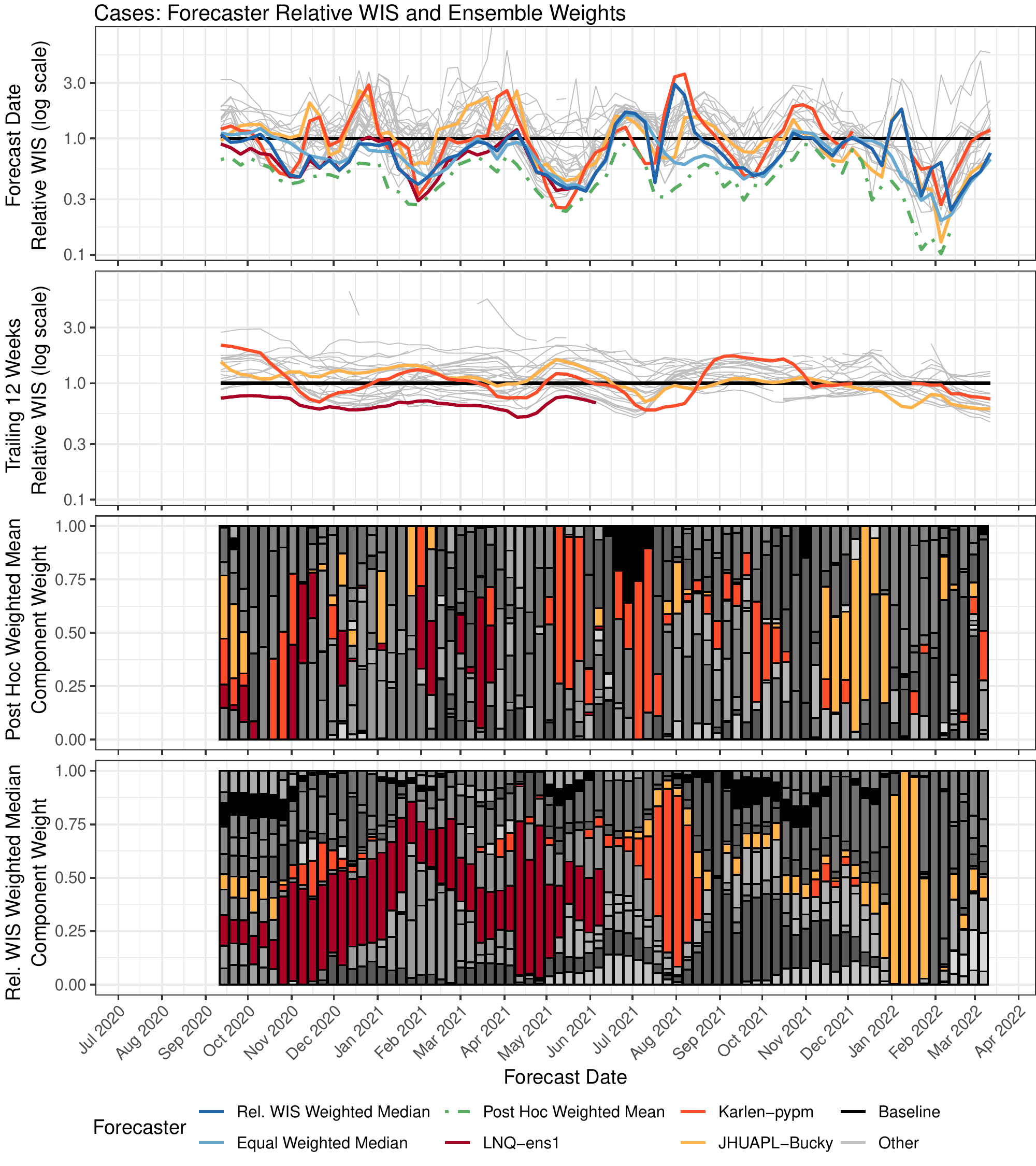}
\caption{Performance of weekly case forecasts from component forecasters and selected ensembles, along with component forecaster weights. Component forecasters that were given high weight at key times are highlighted. The top row shows the relative WIS of forecasts made each week. The second row shows the relative WIS over the 12 weeks before the forecast date, for forecasts of quantities that were observed by the forecast date. These scores, which are used to compute the component weights in the relative WIS weighted median ensemble, are calculated using data available as of the forecast date. The third row shows component forecaster weights for the post hoc weighted mean ensemble, and the bottom row shows the component model weights for the relative WIS weighted median ensemble; each component forecaster is represented with a different color. Over the time frame considered, 31 distinct component forecasters were included in this top-10 ensemble.}
\label{fig:component_weights_cases}
\end{figure}

\begin{figure}
\includegraphics[width=\textwidth]{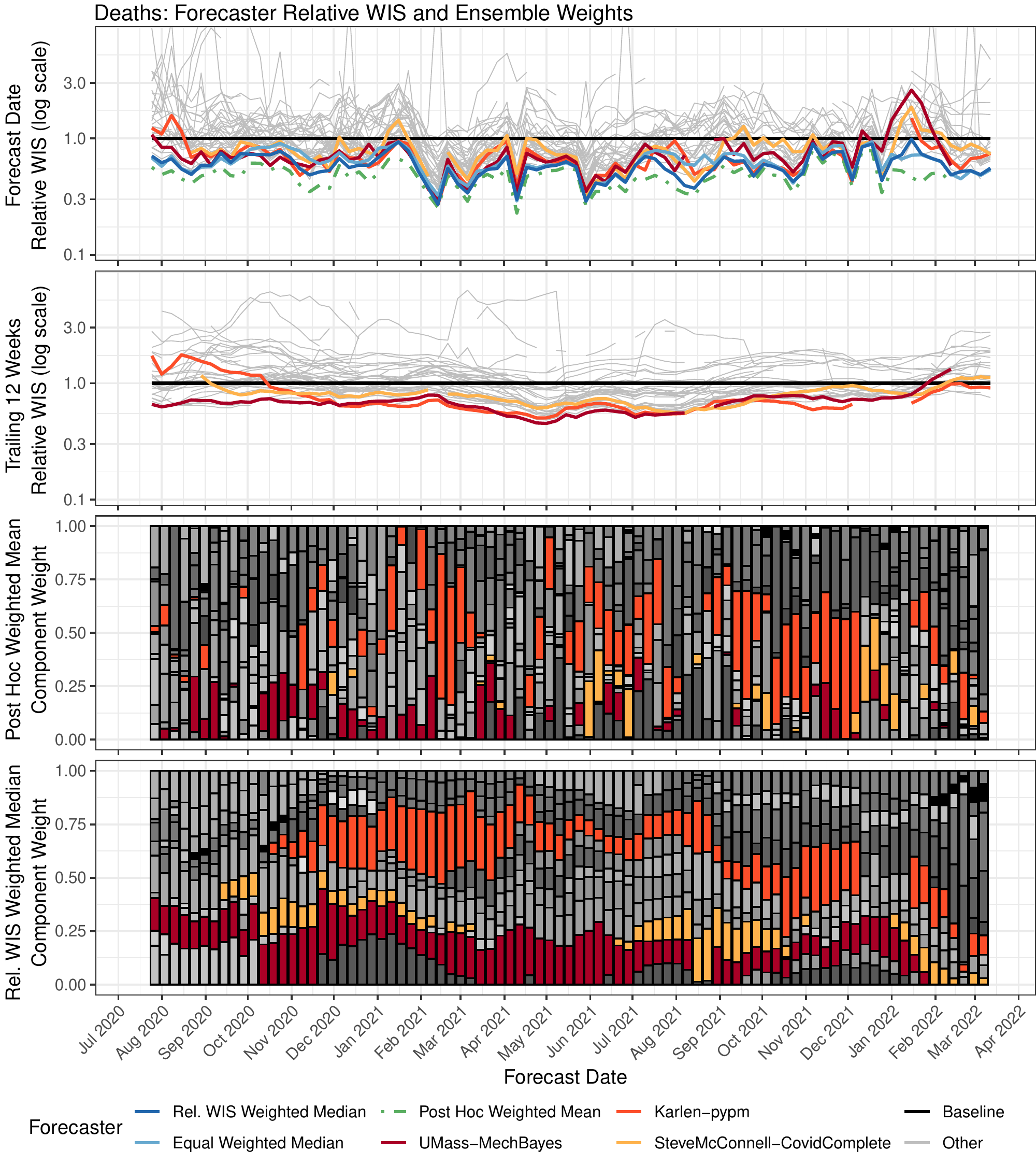}
\caption{Performance of weekly death forecasts from component forecasters and selected ensembles, along with component forecaster weights. Component forecasters that were given high weight at key times are highlighted. The top row shows the relative WIS of forecasts made each week. The second row shows the relative WIS over the 12 weeks before the forecast date, for forecasts of quantities that were observed by the forecast date. These scores, which are used to compute the component weights in the relative WIS weighted median ensemble, are calculated using data available as of the forecast date. The third row shows component forecaster weights for the post hoc weighted mean ensemble, and the bottom row shows the component model weights for the relative WIS weighted median ensemble; each component forecaster is represented with a different color. Over the time frame considered, 34 distinct component forecasters were included in this top-10 ensemble.}
\label{fig:component_weights_deaths}
\end{figure}

Averaging across all forecasts for incident cases and deaths in the model development phase, the weighted median was better than the equally weighted median and the weighted mean was better than the equally weighted mean (Figure~\ref{fig:wis_boxplots_and_calibration_by_phase_US_central_only}).
However, in the prospective evaluation, the trained methods showed improved mean WIS relative to untrained methods when forecasting deaths, but were worse when forecasting cases. In general, the trained ensembles also came closer to matching the performance of a post hoc weighted mean ensemble for deaths than for cases (Figures~\ref{fig:component_weights_cases} and \ref{fig:component_weights_deaths}). This post hoc weighted mean ensemble estimated the optimal weights for each week after the forecasted data were observed; it would not be possible to use this method in real time, but it gives a bound on the ensemble forecast skill that can be achieved using quantile averaging.

We believe that this difference in the relative performance of trained and untrained ensemble methods for cases and deaths is primarily due to differences in component model behavior for forecasting cases and deaths.
A fundamental difference between these outcomes is that cases are a leading indicator relative to deaths, so that trends in cases in the recent past may be a helpful input for forecasting deaths \textemdash but there are not clear candidates for a similar leading indicator for cases (e.g., see \cite{mcdonald_auxiliary_indicators_covid_2021} for an investigation of some possibilities that were found to yield only modest and inconsistent improvements in forecast skill).
Indeed, the best models for forecasting mortality generally do use previously reported cases as an input to forecasting \citep{cramerEvaluationIndividualEnsemble2022}, and it has previously been noted that deaths are an easier target to forecast than cases \citep{2021Reich_covid_predictability_blog, bracher_preregistered_covid_de_pl_2021}.
This is reflected in the performance of trained ensembles, which were often able to identify a future change in direction of trends when forecasting deaths, but generally tended to predict a continuation of recent trends when forecasting cases (Supplemental Section 7, Supplemental Figures 25 and 26).
An interpretation of this is that the component forecasters with the best record of performance for forecasting deaths during the training window were able to capture changes in trend, but the best component forecasters for forecasting cases were often simply extrapolating recent trends.
While all ensemble methods tended to ``overshoot" at local peaks in weekly incidence, this tendency was more pronounced for forecasts of cases than for forecasts of deaths \textemdash and training tended to exacerbate the tendency to overshoot when forecasting cases, but to mitigate this tendency when forecasting deaths (Supplemental Figure 25).


Another difference in component behavior when forecasting cases and deaths is illustrated in Figures~\ref{fig:component_weights_cases} and \ref{fig:component_weights_deaths}, which explore the relationships between component forecaster performance and the relative performance of trained and untrained ensemble methods in more detail.
For deaths, the trained ensemble was able to identify and upweight a few component forecasters that had consistently good performance (e.g., Karlen-pypm and UMass-MechBayes).
This led to consistently strong performance of the trained ensemble; it was always among the best models contributing to the U.S. Hub, and was better than the equally weighted median ensemble in nearly every week.

For cases, the trained ensemble also had strong performance for many months when the LNQ-ens1 forecaster was contributing to the U.S. Hub.
However, when LNQ-ens1 stopped contributing forecasts in June 2021, the trained ensemble shifted to weighting Karlen-pypm, which had less stable performance for forecasting cases.
During July 2021, Karlen-pypm was the only forecaster in the U.S. Hub that predicted rapid growth at the start of the Delta wave, and it achieved the best relative WIS by a substantial margin at that time. However, that forecaster predicted continued growth as the Delta wave started to wane and it had the worst relative WIS a few weeks later.
A similar situation occurred during the Omicron wave in January 2022, when the JHUAPL-Bucky model was one of a small number of forecasters that captured the rise at the beginning of the wave, but it then overshot near the peak.
In both of these instances, the post hoc weighting would have assigned a large amount of weight to the forecaster in question at the start of the wave, when it was uniquely successful at identifying rising trends in cases \textemdash but then shifted away from that forecaster as the peak neared.
Trained ensembles that estimated weights based on past performance suffered, as they started to upweight those component forecasters just as their performance dropped.
This recurring pattern highlights the challenge that nonstationary component forecaster performance presents for trained ensembles.
Reinforcing this point, we note that in the post hoc weighted mean ensemble, the component forecaster weights are only weakly autocorrelated (Figures~\ref{fig:component_weights_cases} and \ref{fig:component_weights_deaths}, Supplemental Figure 27), again suggesting that an optimal weighting may require frequently changing component weights to adapt to nonstationary performance.


During the model development phase, the trained ensembles had better probabilistic calibration than their equally weighted counterparts (Figure \ref{fig:wis_boxplots_and_calibration_by_phase_US_central_only} panel (b)).
During prospective evaluation, the trained median ensemble had generally higher one-sided coverage rates, corresponding to better calibration in the upper tail but slightly worse calibration in the lower tail.
The trained mean ensemble had slightly better calibration than the equally weighted mean when forecasting deaths in the prospective evaluation phase, but inconsistent gains across different quantile levels when forecasting cases.
Supplemental Figures 12 and 13 show that the widths of 95\% prediction intervals from both the equally weighted median ensemble and the relative WIS weighted median ensemble tended to rank near the middle of the widths of 95\% prediction intervals from the component forecasters.
This can be interpreted as an advantage if we are concerned about the possible influence of component forecasters with very narrow or very wide prediction intervals. However, it can also be viewed as a disadvantage, particularly if improved calibration could have been realized if the prediction intervals were wider. We return to this point in the discussion.

\subsection{Post hoc evaluation of weight regularization}
\label{subsec:results_regularization}

Motivated by the assignment of large weights to some component forecasters in the trained ensembles for cases (Figure~\ref{fig:component_weights_cases}), in January 2022 we conducted a post hoc evaluation of trained ensembles that were regularized by imposing a limit on the weight that could be assigned to any one component forecaster (see Section \ref{subsec:methods_ensembles}). In this evaluation, we constructed relative WIS weighted median ensemble forecasts for all historical forecast dates up through the week of January 3, 2022. These ensemble fits included the top 10 component forecasters and were trained on a rolling window of the 12 most recent forecast dates, matching the settings that were selected for the prospective analysis. We considered six values for the maximum weight limit: 0.1, 0.2, 0.3, 0.4, 0.5, and 1.0. A weight limit of 1.0 corresponds to the unregularized method considered in the prospective evaluation, and a weight limit of 0.1 corresponds to an equally weighted median of the top ten forecasters, which was previously considered during the model development phase.

\begin{figure}
\includegraphics[width=\textwidth]{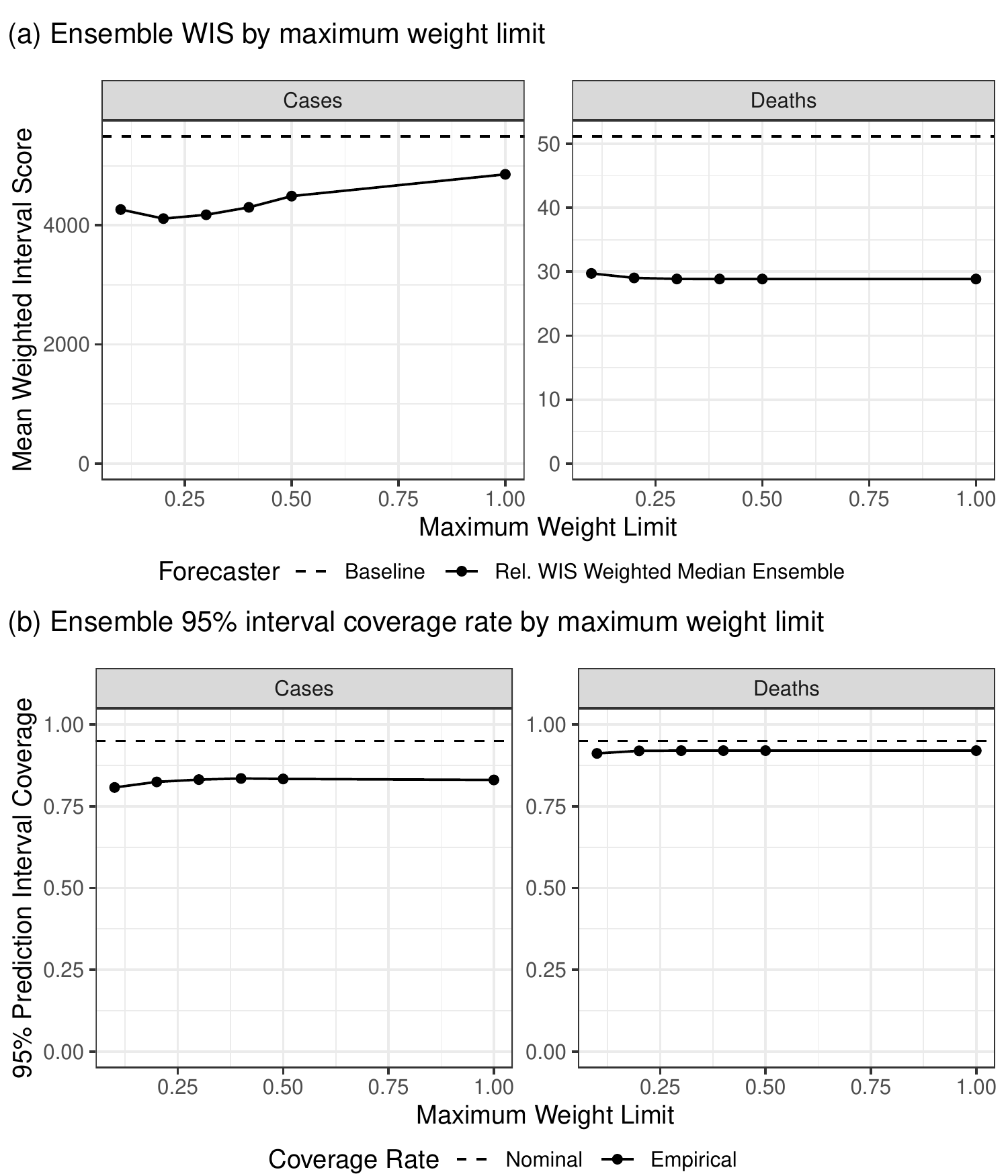}
\caption{Mean WIS and 95\% prediction interval coverage rates for relative WIS weighted median trained ensemble variations with varying sizes of a limit on the weight that could be assigned to any one model. In panel (a), the baseline forecaster is included as a reference. Results are for a post hoc analysis including forecast dates up to January 3, 2022.}
\label{fig:max_weight_limits_overall}
\end{figure}

For both cases and deaths, the results of this analysis indicate that a weight limit as low as 0.1 was unhelpful (Figure~\ref{fig:max_weight_limits_overall}).
When forecasting deaths, this regularization strategy had limited impact on the trained ensemble performance as long as the maximum weight limit was about 0.3 or higher, which is consistent with the fact that the trained ensembles for deaths rarely assigned a large weight to one model (Figure~\ref{fig:component_weights_deaths}).
However, when forecasting cases, the regularization resulted in large improvements in mean WIS, with the best WIS at limits near 0.2 or 0.3.
These improvements were concentrated in short periods near local peaks in the epidemic waves (Supplemental Figure 28).
For both cases and deaths, smaller limits on the maximum weight were associated with a slight reduction in the empirical coverage rates of 95\% prediction intervals.
Based on these results, the U.S. Hub used a weight limit of 0.3 in trained ensemble forecasts starting in January 2022.

\subsection{Results in the European application}
\label{subsec:results_eu}

\begin{figure}
\includegraphics[width=\textwidth]{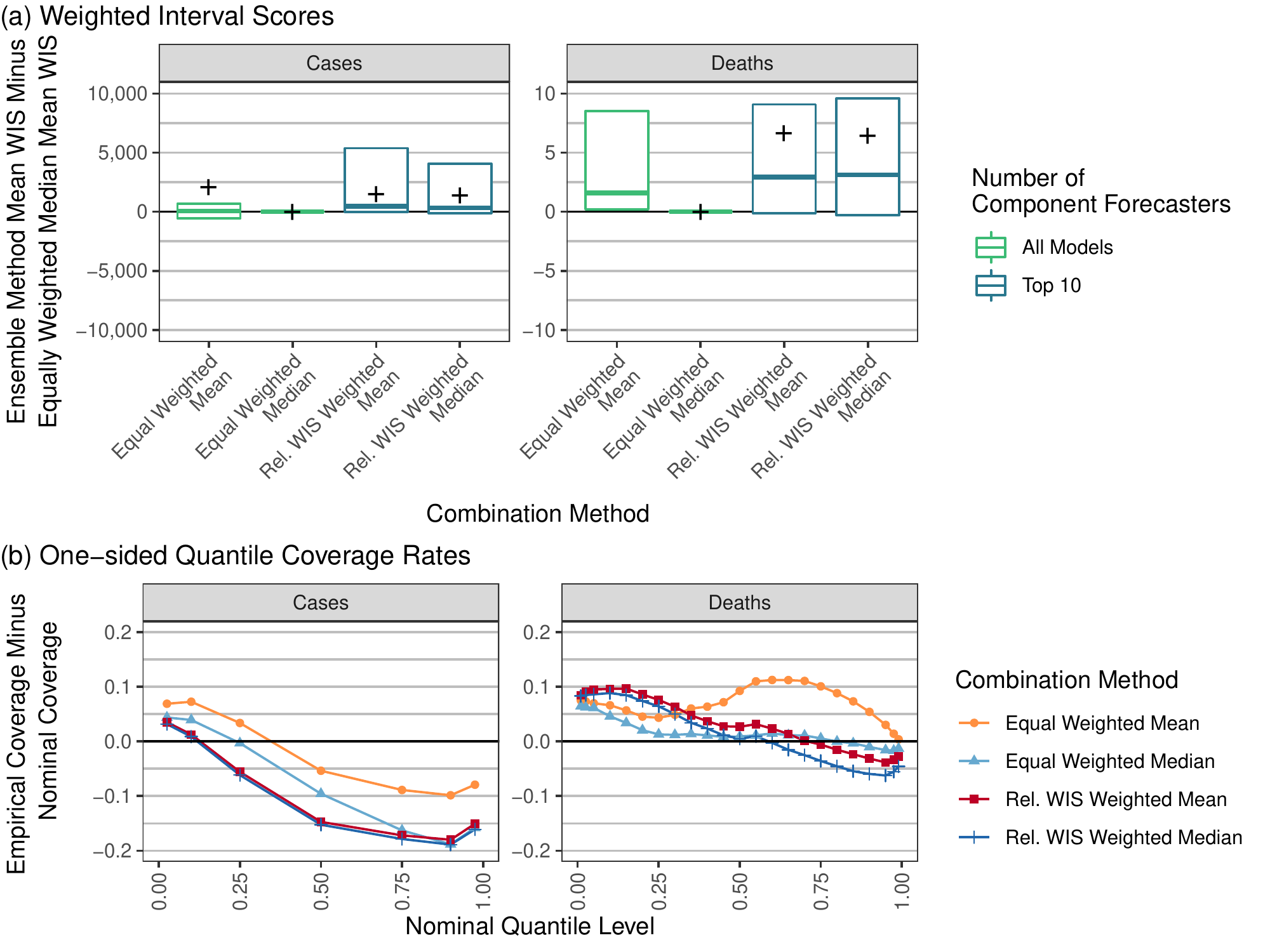}
\caption{Performance measures for ensemble forecasts of weekly cases and deaths in Europe. In panel (a) the vertical axis is the difference in mean WIS for the given ensemble method and the equally weighted median ensemble.
Boxes show the 25th percentile, 50th percentile, and 75th percentile of these differences, averaging across all locations for each combination of forecast date and horizon.
For legibility, outliers are suppressed here; Supplemental Figure 9 shows the full distribution.
A cross is displayed at the difference in overall mean scores for the specified combination method and the equally weighted median of all models, averaging across all locations, forecast dates, and horizons.
A large mean score difference of approximately 666 is suppressed for the Equal Weighted Mean ensemble forecasts of deaths.
A negative value indicates that the given method had better forecast skill than the equally weighted median.
Panel (b) shows the probabilistic calibration of the forecasts through the one-sided empirical coverage rates of the predictive quantiles.
A well-calibrated forecaster has a difference of 0 between the empirical and nominal coverage rates, while a forecaster with conservative (wide) two-sided intervals have negative differences for nominal quantile levels less than 0.5 and positive differences for quantile levels greater than 0.5.
}
\label{fig:wis_boxplots_and_calibration_by_phase_EU_central_only}
\end{figure}

Figure~\ref{fig:wis_boxplots_and_calibration_by_phase_EU_central_only} summarizes weighted interval scores and calibration for the four selected ensemble methods when applied prospectively to forecast data collected in the European Forecast Hub. Consistent with what we observed for the U.S. above, the equally weighted median ensemble was generally better than the equally weighted mean. However, in the European evaluation, the trained methods had worse performance than the equally weighted median for forecasting both cases and deaths.

\begin{figure}
\includegraphics[width=\textwidth]{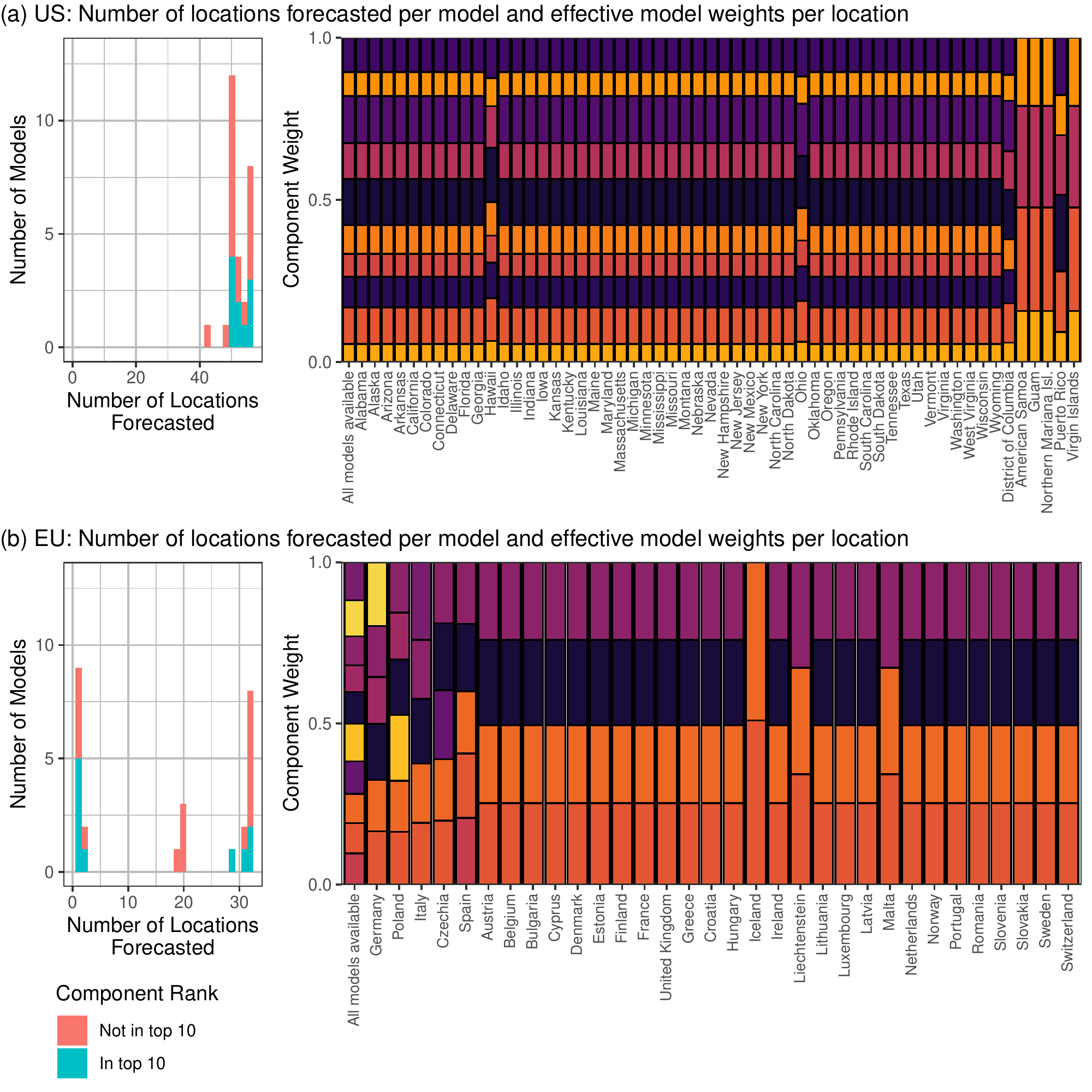}
\caption{A comparison of the impacts of forecast missingness in the applications to the U.S. (panel (a)) and Europe (panel (b)). Within each panel, the histogram on the left shows the number of locations forecasted by each contributing forecaster in the week of October 11, 2021, colored by whether or not the forecaster was among the top 10 forecasters eligible for inclusion in the relative WIS weighted ensemble selected for prospective evaluation. The plot on the right shows the estimated weights that would be used if all of the top 10 models (each represented by a different color) were available for a given location (at left side), and the effective weights used in each location after setting the weights for models that did not provide location-specific forecasts to 0 and rescaling the other weights proportionally to sum to 1.}
\label{fig:component_missingness}
\end{figure}

In a post hoc exploratory analysis, we noted that patterns of missingness in forecast submissions are quite different in the U.S. and in Europe (Figure~\ref{fig:component_missingness}, Supplemental Figures 29 through 36). In the U.S. Hub, nearly all models submit forecasts for all of the 50 states, and many additionally submit forecasts for at least one of the District of Columbia and territories. However, in the European Hub, roughly half of contributing models submit forecasts for only a small number of locations. Because the trained ensembles selected for prospective evaluation select the top 10 individual forecasters by relative WIS, this means that in practice the trained ensembles only included a few component forecasters for many locations in Europe.

\section{Discussion}

In this work, we have documented the analyses that have informed the selection of methods employed by the official U.S. Hub ensemble that is used by the CDC for communication with public health decision makers and the public more generally. In this context, our preference is for methods that have stable performance across different locations and different points in time, and good performance on average.

Our most consistent finding is that robust ensemble methods (i.e., based on a median) are helpful because they are more stable in the presence of outlying forecasts than methods using a mean. Ensemble methods based on means have repeatedly produced extreme forecasts that are dramatically misaligned with the observed data, but median-based approaches have not suffered from this problem as much. This stability is of particular importance in the context of forecasts that will be used by public health decision makers. These observations informed our decision to use an equally weighted median ensemble for the official U.S.\ Hub ensemble early on.

We have seen more mixed success for trained ensemble methods.
Overall, trained ensemble methods did well when they were able to identify and upweight component forecasters with stable good performance, but struggled when component forecaster skill varied over time.
In the U.S., trained ensembles have a long record of good performance when forecasting deaths, and the U.S. Hub adopted the relative WIS weighted median ensemble as its official method for forecasting deaths in November 2021.
However, trained methods have been less successful at forecasting cases in the U.S., both near peaks in weekly incidence (when they tend to overshoot) and at points where the performance of the component forecasters is inconsistent.
Additionally, the trained methods we adopted did not translate well to a setting with a large number of missing component forecasts, as in the European Hub.
To preserve the prospective nature of our analyses, we have not examined additional ensemble variations in the European application, but we hypothesize that these problems might be mitigated by including all component forecasters rather than the top 10, or by performing weight estimation separately in clusters of locations where the same component forecasters are contributing.
Allowing for different weights in different locations may also be an effective strategy for addressing the impacts of differences in data availability and quality across different locations.


In this manuscript, we have focused on relatively simple approaches to building ensemble forecasts. There are several opportunities for other directions that we have not considered here, and the gap in performance between the ensemble methods we have considered and an ensemble using post hoc optimal weights indicates that there may still be room for improvement in ensemble methods.
In our view, the most central challenge for trained ensembles is the inconsistency of the relative performance of many component forecasters, which may in turn be responsible for the lack of strong short-term temporal correlation in the component forecaster weights that were estimated by the post hoc weighted mean ensemble. For models with a relatively long history of performance over multiple epidemic waves, we believe that the most promising approach to addressing this is by using weights that depend on covariates like recent trends in incidence. This might allow the ensemble to learn the conditions in which component forecasters have been more or less reliable, and upweight models locally during phases similar to those in which they have done well in the past. Similar approaches have been used for other infectious disease systems such as influenza in the past \citep[e.g.,][]{ray_feature_weighted_ensembles}, but they used a substantial amount of training data over multiple years.

There are several other possible directions for further exploration.
We have addressed the challenge posed by outlying component forecasts by using median-based combination mechanisms, but another approach would be to pre-screen the component forecasts and remove outlying forecasts. This is a difficult task because there are times when weekly cases and deaths grow exponentially, and occasionally only one or two models have captured this growth accurately (Supplemental Figures 1 and 2).
A component screening method would have to be careful to avoid screening out methods that looked extreme relative to the data or other component forecasts, but in fact accurately captured exponential growth (see Supplemental Section 1 for more discussion).

Another challenge is that the ensemble forecasts have not always been well calibrated.
We are actively developing approaches to address this by post hoc recalibration of the ensemble forecasts.
Another possible route forward would be to use a different method for ensemble construction.
As we discussed earlier, the ensemble methods that we have considered work by combining the predictions from component forecasters at each quantile level, and therefore tend to have a dispersion that ranks in the middle of the dispersions of the component forecasters.
In contrast, an ensemble forecast obtained as a distributional mixture of component forecasts would exhibit greater uncertainty at times when the component forecasts disagreed with each other.
However, such an approach would be impacted by extreme component forecasts, and would likely require the development of strategies for screening outlying forecasts as discussed above.

Additionally, our methods for constructing ensemble forecasts do not directly account for the fact that some component forecasters are quite similar to each other and may provide redundant information about the future of the pandemic.
Ensembles generally benefit from combining diverse component forecasters, and it could be helpful to encourage this \textemdash for example, by clustering the forecasters and including a representative summary of the forecasts within each cluster as the ensemble components.
There are also related questions about the importance of different component forecasters to ensemble skill; we plan to explore this direction in future work by using tools such as the Shapley value to describe the contribution of individual components to the full ensemble.

We have used the WIS and probabilistic calibration to measure the extent to which forecasts are consistent with the eventually observed data. These summaries of performance are commonly used and provide useful insights into forecast performance, but it is worth noting that they do not necessarily reflect the utility of the forecasts for every particular decision-making context. Aggregated summaries of performance, such as overall quantile coverage rates could obscure finer-scale details \textemdash for instance, a method with good coverage rates on average could have high coverage at times that are relatively unimportant and low coverage when it matters. Additionally, for some public health decision-making purposes, one or another aspect of a forecast may be more important; for example, some users may prioritize accurate assessments about when a new wave may begin, but other users may find accurate forecasts of peak intensity to be more important. Our evaluation metrics do not necessarily reflect the particular needs of those specific end users, and it is possible that different ensemble methods would be more or less appropriate to generate forecasts that serve different purposes.

Careful consideration and rigorous evaluation are required to support decisions about what ensemble methods should be used for infectious disease forecasting. As we discussed earlier, to obtain an accurate measure of a forecaster's performance, it is critical that the versions of ground truth data that would have been available in real time are used for parameter estimation. This applies as much to ensemble forecasters as it does to individual models. Additionally, it is important to be clear about what methods development and evaluation were done retrospectively and what forecasts were generated prospectively in real time. We believe that to avoid disruptions to public health end users, a solid evidence base of stable performance in prospective forecasts should be assembled to support a change in ensemble methods. We have followed these principles in this work, and have followed the EPIFORGE guidelines in describing our analysis (\cite{pollett_2021_epiforge}; Supplemental Section 11).

The COVID-19 pandemic has presented a unique challenge for infectious disease forecasting. The U.S.\ and European Forecast Hubs have collected a wealth of forecasts from many contributing teams \textemdash far more than have been collected in previous collaborative forecasting efforts for infectious diseases such as influenza, dengue, and Ebola. These forecasts have been produced in real time to respond to an emerging pathogen that has been one of the most serious public health crises in the last century. This setting has introduced a myriad of modeling difficulties, from data anomalies due to new reporting systems being brought online and changing case definitions, to uncertainty about the fundamental epidemiological parameters of disease transmission, to rapidly changing social factors such as implementation and uptake of non-pharmaceutical interventions. The behavior of individual models in the face of these difficulties has in turn affected the methods that were suitable for producing ensemble forecasts. We are hopeful that the lessons learned about infectious disease forecasting will help to inform effective responses from the forecasting community in future infectious disease crises.




\bibliography{bibfile}

\end{document}


\title{Supplemental Materials for Comparing trained and untrained probabilistic ensemble forecasts of COVID-19 cases and deaths in the United States}

\author{Evan L. Ray
\and
Logan C. Brooks
\and
Jacob Bien
\and
Matthew Biggerstaff
\and
Nikos I. Bosse
\and
Johannes Bracher
\and
Estee Cramer
\and
Sebastian Funk
\and
Aaron Gerding
\and
Michael A. Johansson
\and
Aaron Rumack
\and
Yijin Wang
\and
Martha Zorn
\and
Ryan J. Tibshirani
\and
Nicholas G. Reich}


\date{}

\maketitle

\linenumbers

\section{Illustration of component forecaster predictive medians}

\begin{figure}[H]
    \centering
    \includegraphics[width=5.5in]{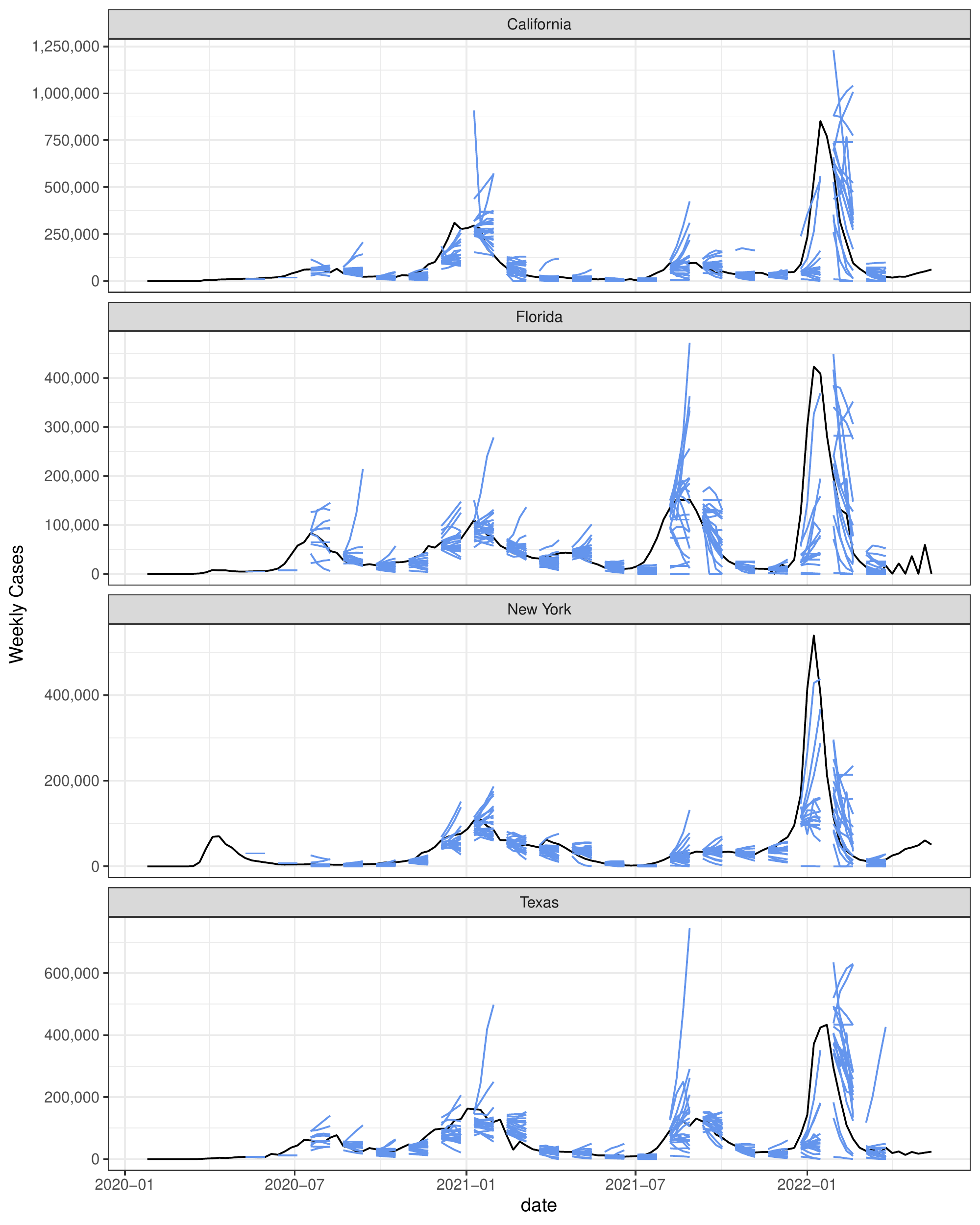}
    \caption{Component forecasters' predictive medians (blue) for weekly cases (black). For legibility, only the predictive medians originating from every fifth forecast date are shown.}
    \label{fig:component_medians_cases}
\end{figure}

\begin{figure}[H]
    \centering
    \includegraphics[width=5.5in]{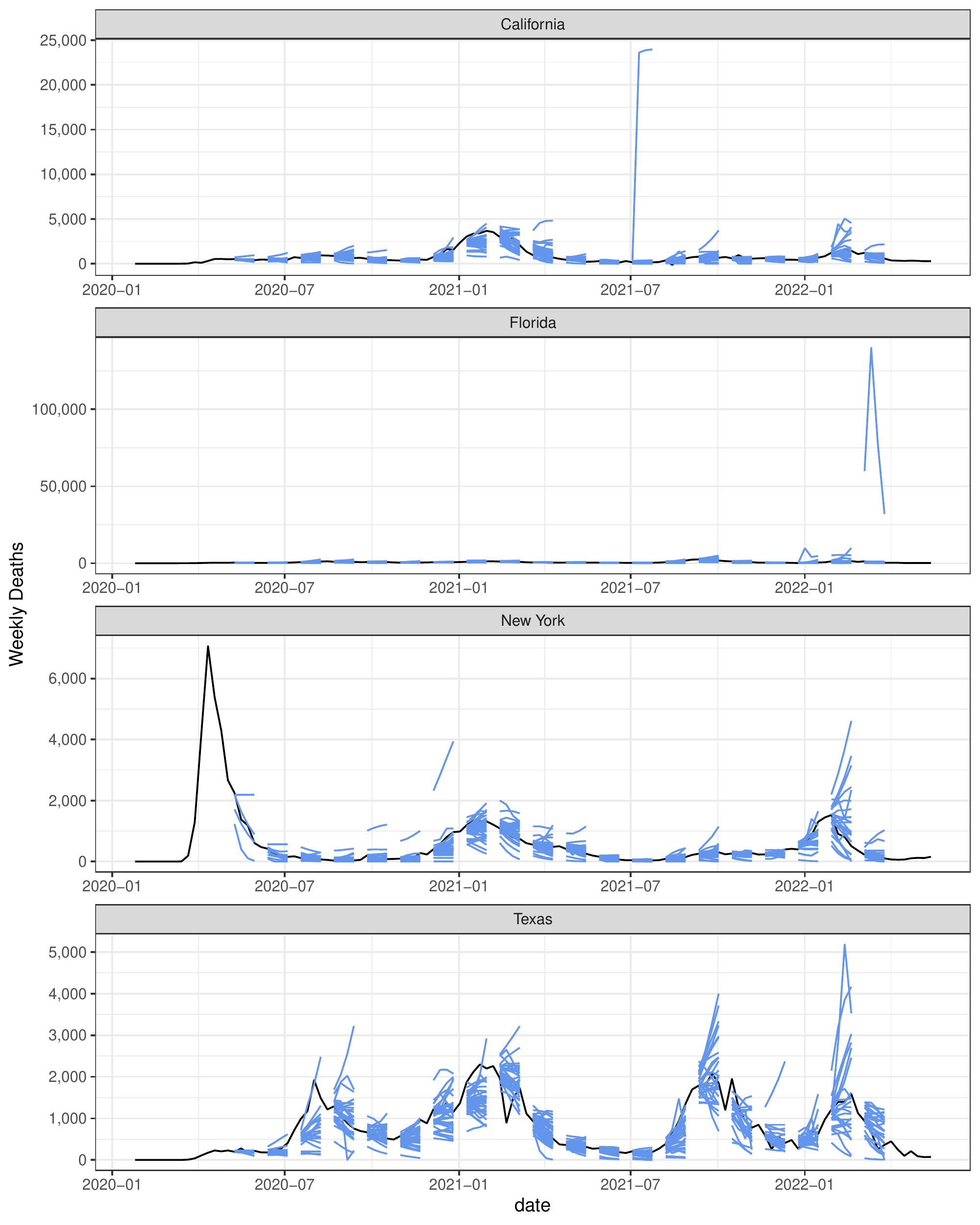}
    \caption{Component forecasters' predictive medians (blue) for weekly deaths (black). For legibility, only the predictive medians originating from every fifth forecast date are shown.}
    \label{fig:component_medians_deaths}
\end{figure}

Supplemental Figures~\ref{fig:component_medians_cases} and \ref{fig:component_medians_deaths} show predictive medians for component forecasts of weekly cases and deaths respectively in selected states with large counts of cases and deaths. We note several important characteristics of the forecasts. For cases, in periods of exponential growth at the start of new waves, it has consistently been the case that two or fewer forecasters accurately predicted that growth. On the other hand, more forecasters have captured periods of growth in deaths.

We also note that for both cases and deaths, there are many outlying forecasts.
These can take the form of forecasts that bear little relation to the observed data, such as forecasts of nearly zero cases near peaks as shown in all facets of Supplemental Figure~\ref{fig:component_medians_cases}, or forecasts that are uniformly too high, such as the outlying forecast of deaths in Florida that is visible in the second panel of Supplemental Figure~\ref{fig:component_medians_deaths}.

Another type of outlying forecast is one that predicts exponential growth that does not materialize, as illustrated at several points in forecasts of cases in Florida and Texas (Supplemental Figure~\ref{fig:component_medians_cases}). We note that these forecasts were unsuccessful, but closely match the trajectories of the few successful forecasts that were made during the Omicron wave in January 2022. This illustrates a potential challenge with automated outlier detection schemes in the context of a process where exponential growth is possible.

\newpage

\section{Relative WIS of component forecasters}

\begin{figure}[H]
    \centering
    \includegraphics[width=5.5in]{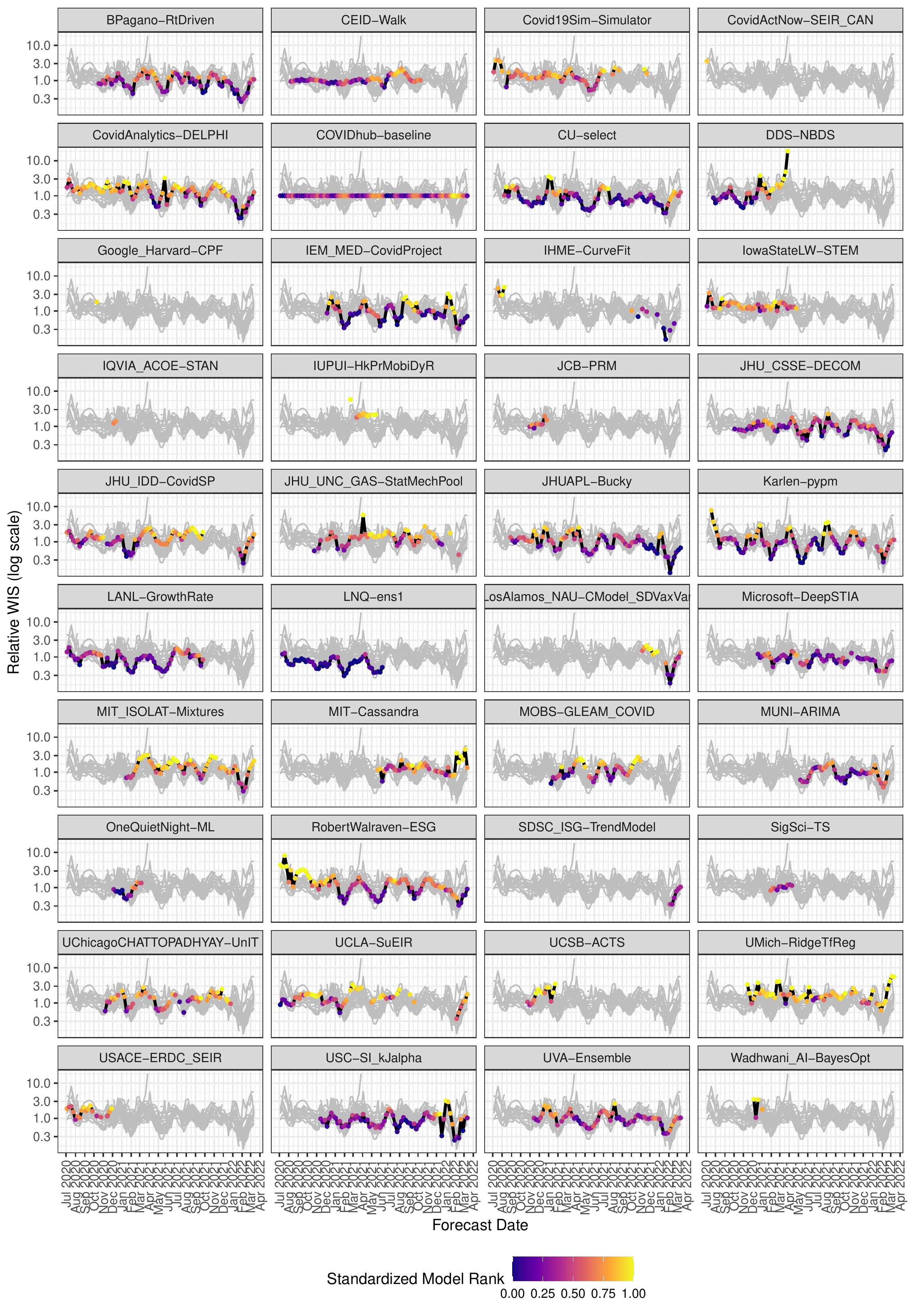}
    \caption{Weekly component ranks according to relative WIS for forecasts of cases in the U.S. A rank of 0 indicates that the model has the best performance in a given week, and a rank of 1 indicates that it has the worst performance. There is a facet for each component forecaster, and the colored line shows the standardized rank of that forecaster.}
    \label{fig:rel_wis_ranks_cases}
\end{figure}

\begin{figure}[H]
    \centering
    \includegraphics[width=5.5in]{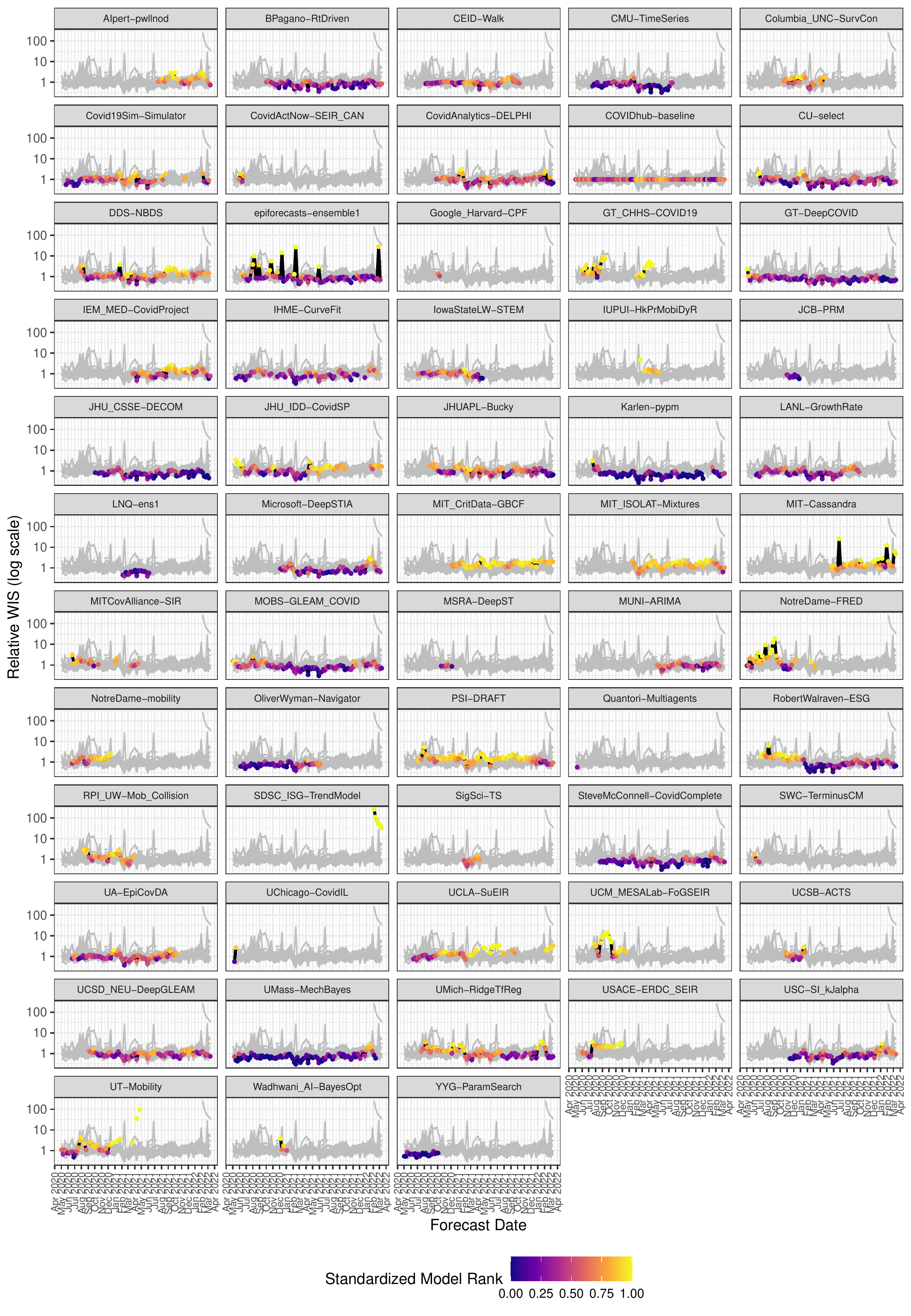}
    \caption{Weekly component ranks according to relative WIS for forecasts of deaths in the U.S. A rank of 0 indicates that the model has the best performance in a given week, and a rank of 1 indicates that it has the worst performance. There is a facet for each component forecaster, and the colored line shows the standardized rank of that forecaster.}
    \label{fig:rel_wis_ranks_deaths}
\end{figure}

Supplemental Figures~\ref{fig:rel_wis_ranks_cases} and \ref{fig:rel_wis_ranks_deaths} show the relative WIS of each component forecaster as a function of the forecast week for the weekly cases and weekly deaths targets respectively. For each week, we calculated a standardized rank for each model based on where that model's relative WIS fell relative to all other models that had submissions that week. In these rankings, 0 indicates the model with the best performance, and 1 indicates the model with the worst performance as measured by relative WIS. We see that nonstationarity of relative performance is very common, with many models alternating between weeks with top-ranking performance and bottom-ranking performance.

\begin{figure}[H]
    \centering
    \includegraphics[width=5.5in]{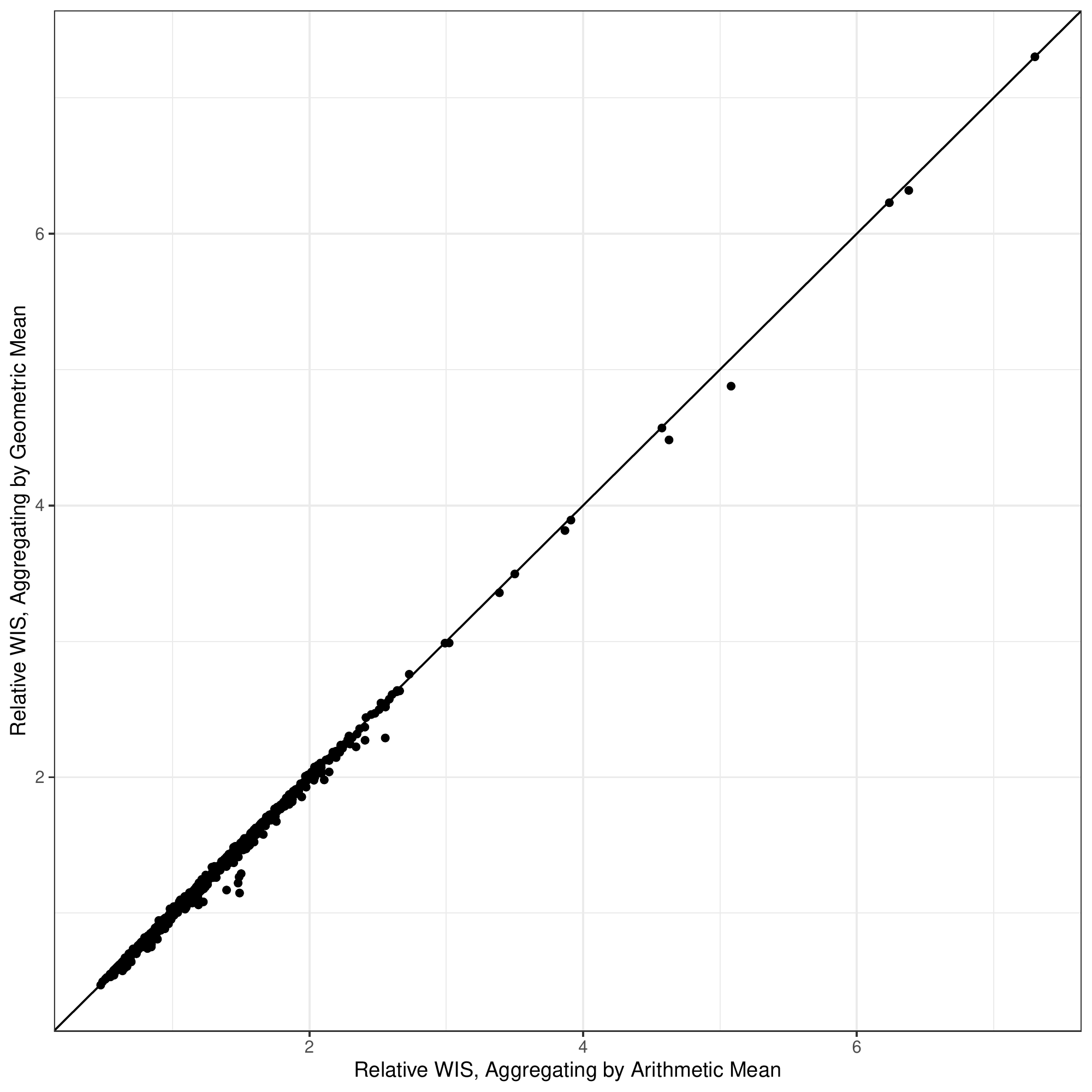}
    \caption{Comparison of aggregation methods used to summarize across models when calculating the relative WIS. Each point corresponds to one combination of component forecaster and forecast date, and shows the relative WIS based on a trailing window of 12 weeks using the arithmetic mean to summarize across models (horizontal axis) or the geometric mean to summarize across models (vertical axis). The figure shows these results for all component model forecasts of incident cases at the state level in the United States made on dates between July 27, 2020 and March 14, 2022 using the truth data available as of the forecast date; the relative WIS scores shown are those that would have been used as input to the relative WIS weighted ensemble methods for real-time forecasts.}
    \label{fig:compare_rel_wis_agg_methods}
\end{figure}

Supplemental Figure~\ref{fig:compare_rel_wis_agg_methods} compares two methods for aggregating across forecasters. Recall that as defined in the main text, the relative WIS uses a geometric mean to aggregate WIS ratios across pairs of forecasters. To refresh the notation, we let $\mathcal{I}$ denote a set of combinations of location $l$ and forecast creation date $s$ over which we desire to summarize model performance, and $\mathcal{I}_{m, m'} \subseteq \mathcal{I}$ be the subset of those locations and dates for which both models $m$ and $m'$ provided forecasts. The relative WIS of model $m$ over the set $\mathcal{I}$ is calculated as
\begin{align*}
\rWIS^{m,geom}_{\mathcal{I}} &= \frac{\theta^m}{\theta^{\text{baseline}}} \text{, where } \\
\theta^m &= \left(\prod_{m'=1}^M \frac{(4 \cdot \vert \mathcal{I}_{m, m'} \vert)^{-1} \sum_{(l, s) \in \mathcal{I}_{m, m'}} \sum_{t=s+1}^{s+4} \WIS(q^m_{l,s,t,1:K},y_{l,t}) }{ (4 \cdot \vert \mathcal{I}_{m, m'} \vert)^{-1} \sum_{(l, s) \in \mathcal{I}_{m, m'}} \sum_{t=s+1}^{s+4} \WIS(q^{m'}_{l,s,t,1:K},y_{l,t})} \right)^{\frac{1}{M}}.
\end{align*}
The figure compares to the alternate strategy of using an arithmetic mean to aggregate across model pairs:
\begin{align*}
\rWIS^{m,arith}_{\mathcal{I}} &= \frac{\theta^m}{\theta^{\text{baseline}}} \text{, where } \\
\theta^m &= \frac{1}{M}\sum_{m'=1}^M \frac{(4 \cdot \vert \mathcal{I}_{m, m'} \vert)^{-1} \sum_{(l, s) \in \mathcal{I}_{m, m'}} \sum_{t=s+1}^{s+4} \WIS(q^m_{l,s,t,1:K},y_{l,t}) }{ (4 \cdot \vert \mathcal{I}_{m, m'} \vert)^{-1} \sum_{(l, s) \in \mathcal{I}_{m, m'}} \sum_{t=s+1}^{s+4} \WIS(q^{m'}_{l,s,t,1:K},y_{l,t})}.
\end{align*}
There is no substantive difference between the relative WIS values obtained using these aggregation strategies.

\newpage

\section{Comparison of logarithmic score and weighted interval score}

In the Forecast Hubs, all forecasts are represented by a set of predictive quantiles at specified probability levels. This has motivated our decision to use the weighted interval score (WIS) for forecast evaluation, as the WIS is a proper score for quantile forecasts. In this section, we discuss challenges with using the logarithmic score when only predictive quantiles are available, and give a qualitative illustration of the difference between the logarithmic score and the WIS. We give a general discussion in Section~\ref{subsec:log_score_wis_example}, and defer some technical details to Section~\ref{subsec:log_score_wis_extrapolation_methods}.

\subsection{Illustrative example comparing the logarithmic score and WIS}
\label{subsec:log_score_wis_example}

Suppose that we have a forecast submission consisting of quantiles of a predictive distribution at the 23 probability levels used in the Forecast Hub, and our goal is to calculate a score that measures the accuracy of this forecast for the observed value $y_{obs}$. One common choice of score for probabilistic forecasts is the logarithmic score; here we consider the negative logarithmic score so that its orientation matches that of the WIS. If a predictive density $f_Y(y)$ is available, the negative logarithmic score is defined as

$$NLS(f_Y, y_{obs}) = -\log[f_Y(y_{obs})]$$

In a setting where we do not have the predictive density $f_Y$, but only predictive quantiles, the logarithmic score cannot be calculated. One possible route forward is to attempt to reconstruct the predictive density from the provided predictive quantiles. We illustrate here that within the limits of the predictive quantiles, it is possible to obtain a reasonable approximation of the predictive density. Here, we do so by fitting a monotonic spline to the provided quantiles to approximate the cumulative distribution function, and then differentiating to approximate the probability density function. However, extrapolating beyond the predictive quantiles is more of a challenge, and requires some assumption to be made about the behavior of the tails of the predictive distribution. This assumption can have a large impact on the log score when observations fall in the tails.

\begin{figure}
    \centering
    \includegraphics[width=\textwidth]{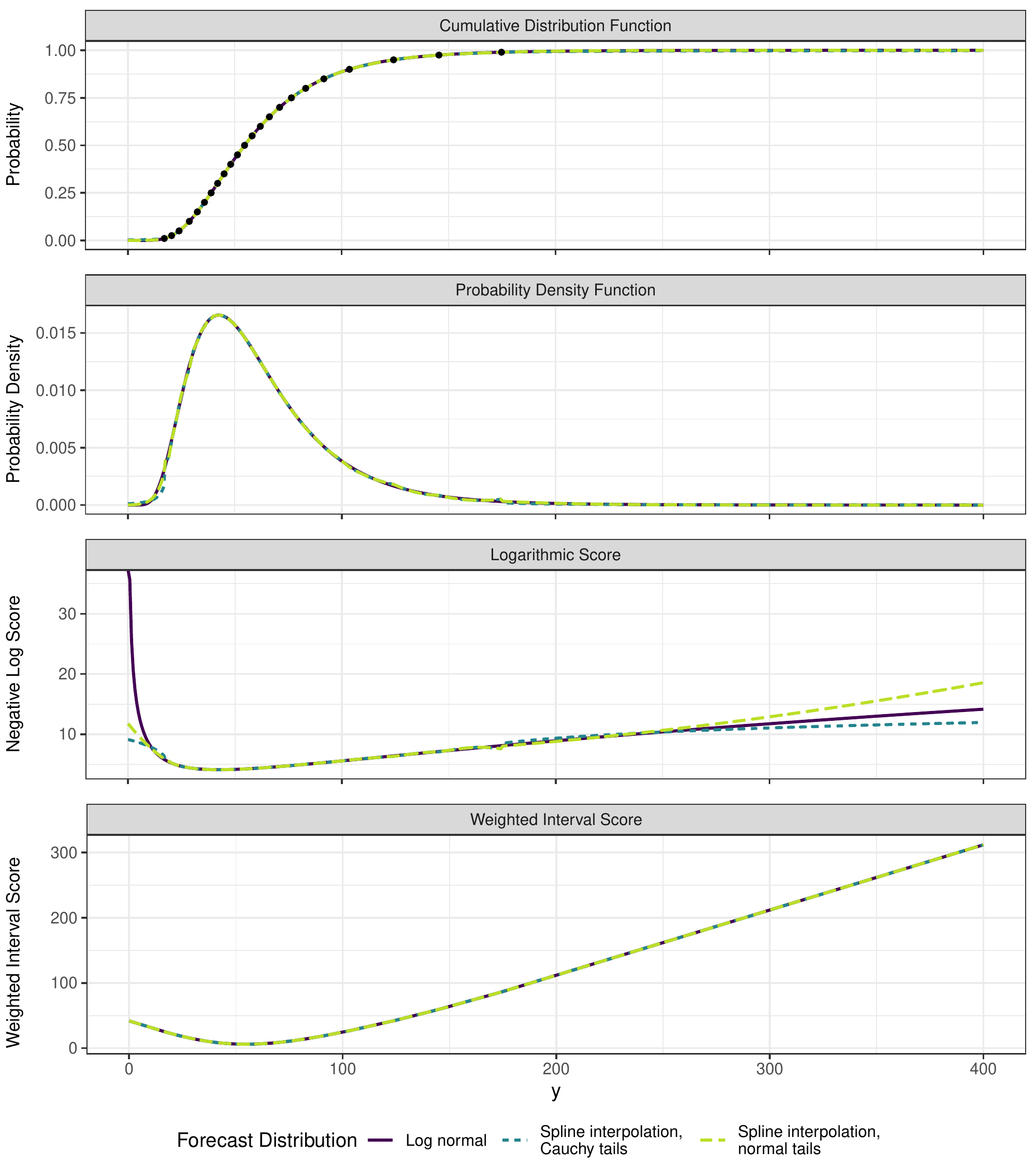}
    \caption{CDFs, PDFs, log scores, and WIS for a $\text{log normal}(4, 0.5)$ forecast and two approximations to it that match 23 specified quantiles but have different behavior in the tails. The predictive quantiles are shown with black points in the top panel.}
    \label{fig:log_score_and_wis}
\end{figure}

Supplemental Figure~\ref{fig:log_score_and_wis} illustrates this using an example of a hypothetical forecast submission consisting of quantiles of the predictive distribution $Y \sim \text{log normal}(4, 0.5)$. We compare this predictive distribution with two approximations of it that are derived from the quantiles: one assuming a normal distribution for the tails, and the second assuming a Cauchy distribution for the tails. Although there are apparently only minor differences in the CDFs and PDFs of these distributions, the log scores diverge substantially in the tails. By construction, all three distributions have the same quantiles at the 23 points that were included in the submission, and so the WIS is identical for all three distributions. To summarize, when only a set of predictive quantiles are provided, there is not enough information to characterize the behavior of the forecast distribution in the tails, and so a log score cannot be calculated. On the other hand, the WIS is defined only in terms of the specified predictive quantiles, and so it does not suffer from this problem.

Qualitatively, the figure shows that both the negative log score and the WIS are minimized when the observed value falls near the center of the predictive distribution, and are larger when the observation falls in the tails. More formally, the negative log score is optimized when the observed value falls at a predictive mode, while the weighted interval score is optimized when the observed value falls at the predictive median. See \cite{bracherEvaluatingEpidemicForecasts2021} for additional discussion of these scores and the continuous ranked probability score.

\subsection{Methods for approximating a predictive density based on predictive quantiles}
\label{subsec:log_score_wis_extrapolation_methods}

Here we describe the methods used in the previous section for obtaining an approximate predictive density $\hat{f}_Y$ based on a set of predictive quantiles $q_1, \ldots, q_K$ at probability levels $\tau_1, \ldots, \tau_K$. The method works in two phases: (1) we estimate the density on the interior of the predictive quantiles as the derivative of a monotonic spline that estimates the CDF; and (2) we approximate the tails with a distribution in a specified location-scale family.

For the interior points, we fit a monotonic spline that interpolates the set of ``observations'' $\{(q_1, \tau_1), \ldots, (q_K, \tau_K)\}$. This spline is an estimate of the predictive CDF, and its derivative estimates the predictive PDF. Because the spline passes through the points $(q_1, \tau_1)$ and $(q_K, \tau_K)$, the integral of its derivative over the interval $[q_1, q_K]$ is equal to $\tau_K - \tau_1$.

We estimate the density for the left and right tails separately, assuming that they come from a specified location-scale family. Setting notation, suppose that $Y = a + b \cdot Z$ where the random variable $Z$ has a specified distribution. Recall that at the probability level $\tau$, a quantile of $Y$ can be calculated in terms of the corresponding quantile of $Z$ via $q_Y(\tau) = a + b \cdot q_Z(\tau)$. Using the quantiles at two probability levels $\tau_i$ and $\tau_j$, we can calculate the value of $b$ using

\begin{align*}
\frac{q_Y(\tau_i) - q_Y(\tau_j)}{q_Z(\tau_i) - q_Z(\tau_j)} &= \frac{a + b \cdot q_Z(\tau_i) - (a + b \cdot q_Z(\tau_j))} {q_Z(\tau_i) - q_Z(\tau_j)} \\
&= b \cdot \frac{q_Z(\tau_i) - q_Z(\tau_j)}{q_Z(\tau_i) - q_Z(\tau_j)} \\
&= b
\end{align*}

Similarly, we can calculate the value of $a$ as

$$q_Y(\tau_i) - b \cdot q_Z(\tau_i) = a + b \cdot q_Z(\tau_i) - b \cdot q_Z(\tau_i) = a$$

In the above expressions, we use the two smallest quantiles when estimating the lower tail and the two largest quantiles when estimating the upper tail. With these choices, by construction the lower tail integrates to $\tau_1$ on the interval $(-\infty, q_1]$ and the upper tail integrates to $1 - \tau_K$ on the interval $[q_K, \infty)$.

\newpage

\section{Quantile ensembles as horizontal combinations of predictive CDFs}

\begin{figure}[H]
    \centering
    \includegraphics[width=\textwidth]{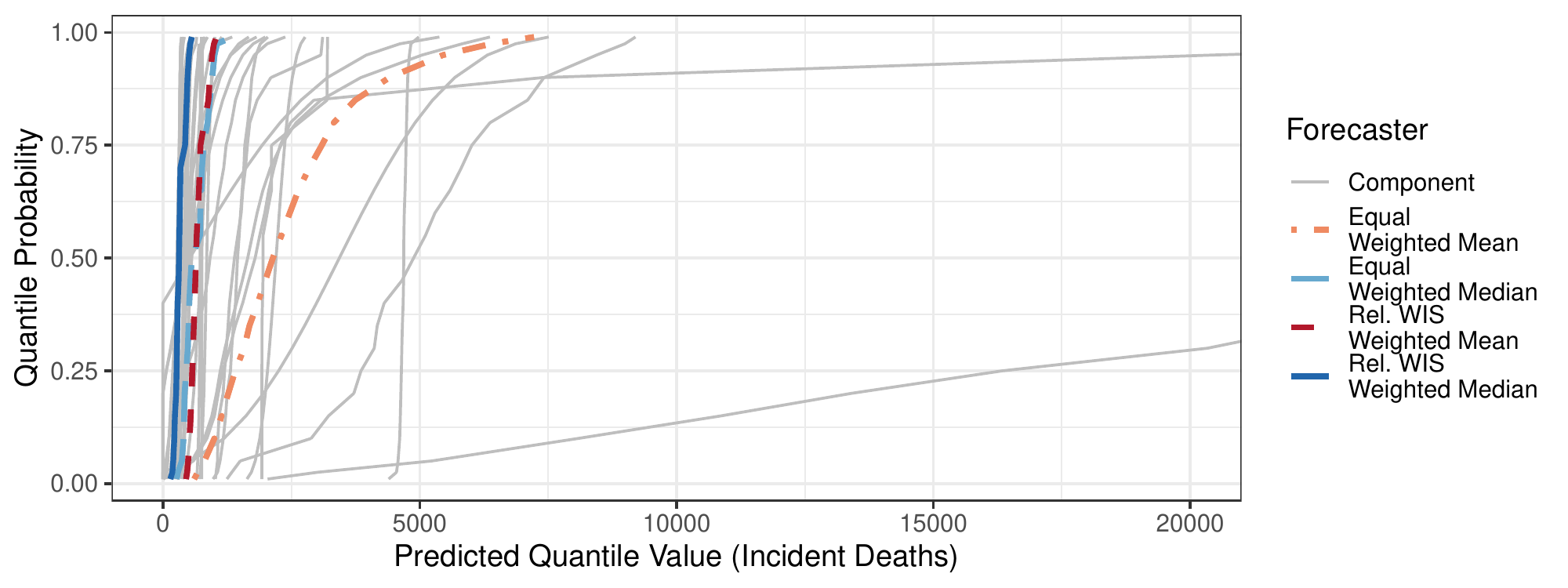}
    \caption{Illustration of four ensemble methods for forecasting incident deaths in Ohio at a forecast horizon of 1 week from February 15, 2021. Each line corresponds to the forecast distribution from one component model or ensemble, and is obtained by interpolating between the 23 predictive quantiles; the resulting curves approximate the predictive CDFs associated with these forecasts. The curves are cut off for two component forecasters with extremely wide predictive distributions. At each quantile level along the vertical axis, the ensemble forecasts are obtained as a combination of the component model forecasts at that quantile level.}
    \label{fig:ensemble_cdf_combination}
\end{figure}

Supplemental Figure \ref{fig:ensemble_cdf_combination} illustrates the ensemble methods considered in this manuscript as horizontal combinations of the cumulative distribution functions of predictive distributions from component forecasters, computing a weighted or unweighted mean or median at each quantile probability level along the vertical axis.

\newpage

\section{Expanded results from primary analysis}

This section includes figures giving additional views of the primary results from Figures 3, 4, 5, and 7 in the article.

\subsection{Full distributions of Weighted Interval Score differences}

\begin{figure}[H]
\includegraphics[width=\textwidth]{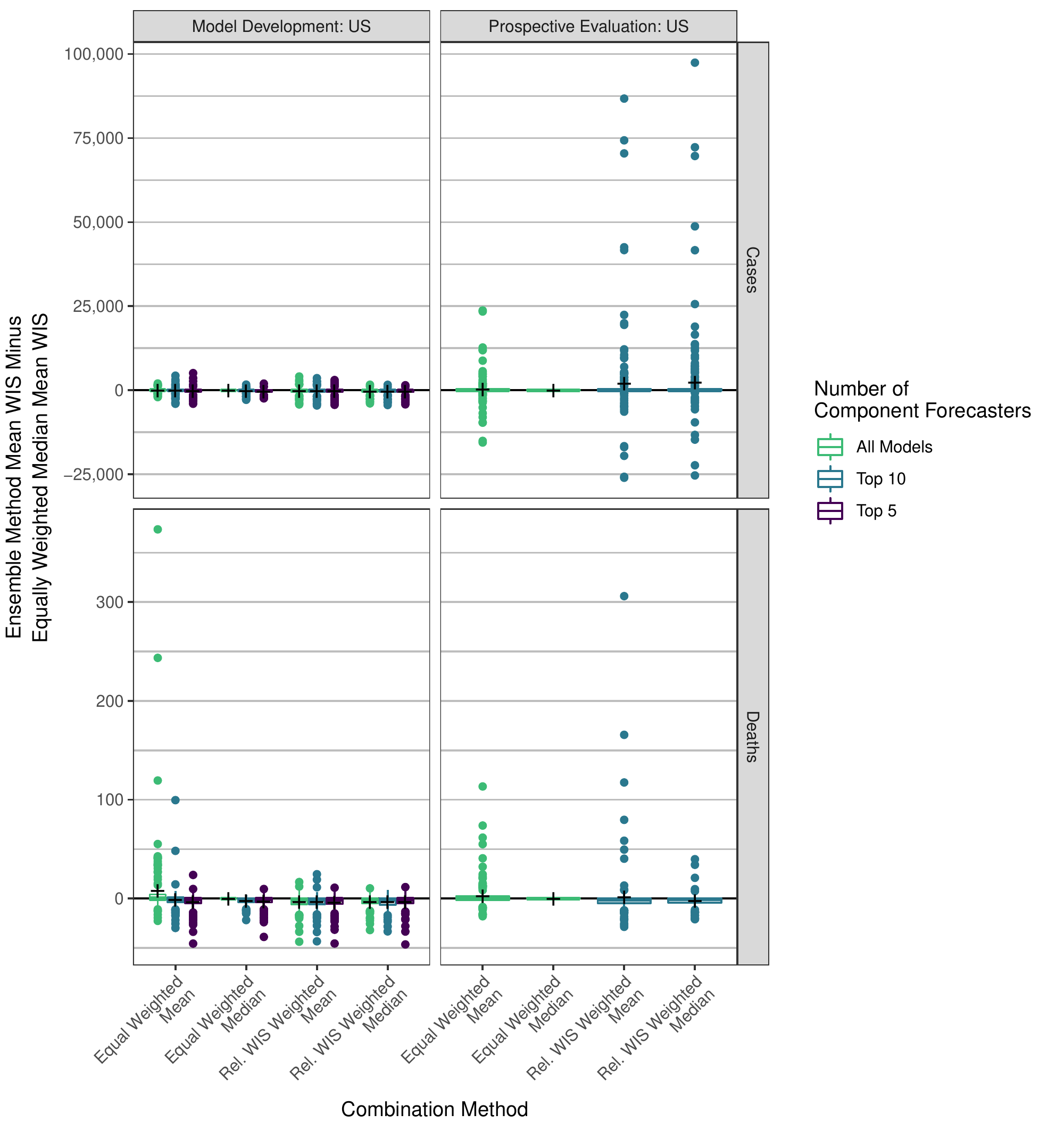}
\caption{Performance measures for ensemble forecasts of weekly cases and deaths in the U.S. The vertical axis is the difference in mean WIS for the given ensemble method and the equally weighted median ensemble.
Boxplots summarize the distribution of these differences in means, averaging across all locations for each combination of forecast date and horizon.
A cross is displayed at the difference in overall mean scores for the specified combination method and the equally weighted median averaging across all locations, forecast dates, and horizons.
A negative value indicates that the given method outperformed the equally weighted median.
}
\label{fig:wis_boxplots_by_phase_all_points_us}
\end{figure}

\begin{figure}[H]
\includegraphics[width=\textwidth]{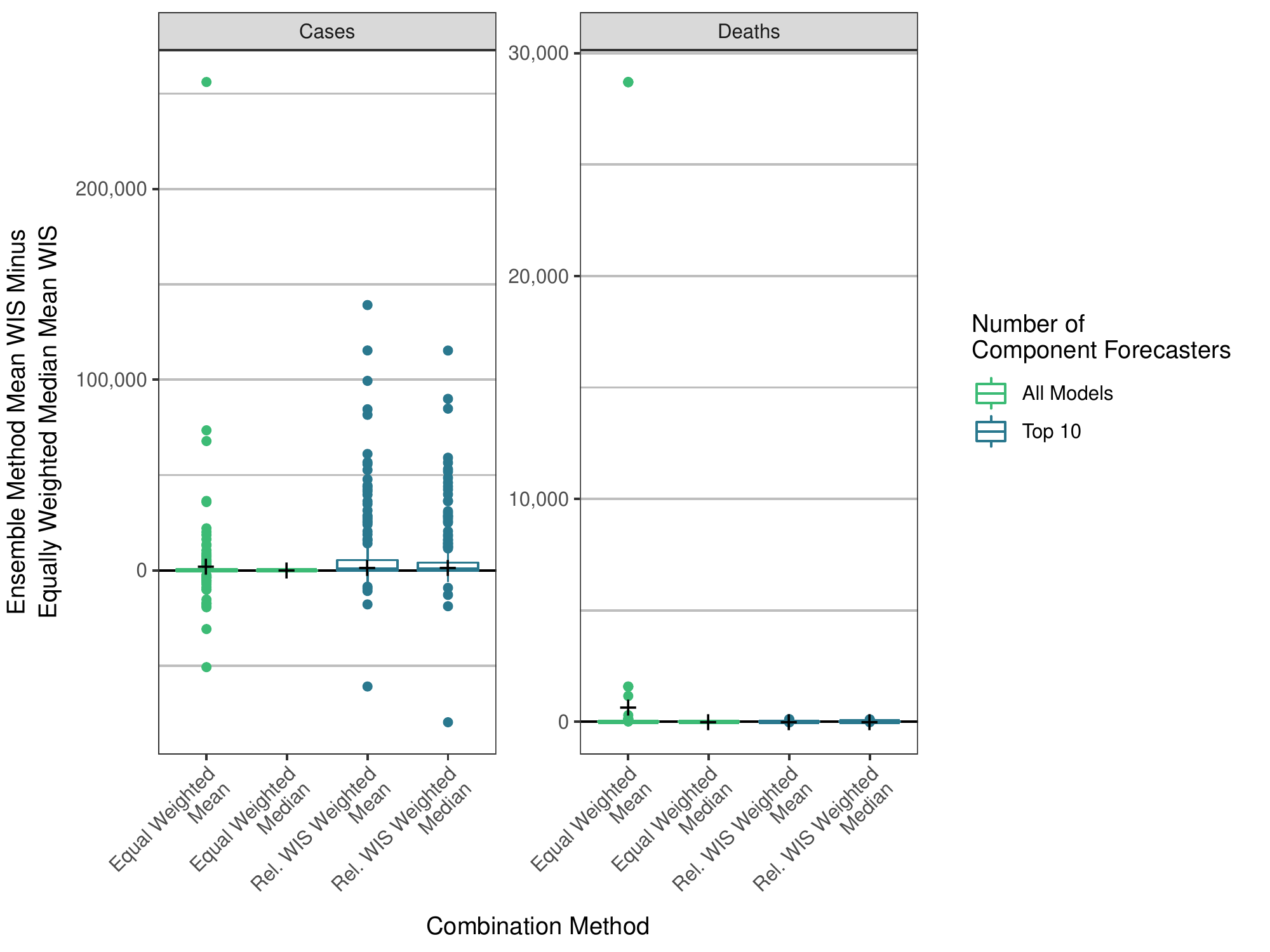}
\caption{Performance measures for ensemble forecasts of weekly cases and deaths in Europe. The vertical axis is the difference in mean WIS for the given ensemble method and the equally weighted median ensemble.
Boxplots summarize the distribution of these differences in means, averaging across all locations for each combination of forecast date and horizon.
A cross is displayed at the difference in overall mean scores for the specified combination method and the equally weighted median averaging across all locations, forecast dates, and horizons.
A negative value indicates that the given method outperformed the equally weighted median.
}
\label{fig:wis_boxplots_by_phase_all_points_eu}
\end{figure}

For legibility, Figures 3 and 7 in the main text displayed the central tendency (central quantiles and means) of differences in weighted interval scores between the different methods, but suppressed outliers corresponding to individual combinations of forecast dates and horizons with large differences between the equal weighted median ensemble and another ensemble method. Supplemental Figures \ref{fig:wis_boxplots_by_phase_all_points_us} and \ref{fig:wis_boxplots_by_phase_all_points_eu} display full box plots including outliers that were suppressed in the main text.

\newpage

\subsection{Scores by forecast creation date}

\begin{figure}
  \includegraphics[width=6in]{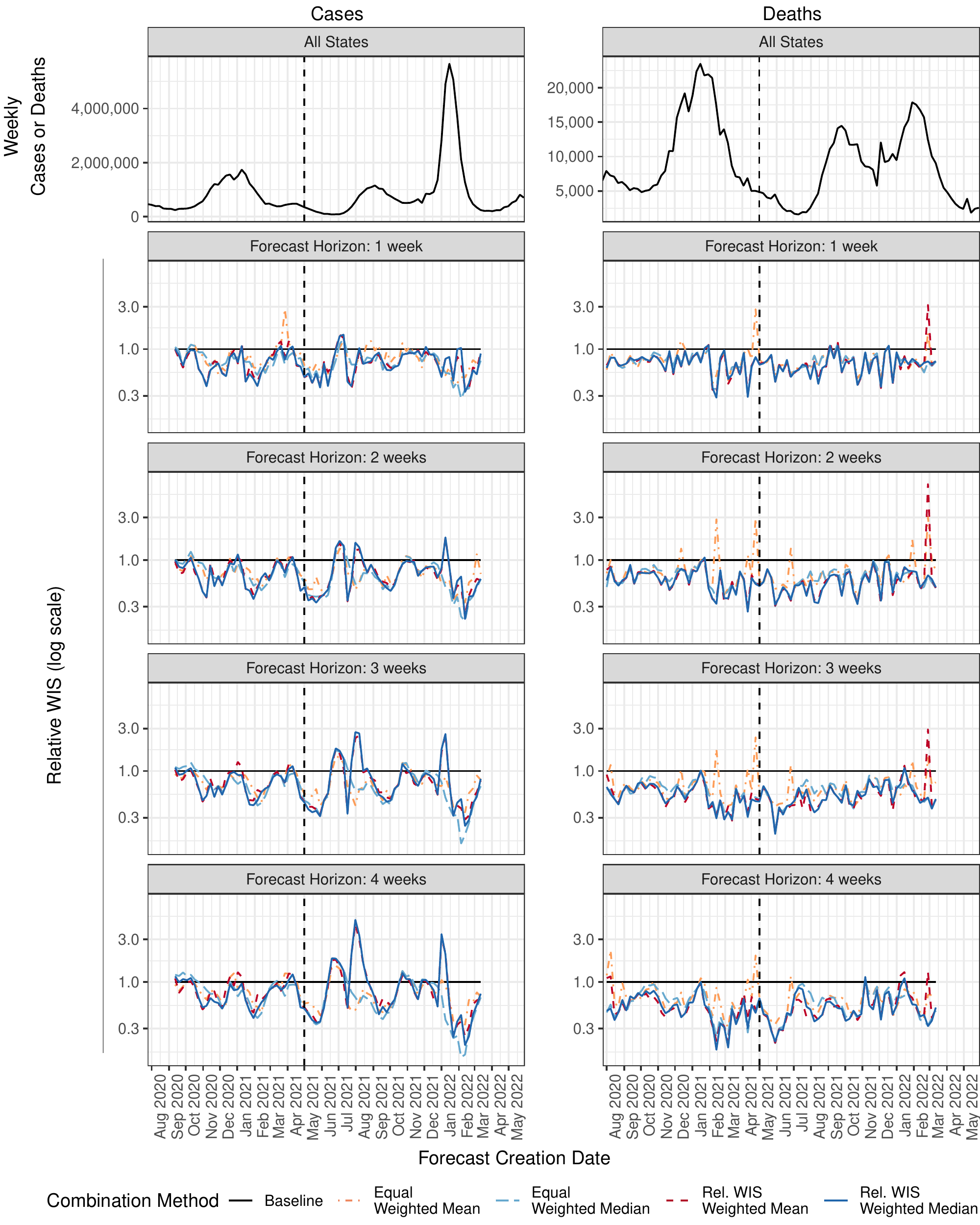}
  \caption{Weekly reported cases and deaths at the national level in the United States and mean weighted interval scores (WIS) relative to the baseline for state-level forecasts over time for four ensembles.
  Mean WIS is calculated separately for each combination of forecast horizon and forecast creation date, averaging across all states and territories, and then normalized relative to the mean WIS for the baseline model.
  Lower scores indicate better forecast performance.
  A vertical dashed line is shown at the start of the prospective evaluation phase.
  }
  \label{fig:rel_WIS_over_time_us}
\end{figure}

\begin{figure}
  \includegraphics[width=6in]{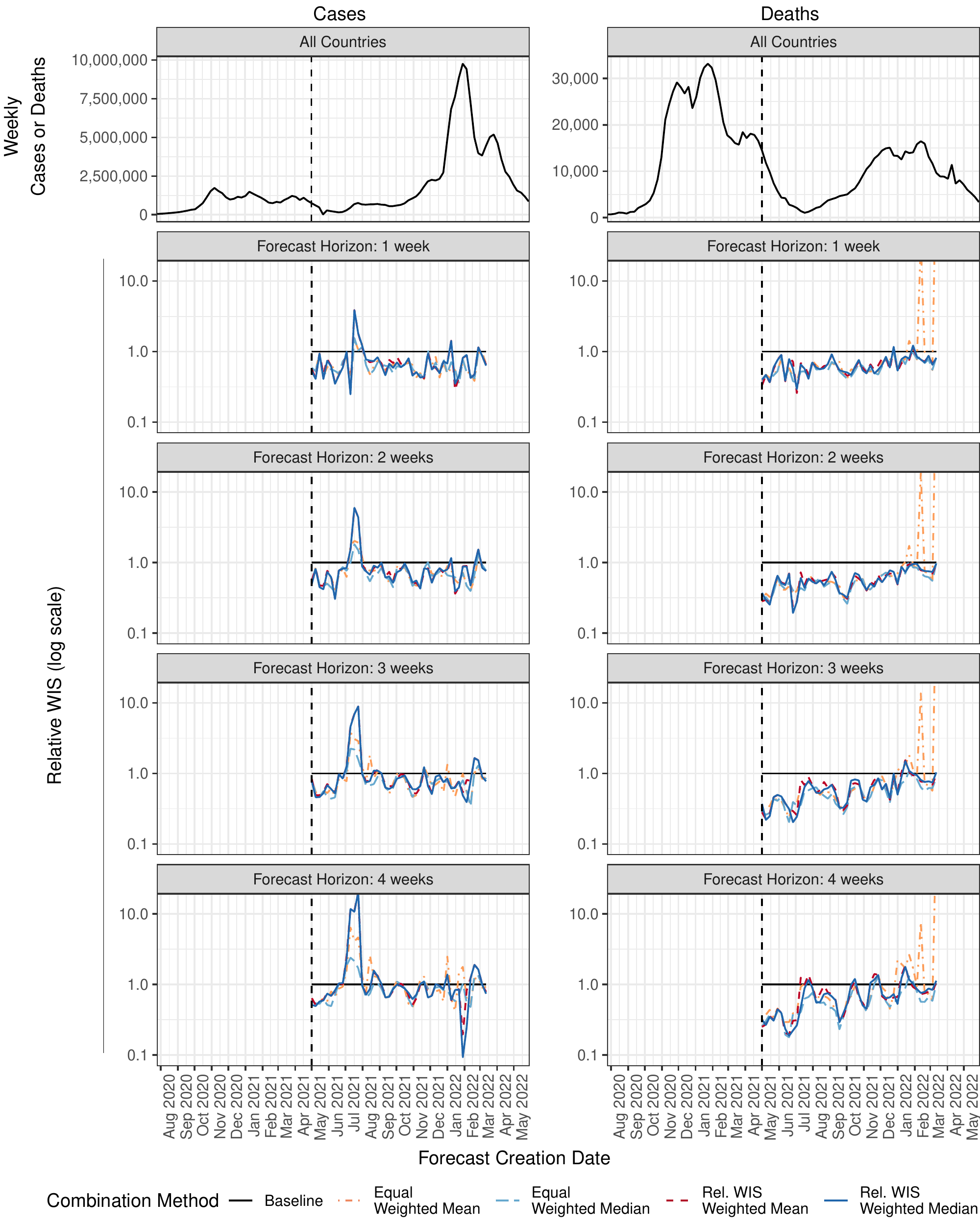}
  \caption{Weekly reported cases and deaths aggregated across all European countries included in the European Forecast Hub and mean weighted interval scores (WIS) relative to the baseline for state-level forecasts over time for four ensembles.
  Mean WIS is calculated separately for each combination of forecast horizon and forecast creation date, averaging across all states and territories, and then normalized relative to the mean WIS for the baseline model.
  Lower scores indicate better forecast performance.
  A vertical dashed line is shown at the start of the prospective evaluation phase.
}
  \label{fig:rel_WIS_over_time_euro}
\end{figure}

Supplemental Figures \ref{fig:rel_WIS_over_time_us} and \ref{fig:rel_WIS_over_time_euro} show relative WIS for the ensemble methods over time for forecasts in the US and in Europe respectively.
The included ensemble methods are 1) an equally weighted mean ensemble, 2) an equally weighted median ensemble, 3) a weighted mean ensemble, and 4) a weighted median ensemble.
Both of the weighted ensembles combine the ten component forecasters with best individual performance as measured by the relative WIS, and are trained on a sliding 12-week window.
The component forecasters included in the trained ensembles are updated each week based on performance during the training window.
  
\newpage

\subsection{95\% prediction interval widths}

\begin{figure}
  \includegraphics[width=6in]{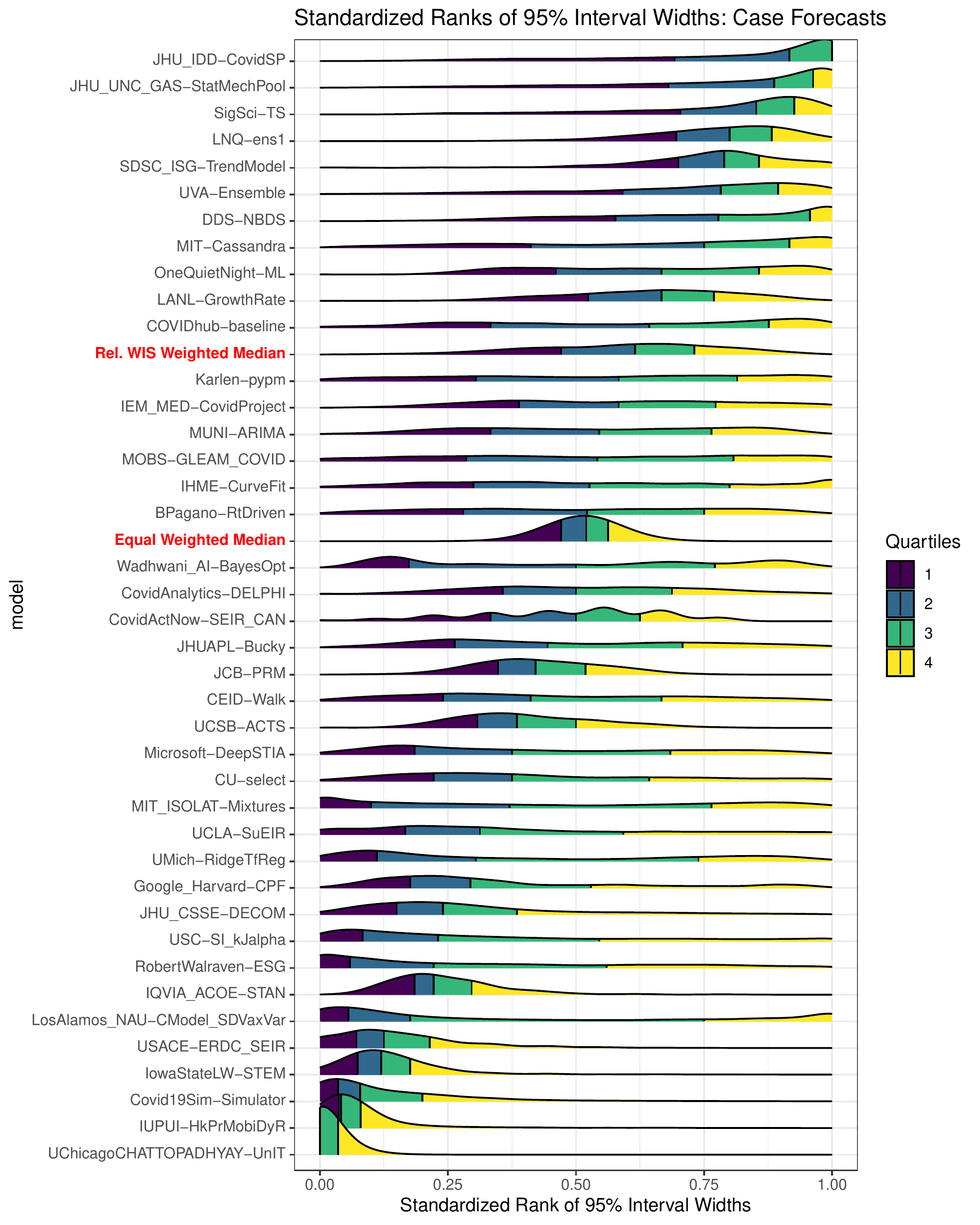}
  \caption{Standardized ranks of 95\% prediction intervals for weekly cases from component forecasters, the equally weighted median ensemble, and the relative WIS weighted median ensemble. For each combination of location, forecast date, and forecast horizon, we rank the widths of 95\% prediction intervals from all forecasters that submitted the relevant forecast on a scale from 0 to 1, where the forecaster with the narrowest interval has rank 0 and the forecaster with the widest interval has rank 1. Density plots summarize the distribution of these ranks for each forecaster; forecasters are sorted by their median rank.}
  \label{fig:95_pi_width_cases}
\end{figure}

\begin{figure}
  \includegraphics[width=6in]{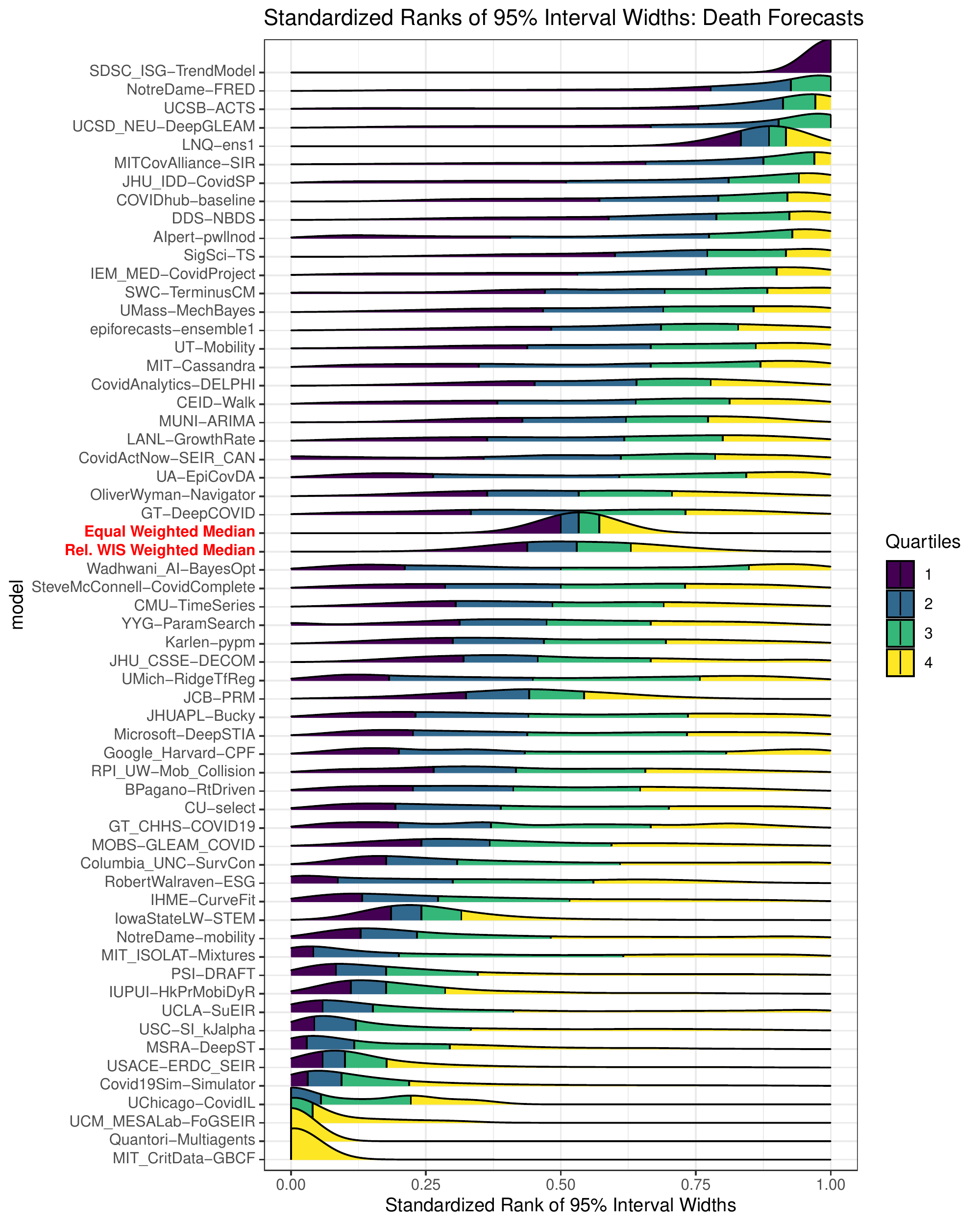}
  \caption{Standardized ranks of 95\% prediction intervals for weekly deaths from component forecasters, the equally weighted median ensemble, and the relative WIS weighted median ensemble. For each combination of location, forecast date, and forecast horizon, we rank the widths of 95\% prediction intervals from all forecasters that submitted the relevant forecast on a scale from 0 to 1, where the forecaster with the narrowest interval has rank 0 and the forecaster with the widest interval has rank 1. Density plots summarize the distribution of these ranks for each forecaster; forecasters are sorted by their median rank.}
  \label{fig:95_pi_width_deaths}
\end{figure}

Supplemental Figures~\ref{fig:95_pi_width_cases} and \ref{fig:95_pi_width_deaths} illustrate that the widths of 95\% prediction intervals for the ensemble forecasters generally fall in the middle of the widths of the intervals from the component forecasters. This is true because the ensemble forecasts are a (weighted) mean or median of predictive quantiles of the component forecasters. In particular, the equal weighted median forecast typically has a 95\% interval width that ranks very close to the middle of the other forecasters' interval widths.

We note that interval coverage rates are not in themselves a measure of forecast skill, and it is important to also consider whether the interval contains the forecasted quantity at the nominal coverage rate (i.e., whether the forecasts are well-calibrated). The figures in the main text include displays of calibration as well as proper scores that measure both calibration and sharpness together.

\newpage

\subsection{Impact of reporting anomalies}

We conducted a supplemental analysis in which we removed forecasts that were affected by reporting anomalies before calculating summaries of forecast performance. We catalogued two types of reporting anomalies, as illustrated in Supplemental Figure~\ref{fig:anomalies_example}:
\begin{enumerate}
    \item \textbf{Outliers} were identified manually by examining plots of the data. Negative weekly counts and other observations that did not appear to match local trends were recorded as outliers.
    \item \textbf{Revisions} were identified automatically. The value for a particular week was identified as a large revision if the difference between the original reported value and the final reported value was at least 20, and that difference was at least 40\% of the initial reported value or the final reported value.
\end{enumerate}

\begin{figure}
  \includegraphics[width=\textwidth]{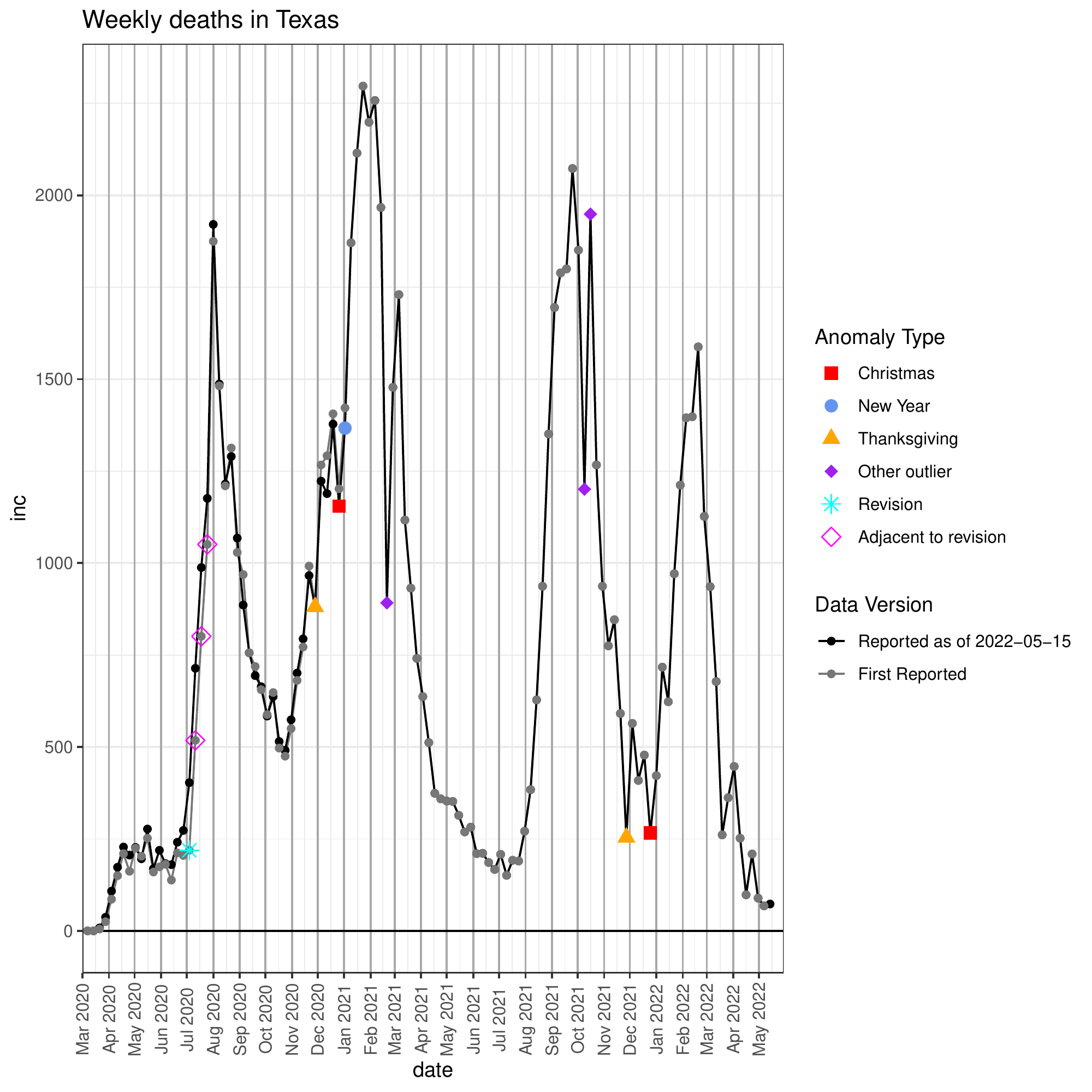}
  \caption{An illustration of data anomalies identified for weekly deaths in the state of Texas. Identified outliers include suppressed reporting during holidays and a winter storm in February 2021, as well as a period of reduced reporting followed by catch-up reporting in October 2021.}
  \label{fig:anomalies_example}
\end{figure}

For the purpose of this analysis, forecasts with a target end date coinciding with an observation that was identified as an outlier were excluded. This is because we would prefer forecasting methods to focus on capturing the epidemiological process rather than aspects of the reporting process that lead to outliers.

Forecasts with a forecast date on the date of a value that was later revised, or within the following 3 weeks unless the revision had already been made by the time of the forecast, were excluded. These forecasts were excluded because the input data used for the forecast were not a reliable indicator of the state of the epidemic at the time of the forecast. One might reasonably expect forecasters to account for the possibility of such data revisions, but this analysis represents a conservative examination of whether these revisions affected the main results in the article.

Together, these criteria led to removal of 531 combinations of location, forecast date, and forecast horizon out of 19,256 such combinations throughout the model development and prospective evaluation phases in the U.S.

\begin{figure}[H]
  \includegraphics[width=\textwidth]{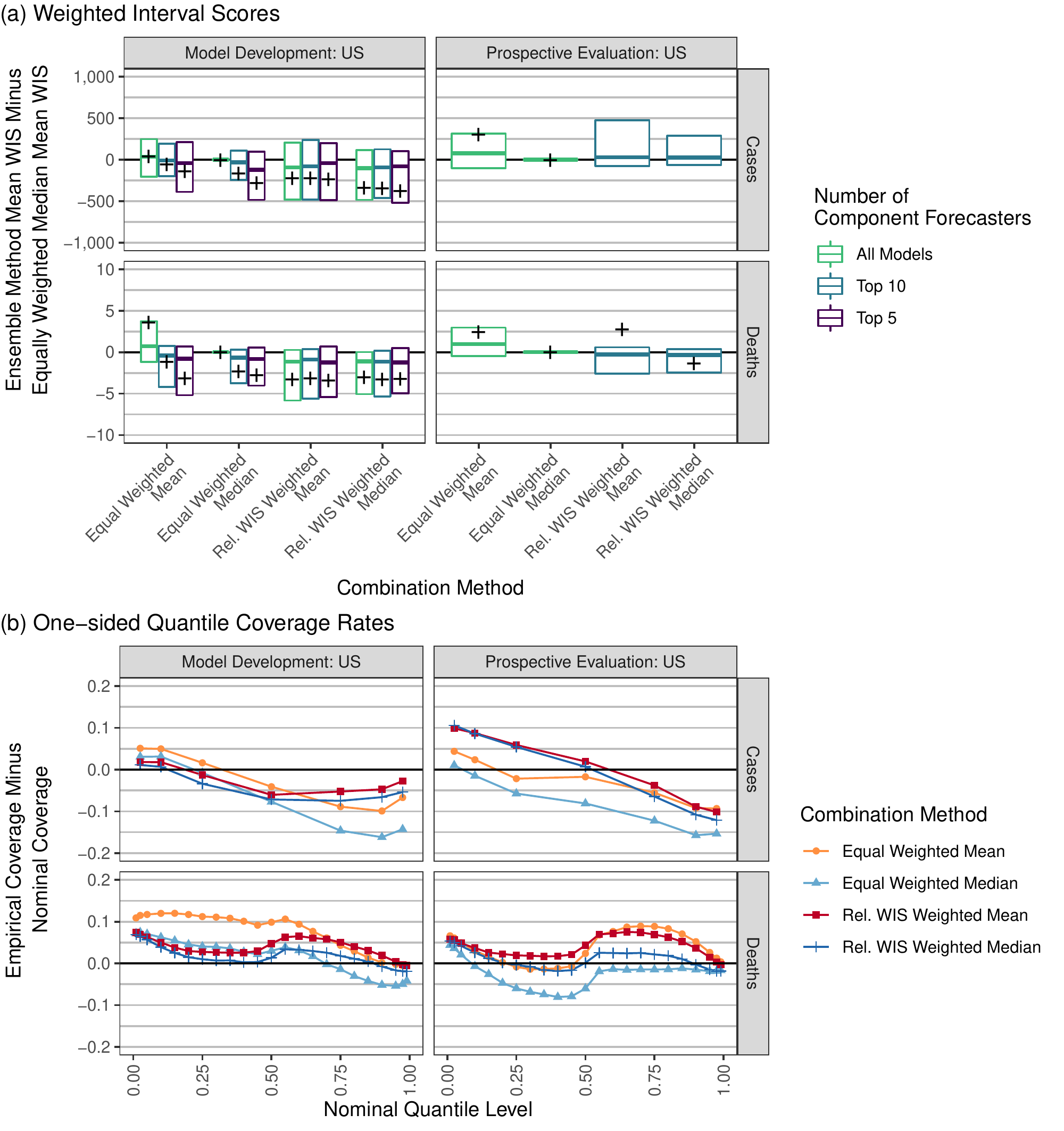}
  \caption{Summaries of forecast performance after removing forecasts affected by reporting anomalies. Panel (a) shows the 25th percentile, median, and 75th percentile of differences in mean WIS between specified ensemble methods and the equally weighted median ensemble, where the means average across locations for each combination of forecast date and forecast horizon. Crosses show the difference in overall mean WIS averaging across all locations, forecast dates, and forecast horizons. Panel (b) shows the calibration of predictive quantiles, with the difference between the empirical coverage rate and the nominal coverage rate on the vertical axis. A well calibrated model will have a difference between the empirical coverage rate and the nominal quantile level that is approximately zero. A method that generates conservative two-sided intervals would have a difference that is negative for nominal quantile levels less than 0.5 and positive for nominal quantile levels greater than 0.5.}
  \label{fig:wis_calibration_us_non_anomalous}
\end{figure}

\begin{figure}[H]
  \includegraphics[width=\textwidth]{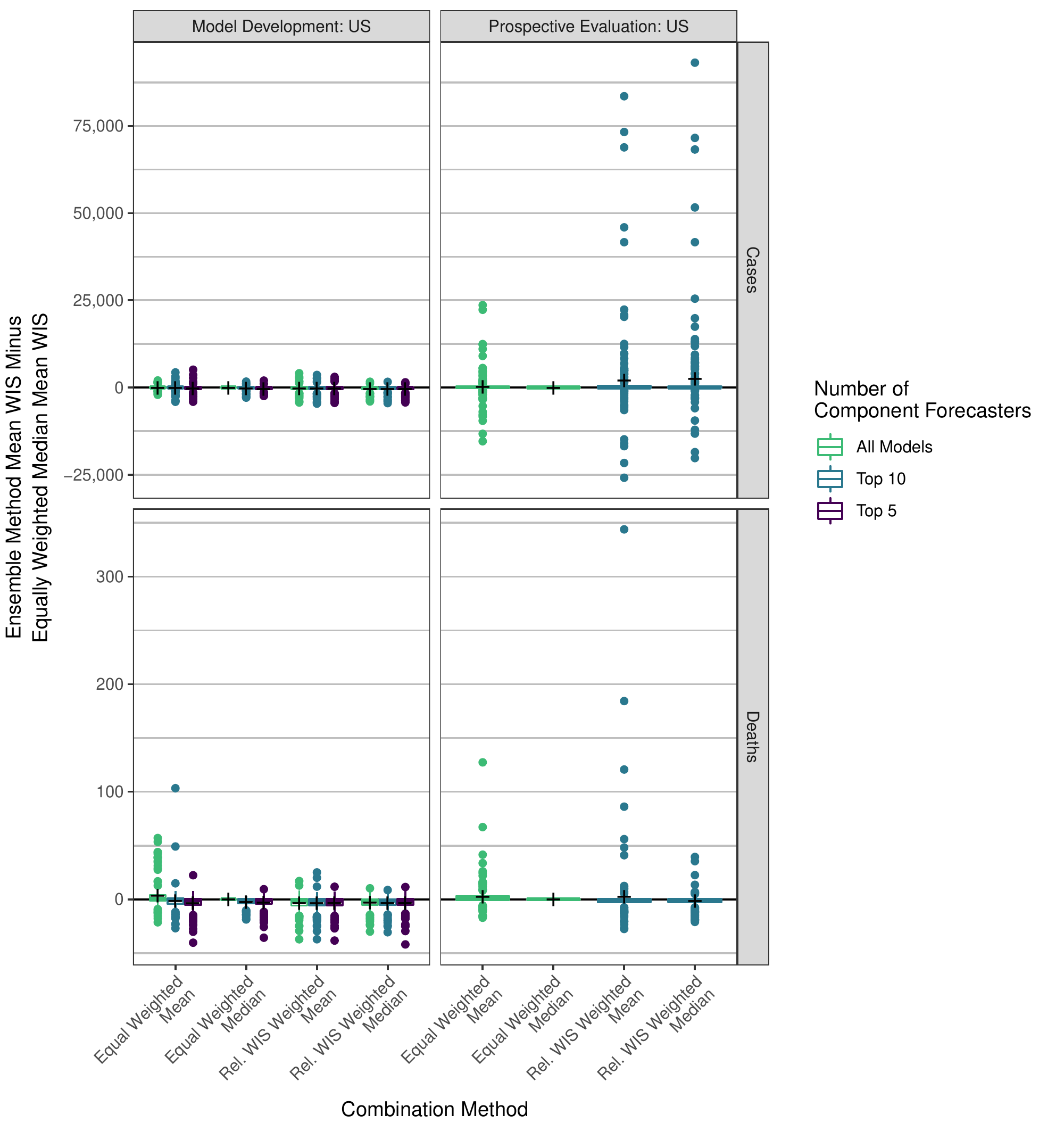}
  \caption{Summaries of forecast performance after removing forecasts affected by reporting anomalies. Boxplots summarize the distribution of differences in mean WIS between specified ensemble methods and the equally weighted median ensemble, where the means average across locations for each combination of forecast date and forecast horizon. Crosses show the overall difference in mean WIS averaging across all locations, forecast dates, and forecast horizons.}
  \label{fig:wis_calibration_us_non_anomalous_all_points}
\end{figure}

Supplemental Figures~\ref{fig:wis_calibration_us_non_anomalous} and \ref{fig:wis_calibration_us_non_anomalous_all_points} mirror Figure 3 in the primary text and Supplemental Figure~\ref{fig:wis_boxplots_by_phase_all_points_us}, summarizing forecast skill after removing scores affected by reporting anomalies. Although the forecasts affected by reporting anomalies generally had higher WIS values than other forecasts, they did not have unusually large \textit{differences} in WIS between forecasting methods. The results about the relative performance of ensemble methods hold stable whether or not the forecasts affected by data anomalies are removed.

\newpage

\newpage

\section{Model variations considered during development phase}

Here we present some results for model variations considered during the model development phase. All figures here represent only scores for forecast dates before May 3, 2021.

\subsection{Additional combination methods and training window sizes}

The manuscript gives results for equally weighted mean and median ensembles, and relative WIS weighted mean and median ensembles, using a fixed training set window size of the 12 weeks prior to the forecast date. Here we show results on the development set for forecasts using a range of training set window sizes including 4 weeks, 8 weeks, 12 weeks, and all available forecast history. We also consider two additional combination methods along with those presented in the main text.

The first of these new combination methods is a weighted mean. As a reminder,  $q^m_{l,s,t,k}$ denotes the predictive quantile at probability level $k$ from component model $m$ at location $l$, forecast date $s$, and target end date $t$. With this notation, the weighted mean ensemble forecast quantiles are calculated as
$$q^\text{ens}_{l,s,t,k} = \sum_{m = 1}^M w^m_{s} q^{m}_{l,s,t,k}.$$
The model weights $w^m_s$ are constrained to be non-negative and sum to one; in case of missing forecasts, the weights for any missing models are set to zero and the remaining weights are rescaled to sum to 1.
As indicated by the subscript $s$, the weights $w^m_s$ are updated each week by optimizing the ensemble WIS over the training window of the specified number of weeks before the forecast date $s$.

The second of the new combination methods is a weighted median ensemble that uses the weights estimated for the weighted mean ensemble. This offers comparable flexibility to the weighted mean ensemble, but has the disadvantage that the weights are not obtained by optimizing the forecast skill of the method that is actually used for forecast combination. Direct estimation of the weights for a weighted median by optimizing ensemble WIS is challenging because the objective function is not differentiable in the weights; the optimization problem is a mixed integer linear program, which is computationally demanding. In other experiments, we also considered a method for computing an approximate weighted median by the smoothing weighted distribution of predictive quantiles from component forecasters. However, this method's performance was not substantively different from the other methods considered here and we omit those results for brevity.

\begin{figure}[H]
  \includegraphics[width=\textwidth]{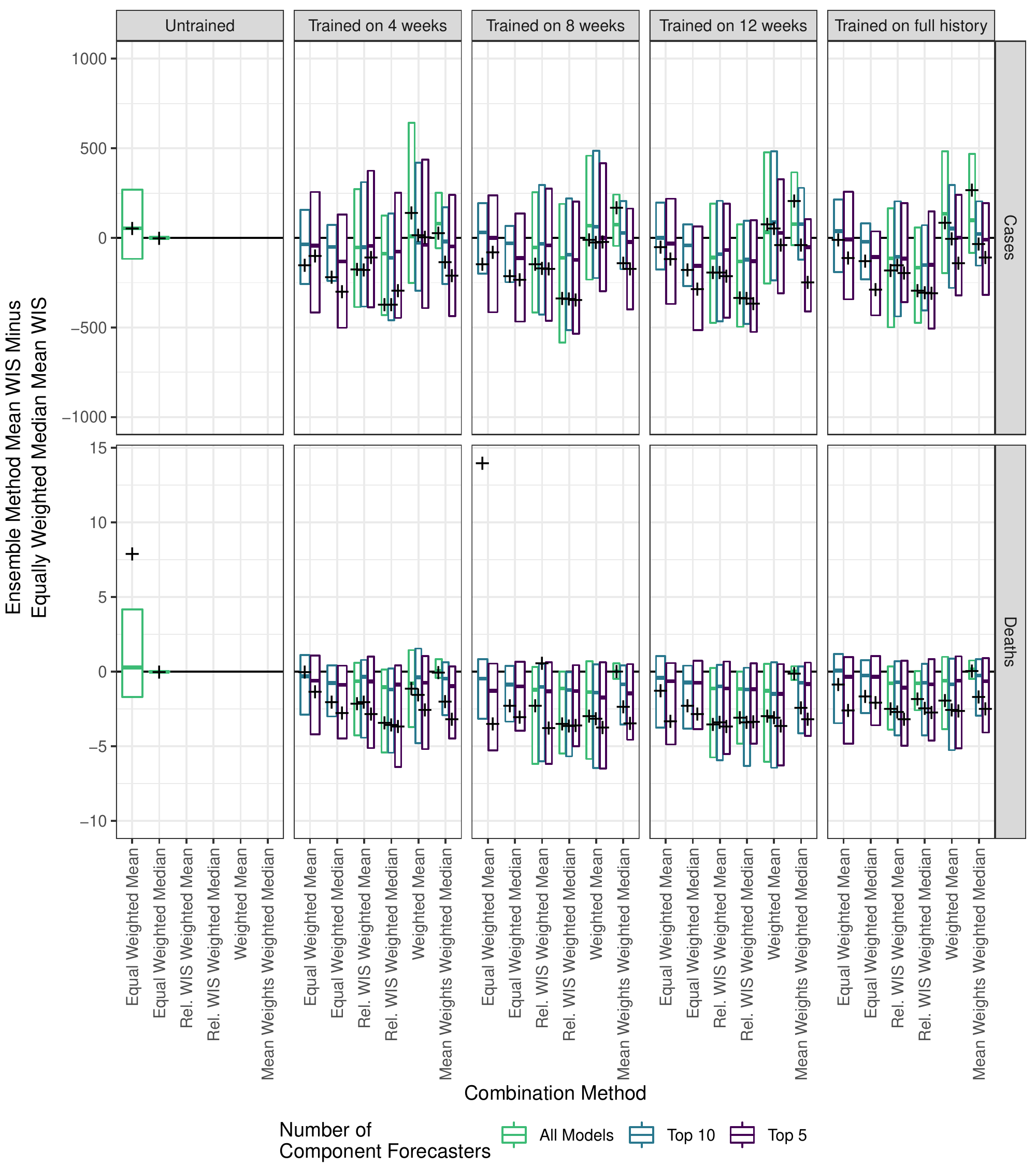}
  \caption{The 25th percentile, median, and 75th percentile of differences in mean WIS between specified ensemble methods and the equally weighted median of all component forcasters, where the means average across locations for each combination of forecast date and forecast horizon. Crosses show the overall difference in mean WIS averaging across all locations, forecast dates, and forecast horizons.
  A negative value indicates that the method corresponding to a particular combination of training set size, number of component forecasters included, and combination method outperformed the ensemble calculated as an equally weighted median of all component forecasts.}
  \label{fig:wis_combination_methods_central_only}
\end{figure}

\begin{figure}[H]
  \includegraphics[width=\textwidth]{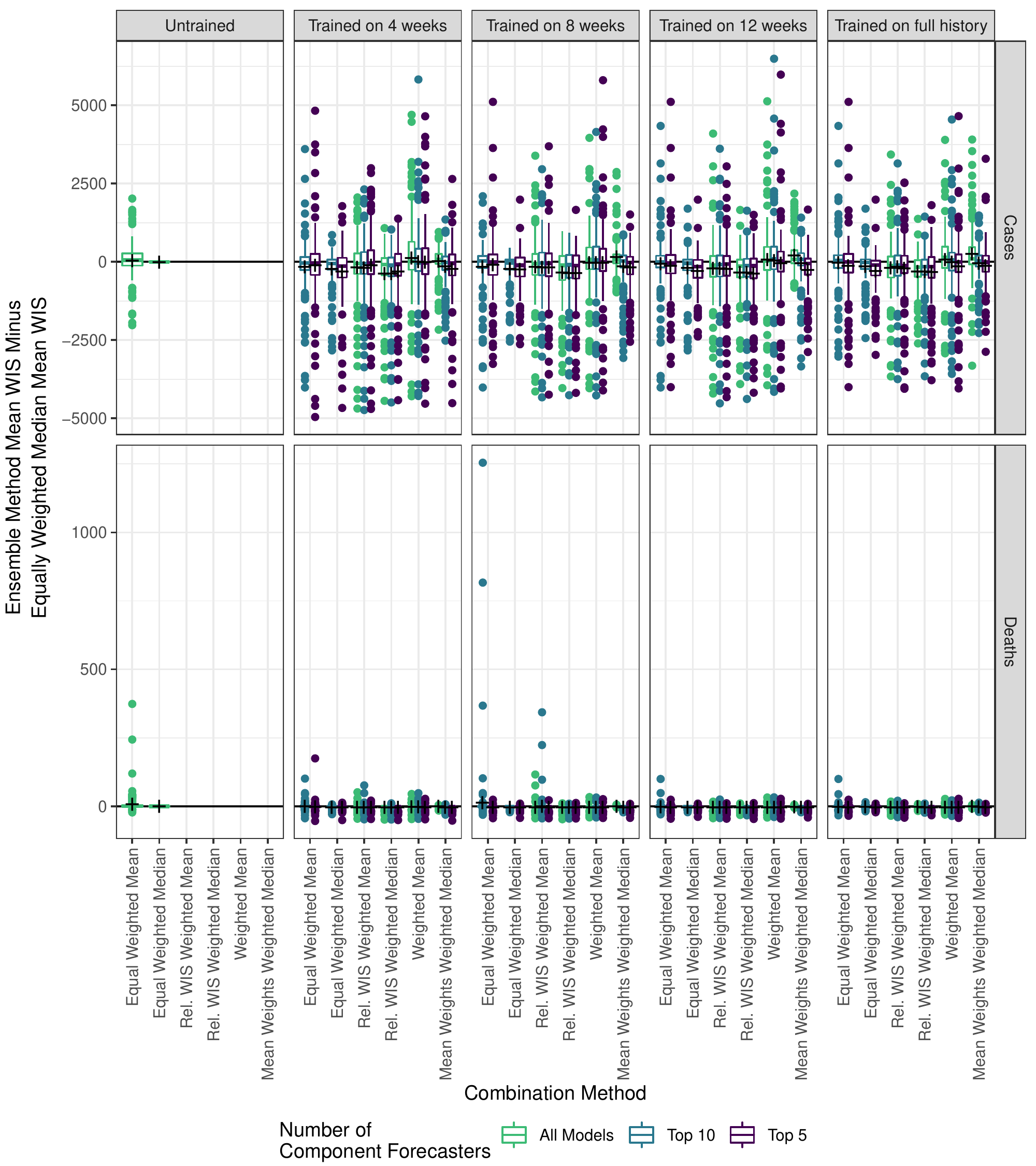}
  \caption{Boxplots summarizing the full distribution of differences in mean WIS between specified ensemble methods and the equally weighted median of all component forcasters, where the means average across locations for each combination of forecast date and forecast horizon. Crosses show the overall difference in mean WIS averaging across all locations, forecast dates, and forecast horizons.
  A negative value indicates that the method corresponding to a particular combination of training set size, number of component forecasters included, and combination method outperformed the ensemble calculated as an equally weighted median of all component forecasts.}
  \label{fig:wis_combination_methods_all_points}
\end{figure}

Supplemental Figures~\ref{fig:wis_combination_methods_central_only} and \ref{fig:wis_combination_methods_all_points} display the results of this expanded comparison including all combinations of the training set window size, the six combination methods, and three variations on the number of top-performing component forecasters included in the ensemble. Across all combinations of training set window size, number of component forecasters included, and target variable (cases or deaths), the relative WIS weighted median ensemble had the most stable performance. For deaths, it had the best mean WIS for all training set window sizes, though it had similar performance to the equally weighted median of the top 5 models. For cases, it was more often matched by other methods, though the performance of the other methods was more inconsistent across different settings for other tuning parameters. Across most settings, using a relative WIS weighted mean or median offered an improvement in mean WIS over taking an equally weighted mean or median of top performing models.

There is perhaps a slight indication that an intermediate training set size of 8 to 12 weeks is better than training on 4 weeks or the full available history, but this signal is not strong. Some of the combination methods were better when fewer top models were included, but the relative WIS weighted median method was not sensitive to this setting.

We selected the relative WIS weighted mean and median ensembles for the prospective evaluation because they were consistently better than both more flexibly weighted methods and equally weighted combinations of top-performing components when forecasting cases, and were comparable to the best of the other approaches when forecasting deaths. We selected an intermediate training set window size of 12 weeks because both the relative WIS weighted mean and median methods did well with that training set size. We selected including the top ten component forecasters as an intermediate setting for that tuning parameter, though we did not see a strong reason to prefer it to the other possibilities we considered.

\newpage

\subsection{Separate weights at different forecast horizons}

We considered a variation on the relative WIS weighted median ensemble that estimated separate weights for each of the one through four week ahead forecast horizons. In this approach, the relative WIS of component forecasters was calculated separately at each forecast horizon, and the estimation of the weighting parameter $\theta$ was performed separately to optimize the forecast skill of the ensemble at each horizon.

\begin{figure}
  \includegraphics[width=\textwidth]{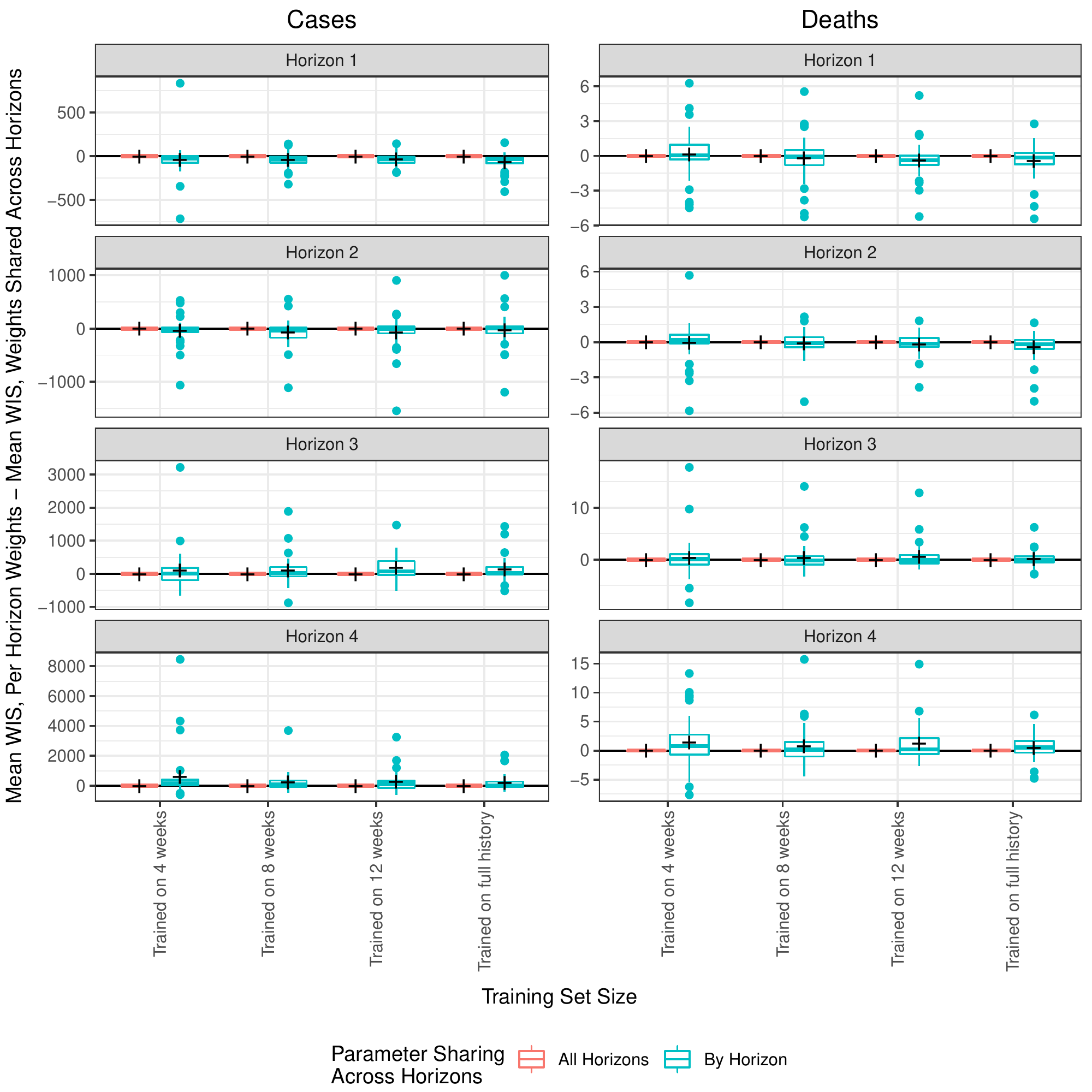}
  \caption{Boxplots summarizing forecast skill for forecasts of weekly cases, varying whether model weights are shared across all forecast horizons or are estimated separately for each forecast horizon.
  The vertical axis is the difference in mean skill for the given ensemble specification when component weights are shared across all horizons and the same specification with separate component weights for each forecast horizon.
  The boxplots summarize the distribution of these differences for each combination of forecast date and horizon, averaging across all locations.
  A cross is displayed at the difference in overall mean scores.
  A negative value indicates that the method with separate component weights for each forecast horizon outperformed the corresponding specification with weights shared across forecast horizons.
  For this analysis, only results for relative WIS weighted ensembles combining the ten best individual component forecasters are presented.}
  \label{fig:wis_grouping_by_horizon}
\end{figure}

Supplemental Figure~\ref{fig:wis_grouping_by_horizon} shows the results of this per-horizon weighting scheme for the relative WIS weighted median ensemble combining the top 10 component forecasters. We found that using separate model weights at each forecast horizon led to small improvements in mean WIS at short-term forecast horizons of one to two weeks ahead, but slightly worse mean WIS at longer forecast horizons of three to four weeks ahead. 

We see two possible contributing factors to these results. First, forecasts at long horizons have scores that are larger in magnitude than forecasts at short horizons, and so tend to dominate the overall score when averaging across horizons. This may result in weights that favor performance at longer term horizons when weights are shared across horizons, thereby harming performance of forecasts at short horizons. Second, sharing weights across horizons may be particularly helpful for longer term forecasts because of the gain in the training set sample size that comes with weight sharing. For example, with a training set size of four weeks and weights estimated separately by horizon, only one week's worth of forecasts are actually included in the training set for weight estimation at a horizon of four weeks, because the target data for four-week ahead forecasts made within the past three weeks have not yet been observed at the time of weight estimation. Sharing weights across horizons means that more information about model performance is available for weight estimation at these longer horizons. In support of this explanation, note that the magnitude of relative losses in forecast skill from estimating per-horizon weights at longer forecast horizons decreases as the training set size increases.

These results suggest the possibility of a blended strategy, where per-horizon estimation is used to obtain the weights for short horizons but the weights for longer horizons are estimated by sharing information across all horizons. In light of the small magnitude of the gains from using a per-horizon weighting at short horizons, we decided to pursue a unified approach of using shared weights across all horizons to reduce methodological and narrative complexity.

\newpage

\subsection{Separate weights at different quantile levels}

We considered strategies for estimation of separate model weights at each quantile level rather than sharing model weights across all quantile levels.
We considered this possibility for both the relative WIS weighted median ensemble and the convex mean ensemble that directly optimizes the component weights rather than setting them to be a sigmoid function of the relative WIS.
In both variations, weights were estimated by optimizing the contribution to the WIS from each quantile level separately, i.e., the pinball loss.
Similarly, for the relative WIS weighted median, the relative WIS for component models that is used as an input for calculating model weights was obtained separately based on the contribution to WIS at each quantile level.

\begin{figure}[H]
  \includegraphics[width=\textwidth]{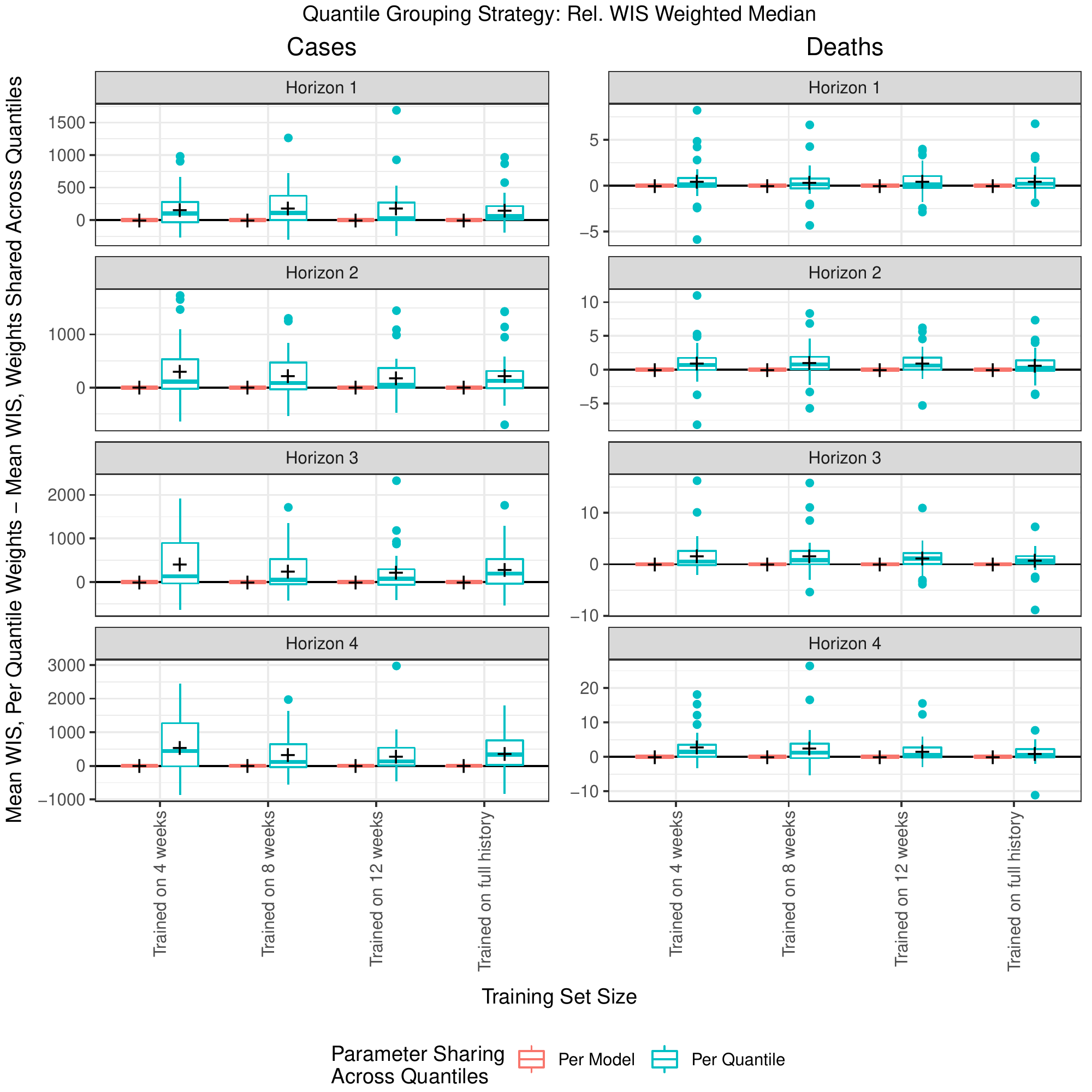}
  \caption{Boxplots summarizing forecast skill for forecasts of weekly cases from a relative WIS weighted median ensemble, varying whether model weights are shared across all quantile levels (``Per Model") or are estimated separately for each quantile level (``Per Quantile").
  The vertical axis is the difference in mean skill for the given ensemble specification when component weights are shared across all quantile levels and the same specification with separate component weights for each quantile level.
  The boxplots summarize the distribution of these differences for each combination of forecast date and horizon, averaging across all locations.
  A cross is displayed at the difference in overall mean scores.
  A negative value indicates that the method with separate component weights for each quantile level outperformed the corresponding specification with weights shared across quantile levels.}
  \label{fig:wis_quantile_grouping_rel_wis}
\end{figure}

\begin{figure}[H]
  \includegraphics[width=\textwidth]{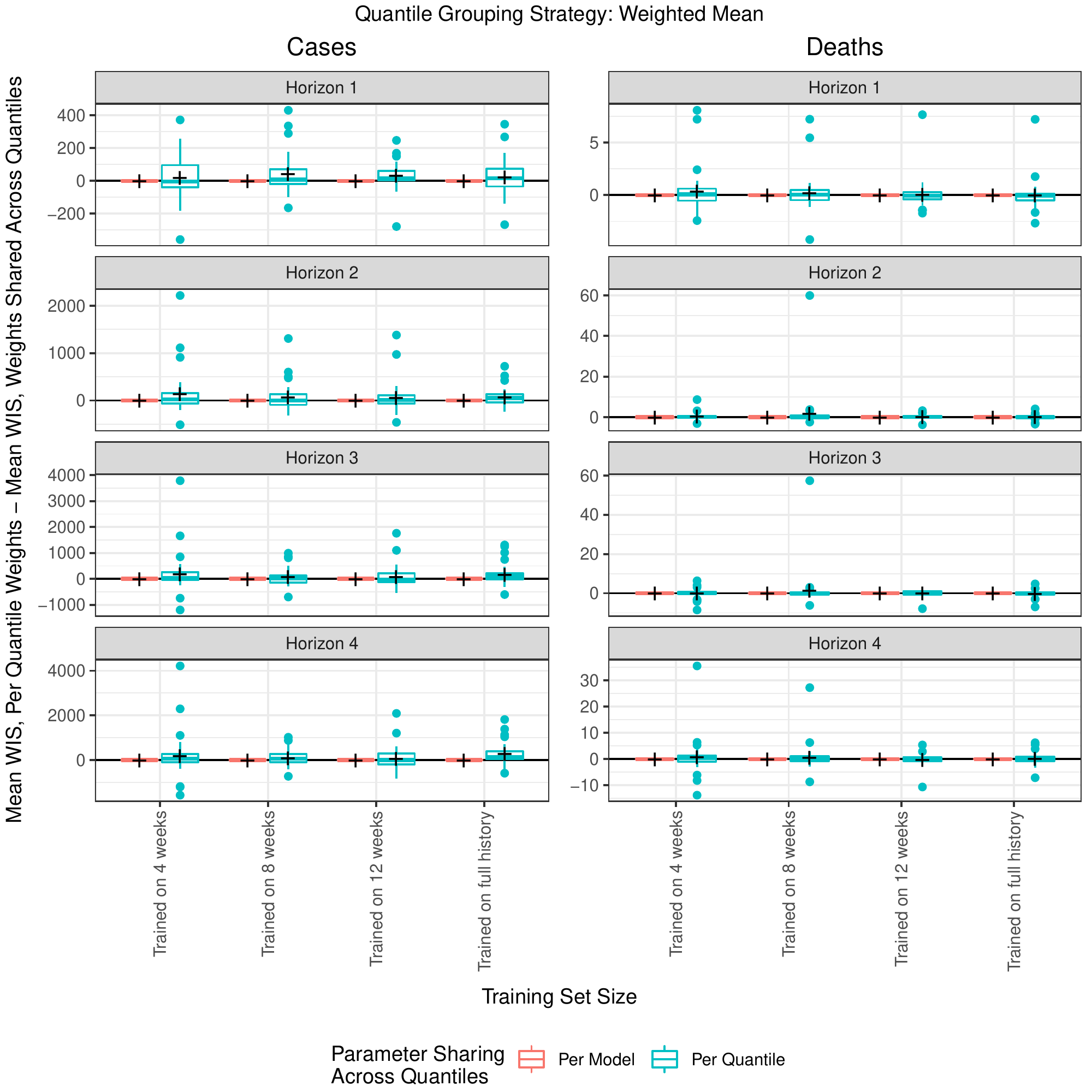}
  \caption{Boxplots summarizing forecast skill for forecasts of weekly cases from a convex weighted mean ensemble with directly estimated weights, varying whether model weights are shared across all quantile levels (``Per Model") or are estimated separately for each quantile level (``Per Quantile").
  The vertical axis is the difference in mean skill for the given ensemble specification when component weights are shared across all quantile levels and the same specification with separate component weights for each quantile level.
  The boxplots summarize the distribution of these differences for each combination of forecast date and horizon, averaging across all locations.
  A cross is displayed at the difference in overall mean scores.
  A negative value indicates that the method with separate component weights for each quantile level outperformed the corresponding specification with weights shared across quantile levels.}
  \label{fig:wis_quantile_grouping_convex}
\end{figure}

\begin{figure}
  \includegraphics[width=\textwidth]{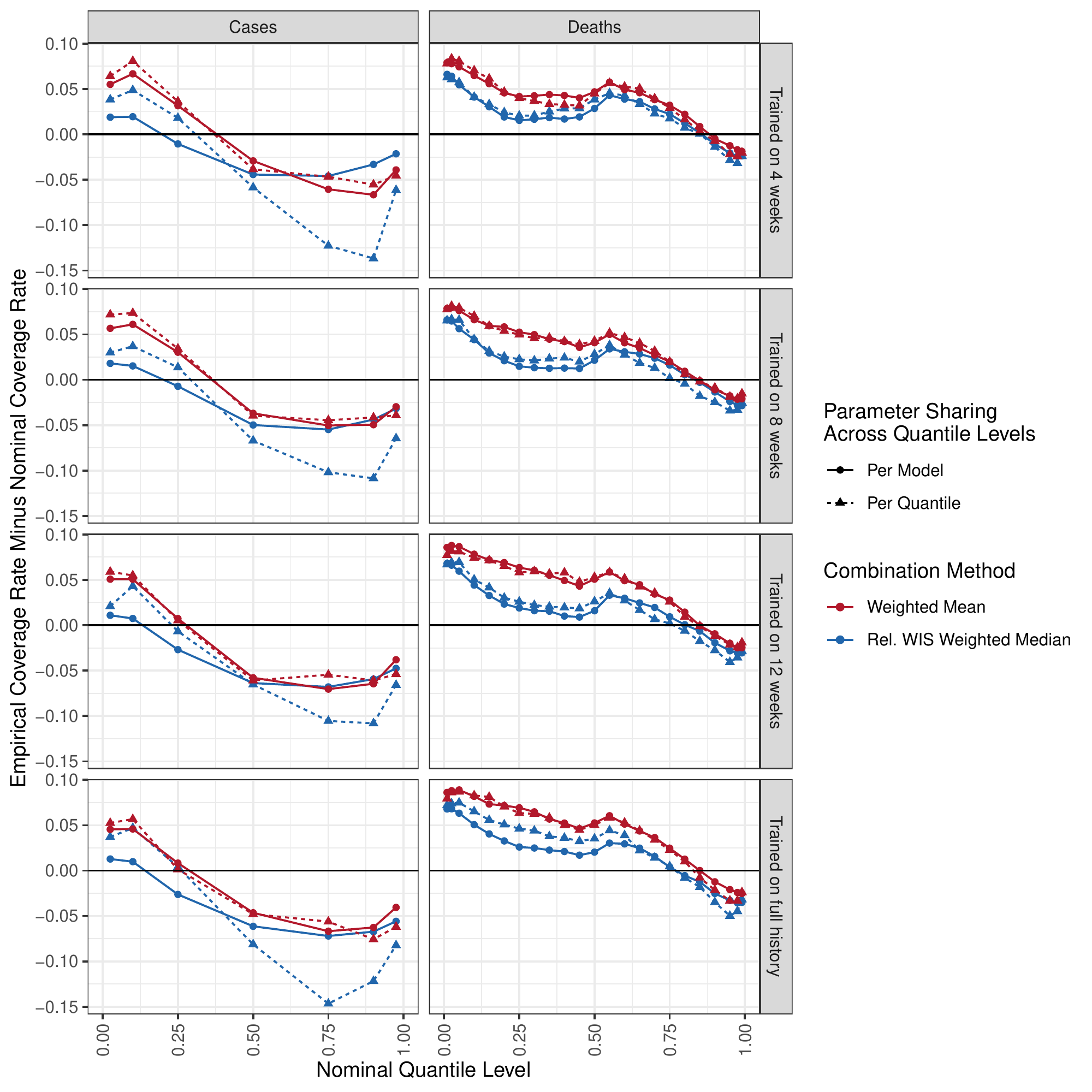}
  \caption{Quantile coverage rates for the convex weighted mean ensemble and the relative WIS weighted median ensemble, varying whether weights are estimated separately per quantile level (``Per Quantile") or shared across all quantile levels (``Per Model"). All methods combine the top 10 component forecasters in the training set window size specified in facet rows. The vertical axis is the difference between the empirical coverage rate and the nominal coverage rate. A well calibrated method would have a difference of 0 between the empirical and nominal coverage rates, and a method that generates conservative (wide) interval forecasts would have a negative difference for quantile levels less than 0.5 and a positive difference for quantile levels greater than 0.5.}
  \label{fig:coverage_quantile_grouping}
\end{figure}

Supplemental Figures~\ref{fig:wis_quantile_grouping_rel_wis} and \ref{fig:wis_quantile_grouping_convex} summarize the WIS of these methods on the model development set, comparing these approaches to the corresponding methods with a single weight per model that is shared across all quantile levels. Supplemental Figure~\ref{fig:coverage_quantile_grouping} displays the probabilistic calibration of these model variations in terms of one-sided coverage rates for predictive quantiles. In this experiment, all methods combine the top ten component forecasters; we consider varying training set window sizes.

Allowing for separate parameters per quantile level led to worse mean WIS. Additionally, for both cases and deaths, the per-quantile weighting schemes led to generally narrower predictive distributions with worse calibration in the tails. This effect was much stronger for forecasts of cases than forecasts of deaths, and it was stronger for the relative WIS weighted median ensemble than the convex weighted mean ensemble.

\newpage

\section{Performance of trained methods near local peaks}

We identified local peaks in state level weekly cases and deaths as weeks that had the maximum incidence within a centered rolling window of 11 weeks (i.e., weeks that had the largest reported weekly counts among the preceding five and following five weeks). By visual inspection, we made manual adjustments to the weeks identified using this rule to remove outliers and weeks that did not correspond to a visually distinct peak. The resulting weeks identified as local peaks are shown in Supplemental Figures~\ref{fig:peak_cases_us_identified} and \ref{fig:peak_deaths_us_identified}. Within our evaluation time frame, there were 159 local peaks for weekly cases and 146 local peaks for weekly deaths across all locations.

\begin{figure}
  \includegraphics[width=\textwidth]{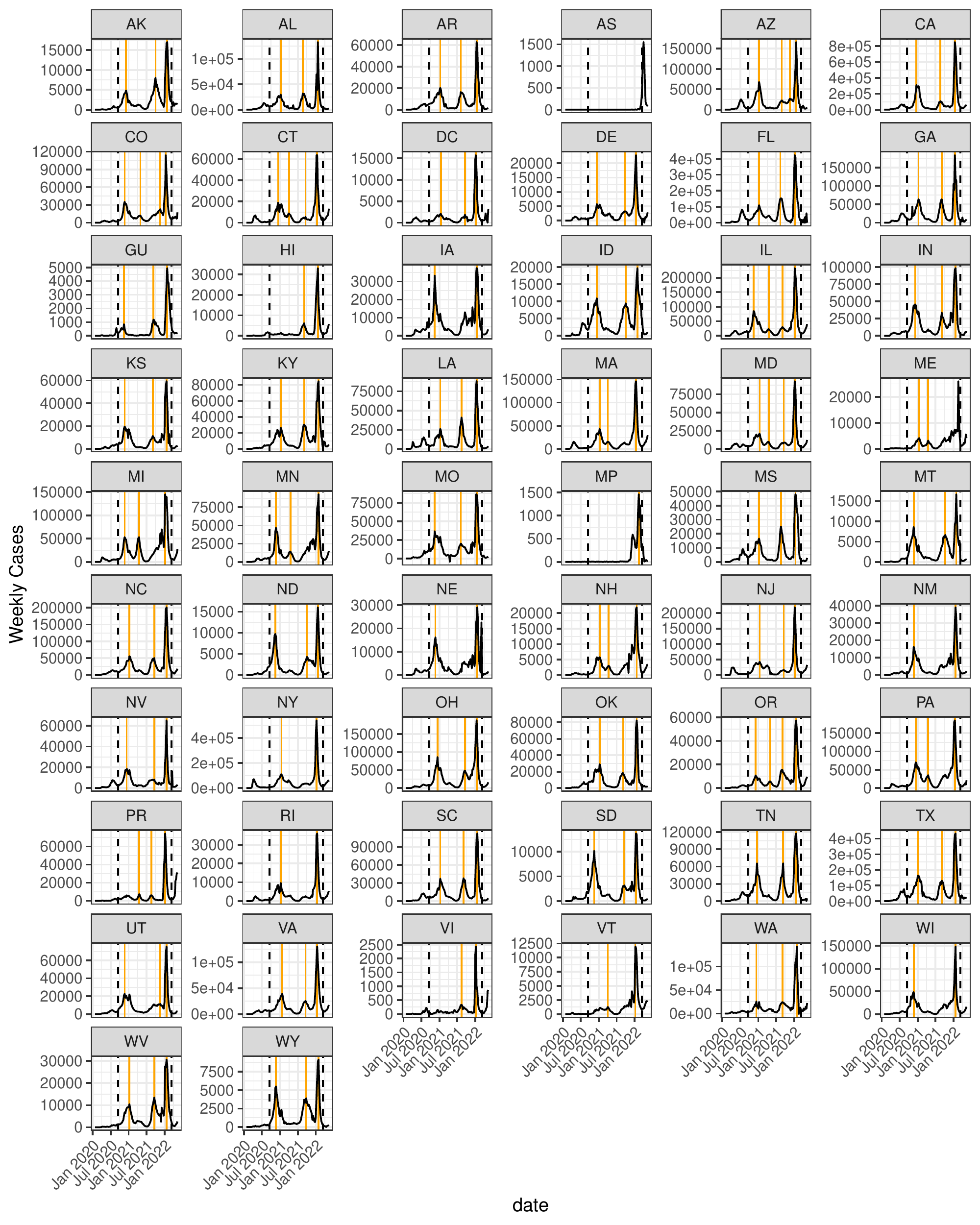}
  \caption{Identified local peaks for weekly cases at the state level. Vertical dashed lines indicate the boundaries of the evaluation phase (the combined model development and prospective evaluation phases). Vertical orange lines indicate the locations of identified local peaks.}
  \label{fig:peak_cases_us_identified}
\end{figure}

\begin{figure}
  \includegraphics[width=\textwidth]{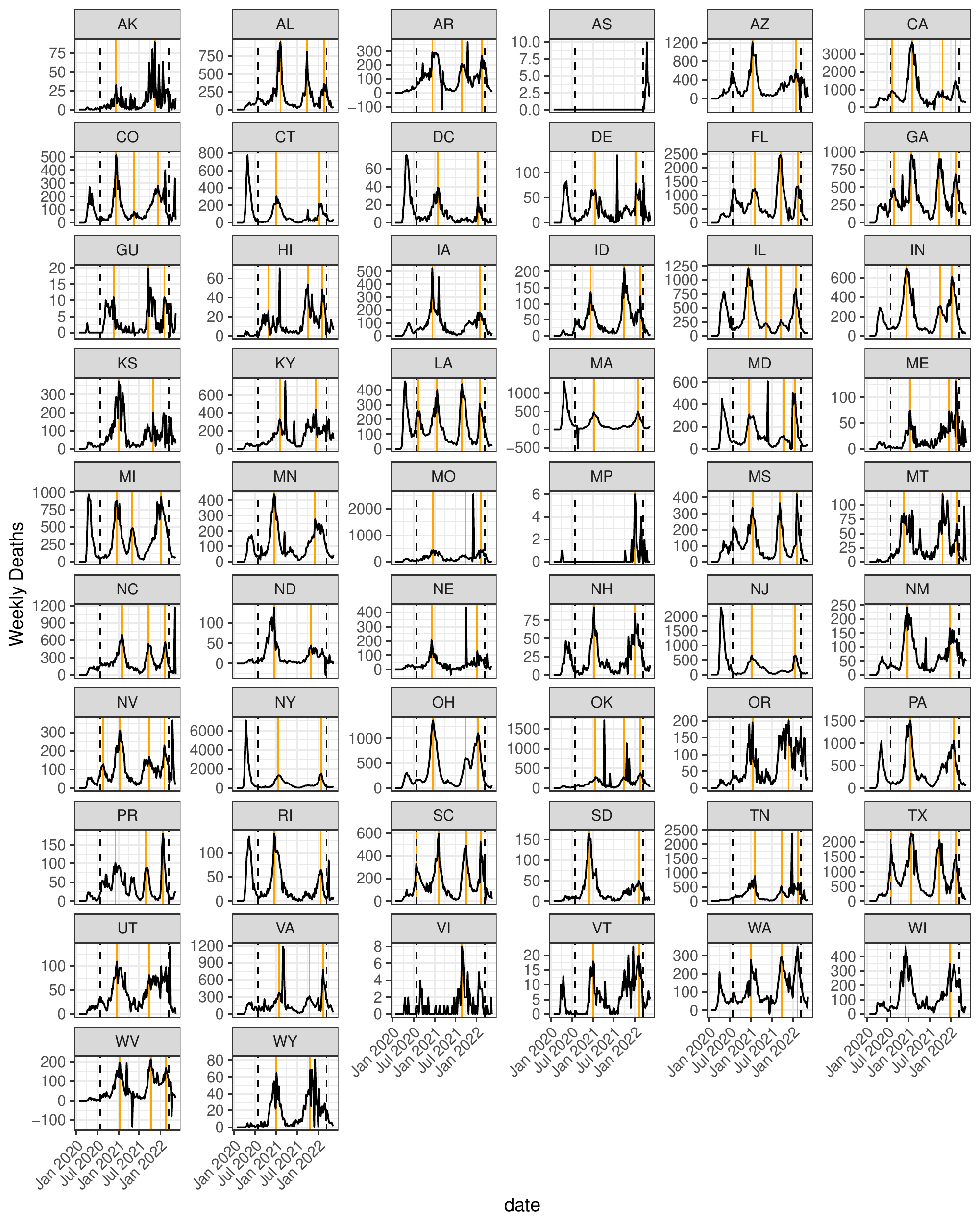}
  \caption{Identified local peaks for weekly deaths at the state level. Vertical dashed lines indicate the boundaries of the evaluation phase (the combined model development and prospective evaluation phases). Vertical orange lines indicate the locations of identified local peaks.}
  \label{fig:peak_deaths_us_identified}
\end{figure}

\begin{figure}[H]
  \includegraphics[width=\textwidth]{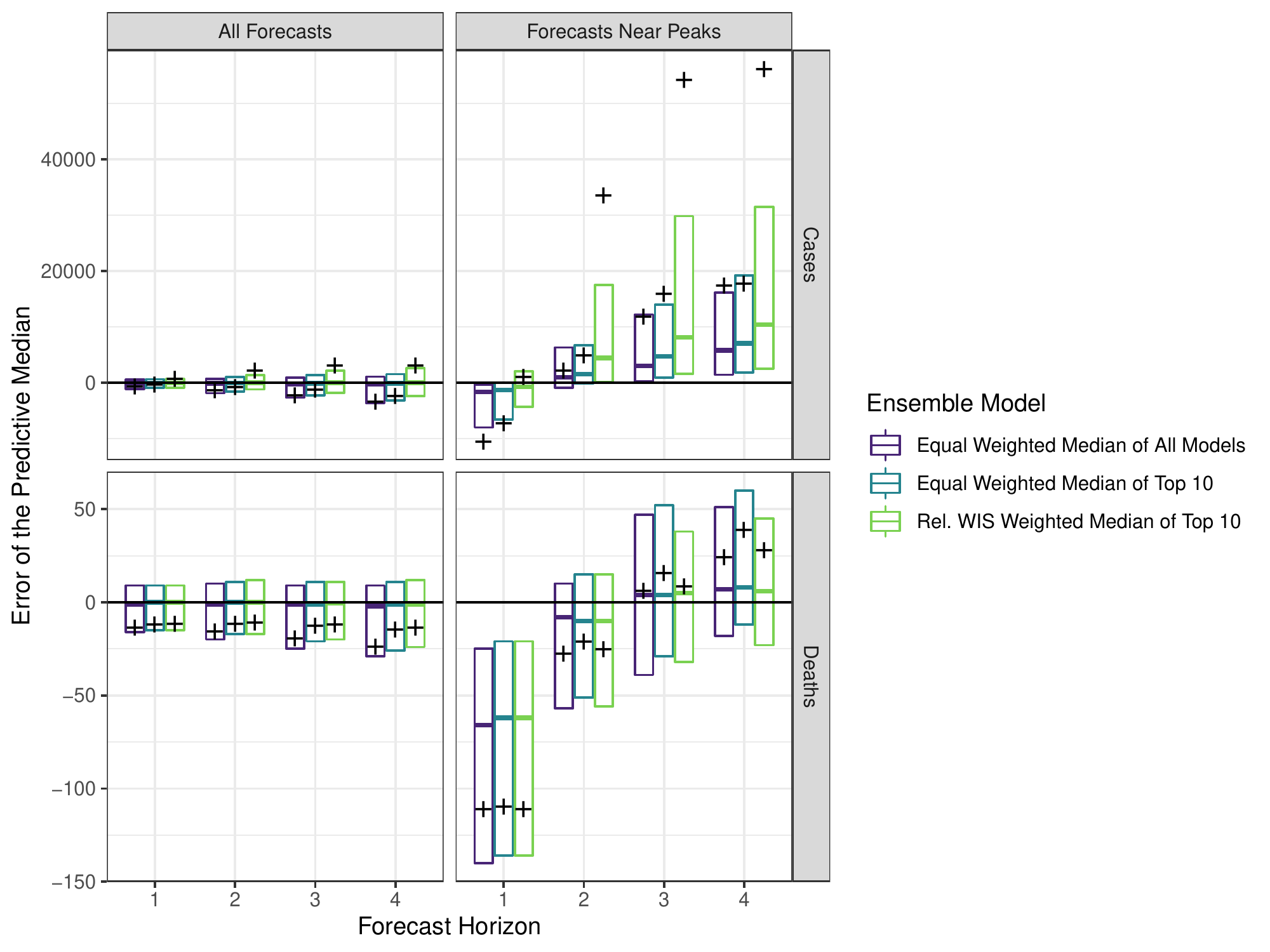}
  \caption{Errors of the predictive median across all forecasts, and for forecasts issued the week before a local peak in weekly cases or deaths. For forecasts issued before a peak, the one week ahead forecasts are for cases in the week of the local peak and forecasts at longer horizons are for cases in weeks after the local peak. The vertical axis is the difference between the predictive median and the observed value. A positive value indicates that the predictive median was larger than the eventually observed value, and a negative value indicates that the predictive median was less than the eventually observed value; a difference of zero is best. Boxes summarize the 25th percentile, median, and 75th percentile of these errors, and crosses show the mean error.}
  \label{fig:peak_case_death_forecast_errors_by_horizon_central_only}
\end{figure}

Supplemental Figure~\ref{fig:peak_case_death_forecast_errors_by_horizon_central_only} summarizes the central tendency of the errors for predictive medians across all forecasts at the state level in the U.S.\ and for just those forecasts issued in the week before a local peak. The errors are calculated as the value of the predictive median minus the observed count of cases or deaths in the target week, so that errors close to zero are preferred. We show results for three ensemble specifications: the equally weighted median of all component forecasters, the equally weighted median of the top 10 forecasters, and the relative WIS weighted median of the top 10 forecasters. Note that the equally weighted median of all components is untrained, the relative WIS weighted median of the top 10 forecasters is trained, and the equally weighted median of the top 10 forecasters represents an intermediate strategy: it is also trained, but it is not as strongly adaptive to component performance as the weighted median. Both trained ensembles use a training set window size of 12 weeks. We summarize our observations about these errors as follows:
\begin{enumerate}
    \item Across all forecasts of cases, the median error is similar for all three strategies, but the magnitude of the average error from the relative WIS weighted median is slightly larger than the magnitude of the average error from the equally weighted median of all components.
    \item Across all forecasts of deaths, the median error is similar for all three strategies, but the magnitude of the average errors from the relative WIS weighted median is slightly smaller than the magnitude of the average error from the equally weighted median of all components.
    \item For forecasts of cases near peaks, the trained methods were better than the untrained approach in the peak week, but have much larger errors at longer horizons. This indicates that forecasts from the trained methods ``overshot" and missed the turning points by a larger margin than the untrained method.
    \item For forecasts of deaths near peaks, the trained methods have comparable performance to the untrained methods. In contrast to forecasts of cases, training did not exacerbate the tendency to overshoot near local peaks when forecasting deaths.
\end{enumerate}

\begin{figure}
  \includegraphics[width=\textwidth]{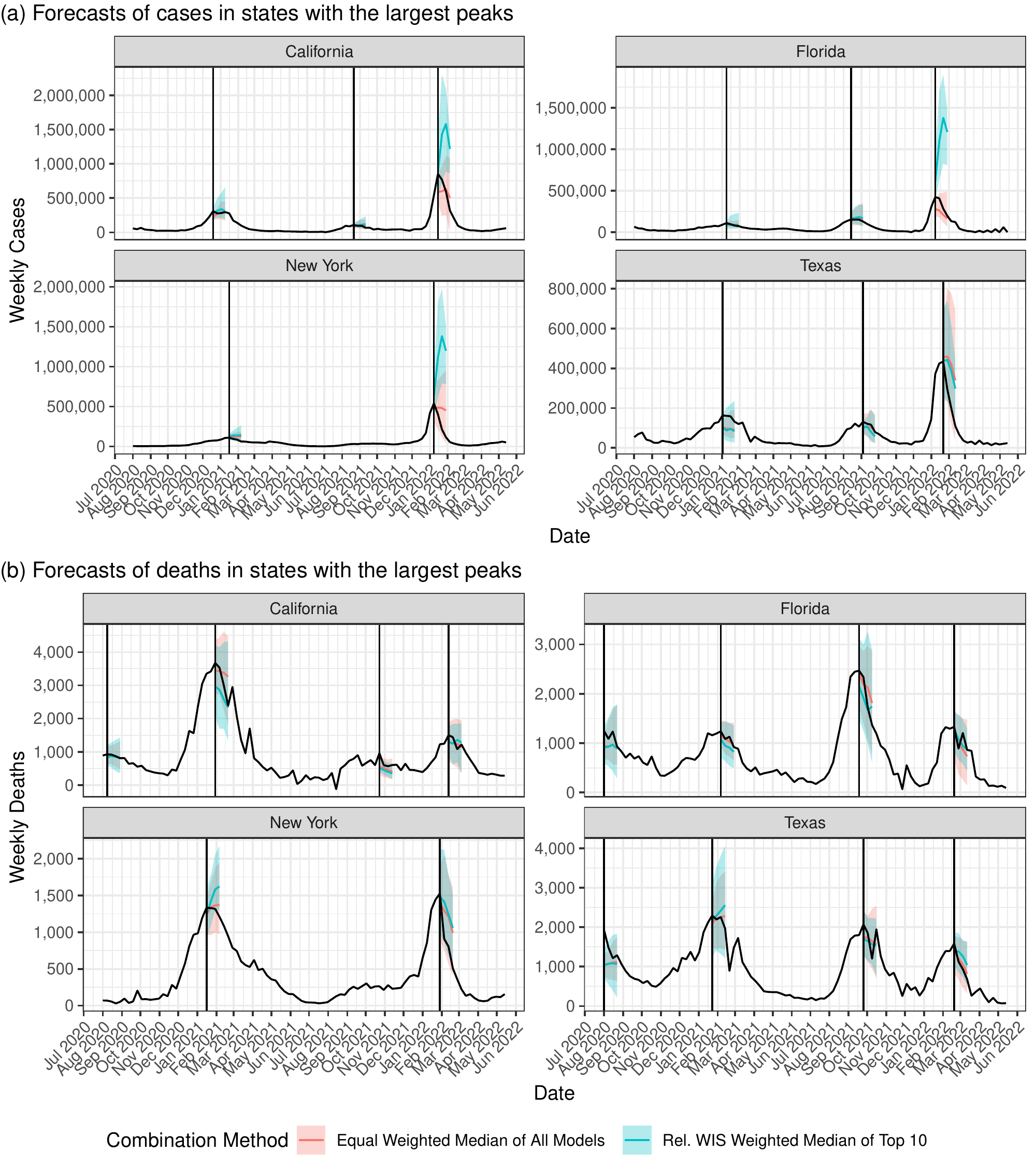}
  \caption{Forecast distributions for weekly cases and deaths issued the week before a local peak. Forecasts are shown for all peaks during the evaluation period in the four states with the largest local peaks. Vertical lines indicate the time of the peak, corresponding to the date of a one-week-ahead forecast.}
  \label{fig:peak_case_death_forecast_examples}
\end{figure}

Supplemental Figure~\ref{fig:peak_case_death_forecast_examples} shows illustrative examples of forecasts made the week before a local peak from the equally weighted median of all components and the relative WIS weighted median of the top 10 components. We can see a systematic tendency for the predictions from the trained method near local peaks to predict a continuation of rising trends for cases, whereas the forecasts of deaths more often capture the coming downturn.

\newpage

\section{Characteristics of component weights in the post hoc weighted mean ensemble}

\begin{figure}
  \includegraphics[width=\textwidth]{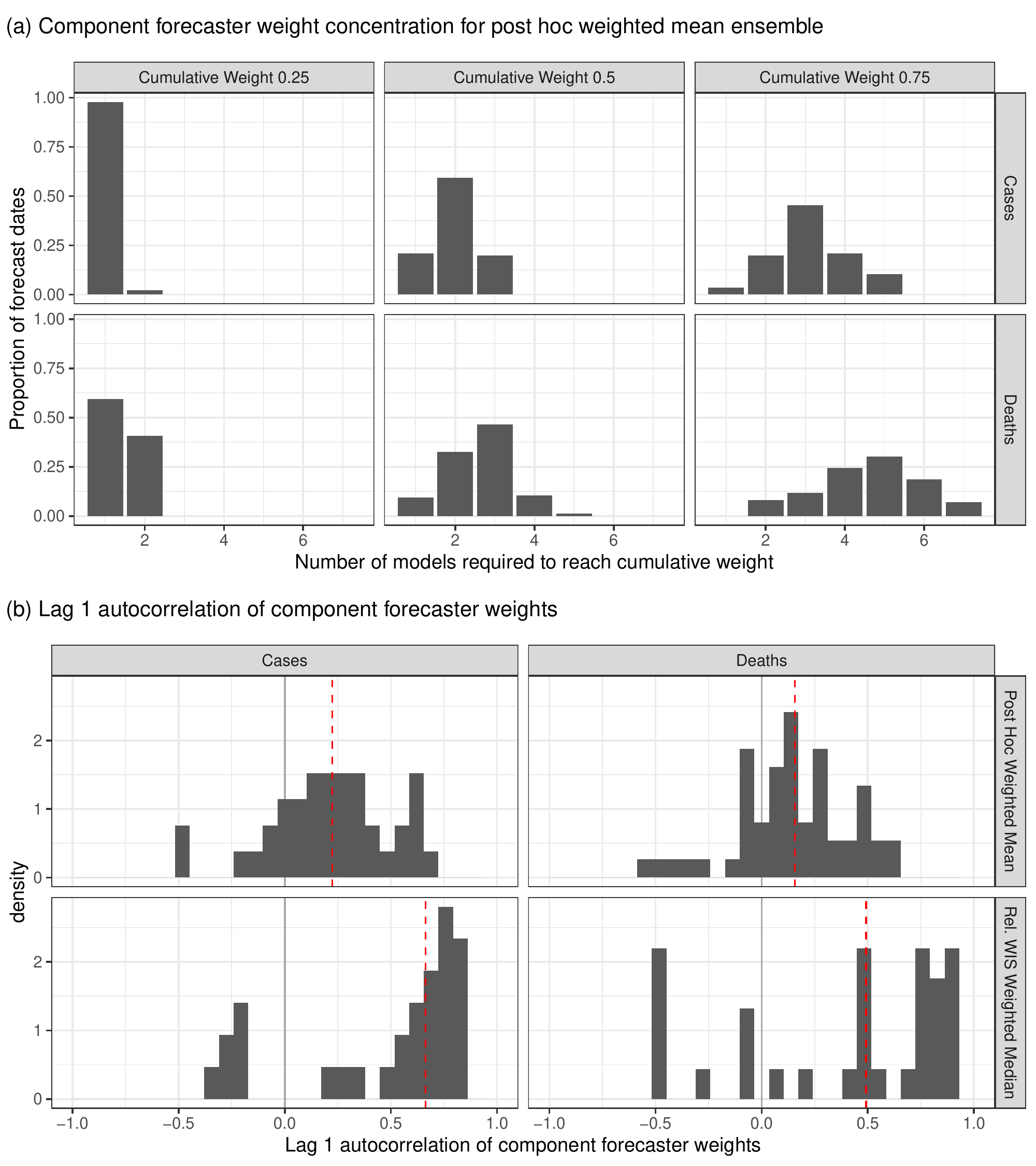}
  \caption{Characteristics of the component forecaster weights in the post hoc weighted mean ensemble and the relative WIS weighted mean ensemble used in the prospective analysis. Panel (a) shows the number of component forecasters that were required to reach a specified cumulative weight in the post hoc weighted mean ensemble. For example, for cases, in about 95\% of forecast dates at least one component forecaster got weight 0.25 or greater. Panel (b) shows the lag 1 autocorrelation of weight assigned to each component forecaster across weeks. A vertical dashed line shows the mean autocorrelation across all component forecasters.}
  \label{fig:post_hoc_weight_characteristics}
\end{figure}

Supplemental Figure~\ref{fig:post_hoc_weight_characteristics} summarizes characteristics of the component forecaster weights that were obtained from the post hoc weighted mean ensemble. Panel (a) illustrates that in general, the weights were distributed across a larger number of component forecasters in forecasts of deaths than in forecasts of cases. Panel (b) illustrates that the component forecaster weights were only weakly autocorrelated in the post hoc weighted mean ensemble, quantifying the observation that the weights changed substantially from week to week; this can be seen in Figures 4 and 5 in the main text. In contrast, the component forecaster weights were more strongly autocorrelated in the relative WIS weighted mean ensemble. In part, this is due to the use of a 12 week rolling window for estimating component weights; much of the training data for weight estimation is shared in consecutive weeks.

\newpage

\section{Post hoc evaluation of component weight regularization}

\begin{figure}
  \includegraphics[width=\textwidth]{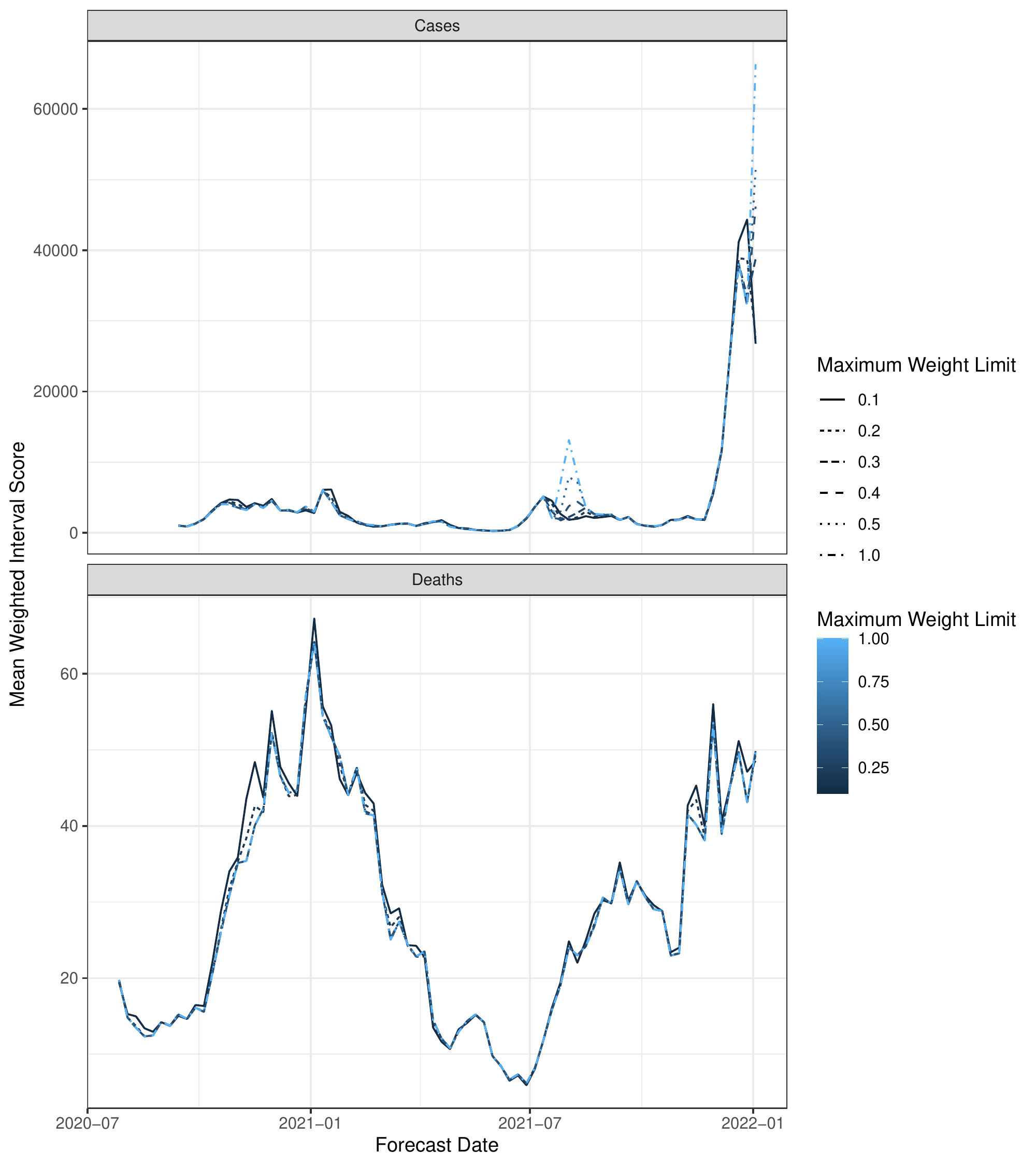}
  \caption{Mean WIS for state level forecasts of weekly cases and deaths in the United States, where averages for each forecast date are calculated across all locations and forecast horizons. Results are shown for six variations on relative WIS weighted median ensembles that combine the top 10 component forecasters based on a rolling 12 week training set, using different limits for the maximum weight that can be assigned to any component forecaster. A maximum weight limit of 1.0 corresponds to the unregularized approach considered in the prospective analysis in the main text, while a maximum weight limit of 0.1 corresponds to an equal weighting of the ten selected forecasters.}
  \label{fig:compare_max_weight_limits_by_date}
\end{figure}

Section 3.4 of the main text describes a post hoc evaluation of regularizing the component forecaster weights in trained ensembles by imposing a limit on the maximum weight that can be assigned to any component. Supplemental Figure~\ref{fig:compare_max_weight_limits_by_date} illustrates that for most forecast dates, imposing this limit on the maximum component weight has negligible effects on the WIS of the ensemble forecasts. However, there are a few forecast dates, particularly concentrated near local peaks for cases, where regularization has a larger impact, with some regularization being helpful for reducing the mean WIS.

\newpage

\section{Differences in forecast missingness in the U.S. and Europe}

\begin{figure}
  \includegraphics[width=\textwidth]{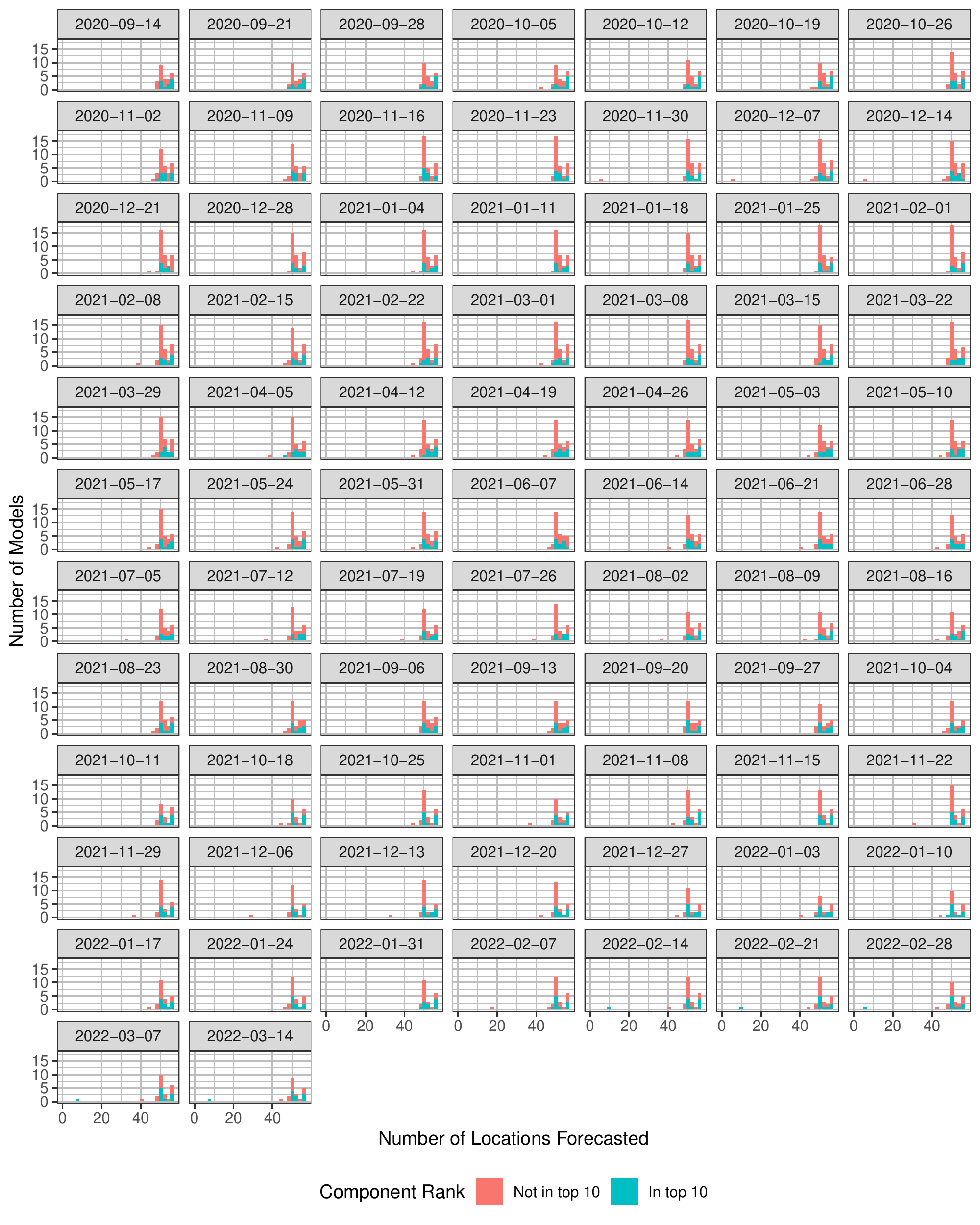}
  \caption{Histograms of the number of locations forecasted by each contributing forecaster for weekly cases in the U.S. The top 10 forecasters, indicated with blue shading, were selected for inclusion in the weighted ensembles used for prospective evaluation.}
  \label{fig:case_us_num_locations}
\end{figure}

\begin{figure}
  \includegraphics[width=\textwidth]{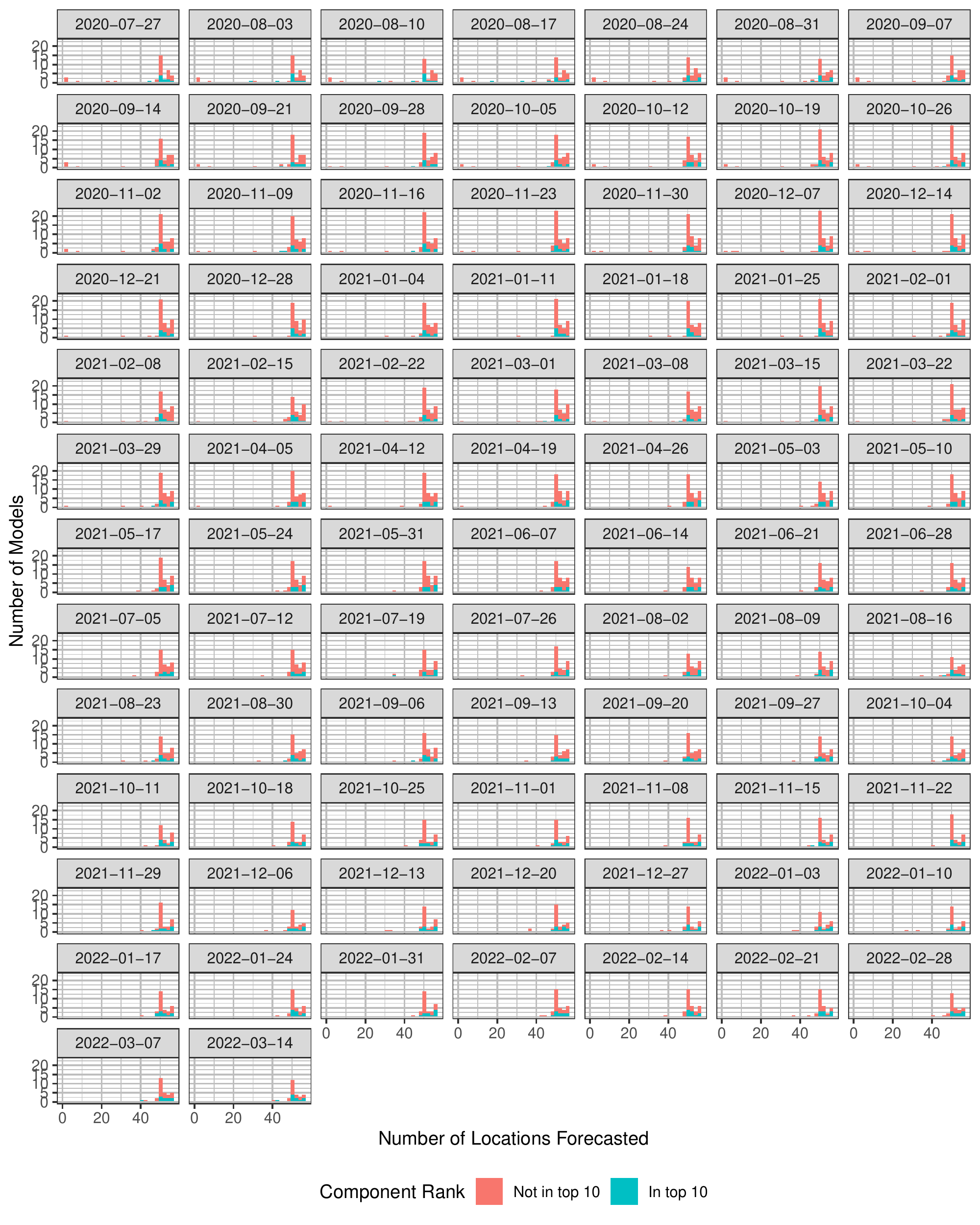}
  \caption{Histograms of the number of locations forecasted by each contributing forecaster for weekly deaths in the US. The top 10 forecasters, indicated with blue shading, were selected for inclusion in the weighted ensembles used for prospective evaluation.}
  \label{fig:death_us_num_locations}
\end{figure}

\begin{figure}
  \includegraphics[width=\textwidth]{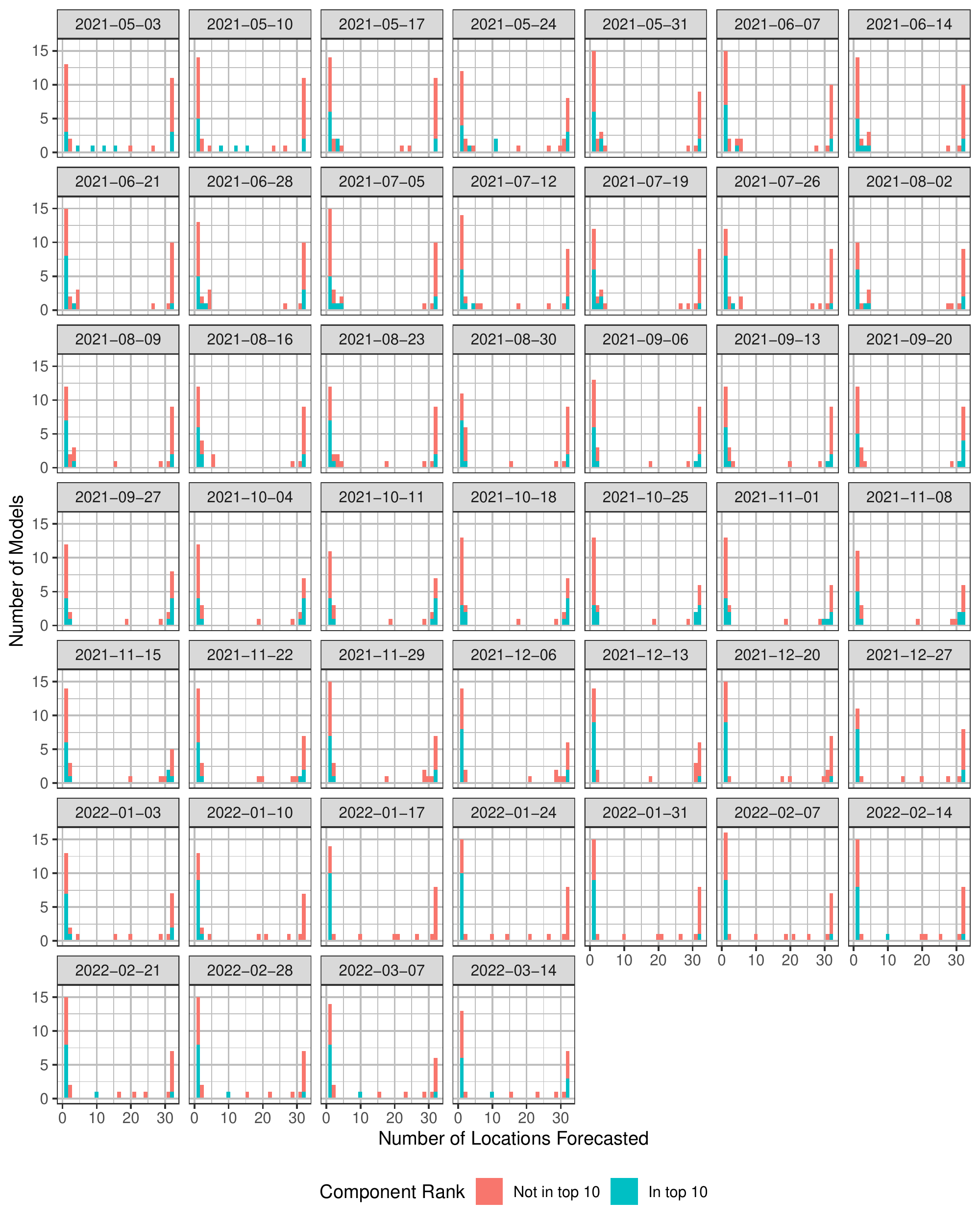}
  \caption{Histograms of the number of locations forecasted by each contributing forecaster for weekly cases in Europe. The top 10 forecasters, indicated with blue shading, were selected for inclusion in the weighted ensembles used for prospective evaluation.}
  \label{fig:case_eu_num_locations}
\end{figure}

\begin{figure}
  \includegraphics[width=\textwidth]{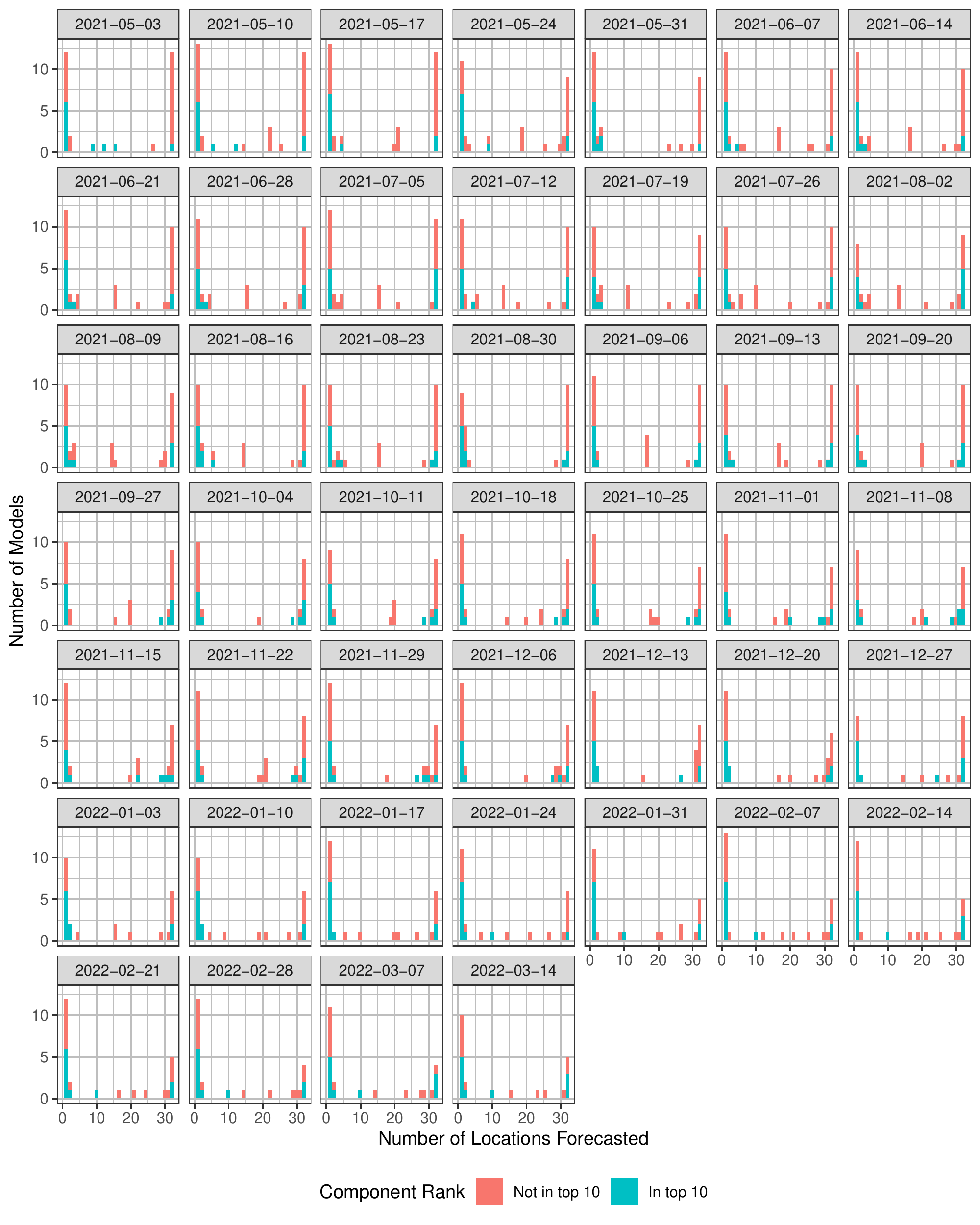}
  \caption{Histograms of the number of locations forecasted by each contributing forecaster for weekly deaths in Europe. The top 10 forecasters, indicated with blue shading, were selected for inclusion in the weighted ensembles used for prospective evaluation.}
  \label{fig:death_eu_num_locations}
\end{figure}

Supplemental Figures~\ref{fig:case_us_num_locations} through \ref{fig:death_eu_num_locations} show histograms of the number of locations forecasted by the component models contributing to the U.S. COVID-19 Forecast Hub and the European COVID-19 Forecast Hub. Patterns of submission are starkly different for the U.S. and the EU. In the U.S., nearly all models submit forecasts for at least the 50 US States, and many additionally submit forecasts for the District of Columbia and US territories. In the EU, roughly half of forecasters provide forecasts for all or most European countries, while the other half provide forecasts for only a few countries.

\begin{figure}
  \includegraphics[width=\textwidth]{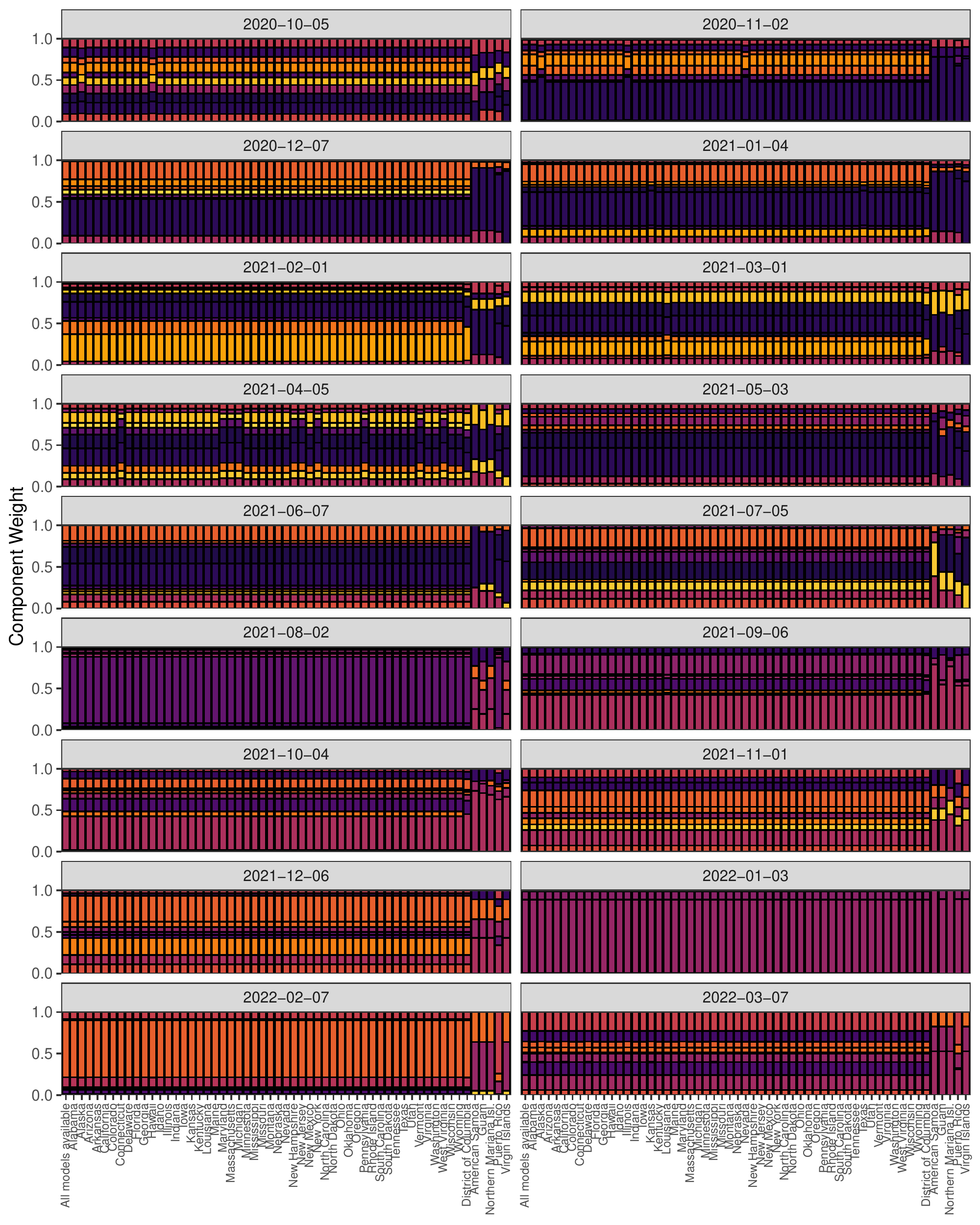}
  \caption{Component weights for forecasts of weekly cases in the U.S., facetted by forecast date. Only the weights for the first week in each month are shown due to space constraints. The estimated weights that would be used if all models were available for a particular location are shown at left within each facet. The weights actually used for each location are obtained by setting the weight for components that are missing forecasts for that location to 0 and rescaling the others proportionally so that they sum to 1.}
  \label{fig:case_us_effective_weights}
\end{figure}

\begin{figure}
  \includegraphics[width=\textwidth]{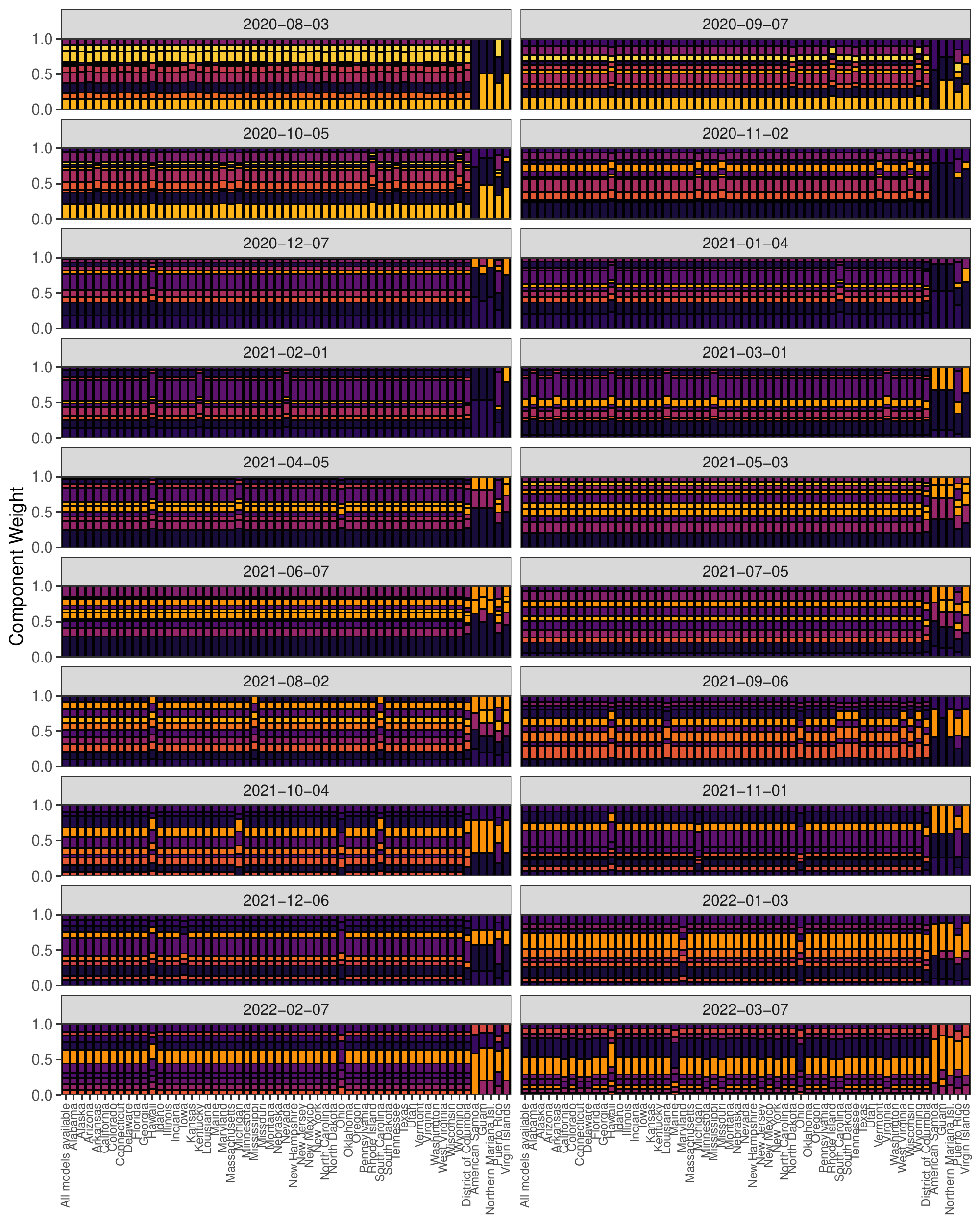}
  \caption{Component weights for forecasts of weekly deaths in the U.S., facetted by forecast date. Only the weights for the first week in each month are shown due to space constraints. The estimated weights that would be used if all models were available for a particular location are shown at left within each facet. The weights actually used for each location are obtained by setting the weight for components that are missing forecasts for that location to 0 and rescaling the others proportionally so that they sum to 1.}
  \label{fig:death_us_effective_weights}
\end{figure}

\begin{figure}
  \includegraphics[width=\textwidth]{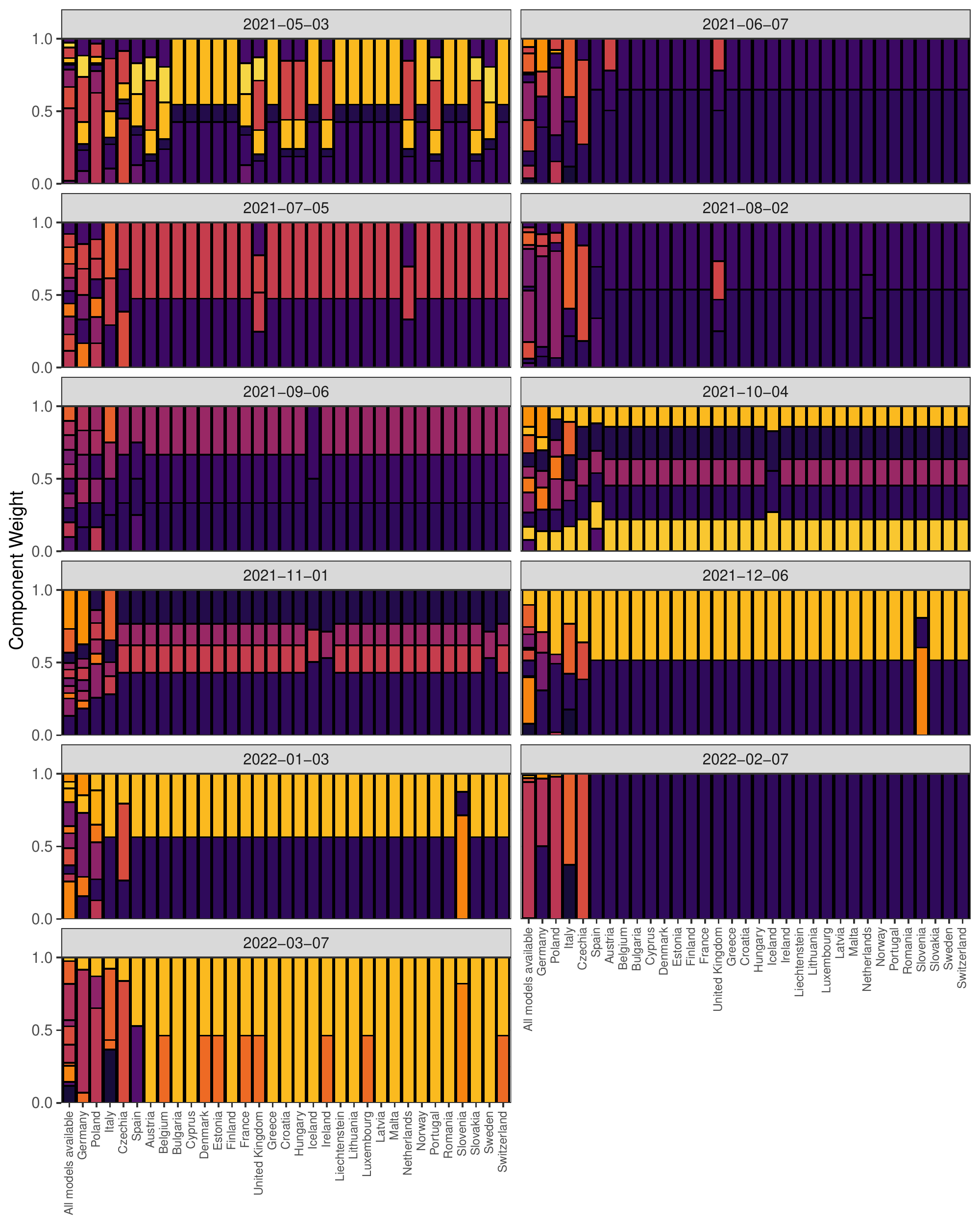}
  \caption{Component weights for forecasts of weekly cases in Europe, facetted by forecast date. Only the weights for the first week in each month are shown due to space constraints. The estimated weights that would be used if all models were available for a particular location are shown at left within each facet. The weights actually used for each location are obtained by setting the weight for components that are missing forecasts for that location to 0 and rescaling the others proportionally so that they sum to 1.}
  \label{fig:case_eu_effective_weights}
\end{figure}

\begin{figure}
  \includegraphics[width=\textwidth]{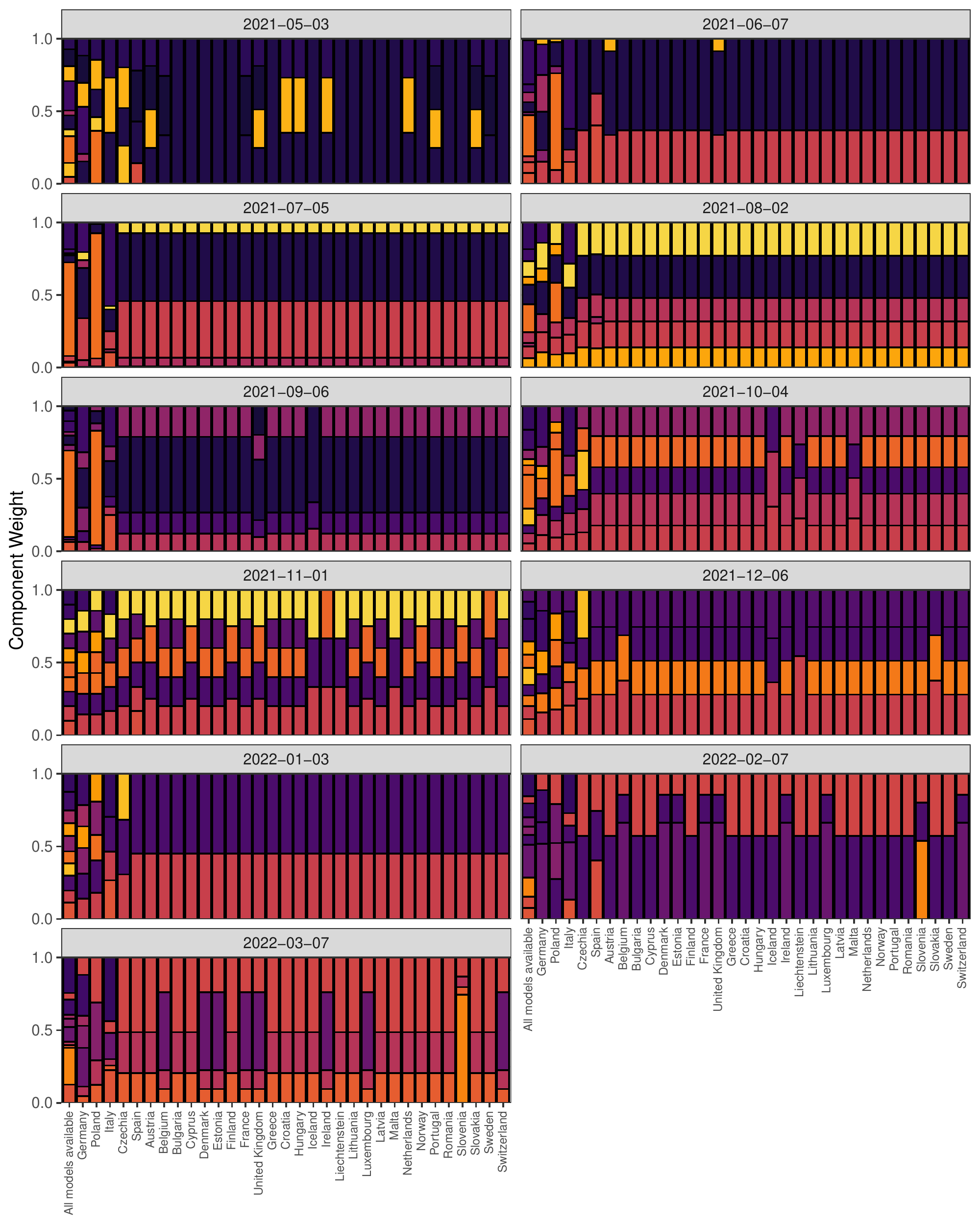}
  \caption{Component weights for forecasts of weekly deaths in Europe, facetted by forecast date. Only the weights for the first week in each month are shown due to space constraints. The estimated weights that would be used if all models were available for a particular location are shown at left within each facet. The weights actually used for each location are obtained by setting the weight for components that are missing forecasts for that location to 0 and rescaling the others proportionally so that they sum to 1.}
  \label{fig:death_eu_effective_weights}
\end{figure}

Supplemental Figures~\ref{fig:case_us_effective_weights} through \ref{fig:death_eu_effective_weights} show the effective weights used in each location after accounting for forecast missingness by rescaling the weights assigned to available models so that they sum to one. In the U.S., the effective weights closely match the nominal estimated weights in nearly all states, differing only slightly in the territories. In the EU, missingness is more prevalent and it is common for only a few of the selected top 10 component forecasters to provide forecasts for many countries.

\newpage

\section{Adherence to EPIFORGE Guidelines}

\begin{longtable}{c  c  p{9cm}  c}
\toprule
Section & Item & Item Description & Pages \\
\midrule
\endhead
Title/Abstract & 1 & Study described as a forecast or prediction research in at least the title or abstract & 1 \\
\midrule
Introduction & 2 & Purpose of study and forecasting targets defined & 4 \\
\midrule
Methods & 3 & Methods fully documented & 6-13 \\
\midrule
Methods & 4 & Identify whether the forecast was performed prospectively, in real-time, and/or retrospectively & 7 \\
\midrule
Methods & 5 & Origin of input source data explicitly described with reference & 6 \\
\midrule
Methods & 6 & Source data made available, or reasons why this was not possible documented & 13 \\
\midrule
Methods & 7 & Input data processing procedures described in detail & 6-9 \\
\midrule
Methods & 8 & Statement and description of model type, with model assumptions documented with references & 11-13 \\
\midrule
Methods & 9 & Model code made available, or reasons why this was not possible documented & 13 \\
\midrule
Methods & 10 & Description of model validation, with justification of approach. & 6-13 \\
\midrule
Methods & 11 & Description of forecast accuracy evaluation method, with justification & 9-11 \\
\midrule
Methods & 12 & Where possible, compare model results to a benchmark or other comparator model, with justification of comparator choice & 9 \\
\midrule
Methods & 13 & Description of forecast horizon, and justification of its length & 6 \\
\midrule
Results & 14 & Uncertainty of forecasting results presented and explained & 8, 15, 16, 20-22 \\
\midrule
Results & 15 & Results briefly summarized in lay terms, including a lay interpretation of forecast uncertainty & - \\
\midrule
Results & 16 & If results are published as a data object, encourage a time-stamped version number & 13 \\
\midrule
Discussion & 17 & Limitations of forecast described, including limitations specific to data quality and methods & 23, 26, 27 \\
\midrule
Discussion & 18 & If the research is applicable to a specific epidemic, comment on its potential implications and impact for public health action and decision making & 28 \\
\midrule
Discussion & 19 & If the research is applicable to a specific epidemic, comment on how generalizable it may be across populations & 26, 28 \\
\bottomrule
\end{longtable}

\newpage

\bibliography{bibfile}